\pdfoutput=1
\RequirePackage{ifpdf}
\ifpdf 
\documentclass[pdftex]{sigma}
\else
\documentclass{sigma}
\fi

\numberwithin{equation}{section}

\numberwithin{theorem}{section}
\numberwithin{proposition}{section}
\numberwithin{lemma}{section}
\numberwithin{corollary}{section}
\numberwithin{definition}{section}
\numberwithin{example}{section}
\numberwithin{remark}{section}
\numberwithin{note}{section}

\usepackage{etex}
\usepackage{bbm}
\usepackage{enumerate}
\usepackage{enumitem}
\usepackage{cancel}
\usepackage{easybmat}
\usepackage[all]{xy}
\usepackage{pstricks-add}

\newcommand{\vs}{\vspace}

\newcommand{\comp}{\prec\!\succ}
\newcommand{\ncomp}{~\!\cancel{\prec\!\succ}\!~}
\newcommand{\nD}{~\!\cancel{\mc D}\!~}

\newcommand{\mc}{\mathcal}
\newcommand{\mf}{\mathfrak}
\newcommand{\bb}{\mathbbm}

\newcommand{\C}{\mathbb C}
\newcommand{\N}{\mathbb N}
\newcommand{\R}{\mathbb R}
\newcommand{\Z}{\mathbb Z}

\renewcommand{\1}{\bb1}

\renewcommand{\a}{\alpha}
\renewcommand{\b}{\beta}

\newcommand{\dd}{\mathrm d}
\newcommand{\D}{\Delta}
\newcommand{\e}{\epsilon}
\newcommand{\ee}{\mathrm e}
\newcommand{\End}{\mathrm{End}}
\newcommand{\I}{\mathbb I}
\renewcommand{\i}{\mathrm i}

\newcommand{\J}{\mathbb J}
\newcommand{\K}{\mathbb K}
\renewcommand{\k}{\kappa}
\renewcommand{\l}{\lambda}
\renewcommand{\L}{\Lambda}

\newcommand{\PP}{\mathbb P}

\newcommand{\s}{\sigma}
\newcommand{\Si}{\Sigma}
\renewcommand{\t}{\theta}

\newcommand{\g}{\gamma}

\newcommand{\m}{\mu}

\begin{document}
\allowdisplaybreaks

\renewcommand{\PaperNumber}{064}

\FirstPageHeading

\ShortArticleName{Classif\/ication of Non-Af\/f\/ine Non-Hecke Dynamical $R$-Matrices}

\ArticleName{Classif\/ication of Non-Af\/f\/ine Non-Hecke\\ Dynamical $\boldsymbol{R}$-Matrices}

\Author{Jean AVAN~$^\dag$, Baptiste BILLAUD~$^\ddag$ and Genevi\`eve ROLLET~$^\dag$}

\AuthorNameForHeading{J.~Avan, B.~Billaud and G.~Rollet}

\Address{$^\dag$~Laboratoire de Physique Th\'eorique et Mod\'elisation,\\
\hphantom{$^\dag$}~Universit\'e de Cergy-Pontoise (CNRS UMR 8089), Saint-Martin 2,\\
\hphantom{$^\dag$}~2, av.~Adolphe Chauvin, F-95302 Cergy-Pontoise Cedex, France}

\EmailD{\href{mailto:avan@u-cergy.fr}{avan@u-cergy.fr}, \href{mailto:rollet@u-cergy.fr}{rollet@u-cergy.fr}}

\Address{$^\ddag$~Laboratoire de Math\'ematiques ``Analyse, G\'eometrie Mod\'elisation'',\\
\hphantom{$^\ddag$}~Universit\'e de Cergy-Pontoise (CNRS UMR 8088), Saint-Martin 2,\\
\hphantom{$^\ddag$}~2, av.~Adolphe Chauvin, F-95302 Cergy-Pontoise Cedex, France}
\EmailD{\href{mailto:bbillaud@u-cergy.fr}{bbillaud@u-cergy.fr}}

\ArticleDates{Received April 24, 2012, in f\/inal form September 19, 2012; Published online September 28, 2012}

\Abstract{A complete classif\/ication of non-af\/f\/ine dynamical quantum $R$-matrices obeying the $\mc Gl_n(\C)$-Gervais--Neveu--Felder equation is obtained without assuming either Hecke or weak Hecke conditions. More general dynamical dependences are observed. It is shown that any solution is built upon elementary blocks, which individually satisfy the weak Hecke condition. Each solution is in particular characterized by an arbitrary partition $\{\I(i),\;\!i\in\{1,\dots,n\}\}$ of the set of indices $\{1,\dots,n\}$ into classes, $\I(i)$ being the class of the index $i$, and an arbitrary family of signs $(\e_\I)_{\I\in\{\I(i), \; i\in\{1,\hdots,n\}\}}$ on this partition. The weak Hecke-type $R$-matrices exhibit the analytical behaviour
$R_{ij,ji}=f(\e_{\I(i)}\L_{\I(i)}-\e_{\I(j)}\L_{\I(j)})$,
where $f$ is a~particular trigonometric or rational function, $\L_{\I(i)}=\sum\limits_{j\in\I(i)}\l_j$, and $(\l_i)_{i\in\{1,\dots,n\}}$ denotes the family of dynamical coordinates.}


\Classification{16T25; 17B37; 81R12; 81R50}

\section{Introduction}
The dynamical quantum Yang--Baxter equation (DQYBE) was originally formulated by Gervais and Neveu in the context of quantum Liouville theory \cite{GN}. It was built by Felder as a quantization of the so-called modif\/ied dynamical classical Yang--Baxter equation \cite{Feh_1, Feh_2, Fe_1, Fe_2}, seen as a~compatibility condition of Knizhnik--Zamolodchikov--Bernard equations \cite{KZB_3, KZB_2, HSTF_2, HSTF_1, KZB_1}. This classical equation also arose when considering the Lax formulation of the Calogero--Moser \cite{CM_1, CM_2} and Ruijsenaar--Schneider model \cite{RS}, and particularly its $r$-matrix \cite{ABB, AT}. The DQYBE was then identif\/ied as a consistency (associativity) condition for dynamical quantum algebras. We introduce $\mf A$ as the considered (dynamical) quantum algebra and $\mc V$ as either a f\/inite-dimensional vector space $V$ or an inf\/inite-dimensional loop space $\mc V=V\otimes\C[[z,z^{-1}]]$. We def\/ine the objects $T \in\End(\mc V\otimes\mf A)$ as an algebra-value
 $d$ matrix encoding the generators of $\mf A$ and  $R\in\End(\mc V\otimes\mc V)$ as the
matrix of structure coef\/f\/icients for the quadratic exchange relations of $\mf A$
  \begin{gather}
R_{12}(\l+\g h_q)T_1(\l-\g h_2)T_2(\l+\g h_1)=T_2 (\l-\g h_1)T_1(\l+\g h_2)R_{12}(\l-\g h_q). \label{DQR}
  \end{gather}

As usual in these descriptions the indices ``$_1$'' and ``$_2$'' in the operators $R$ and $T$ label the respective so-called ``auxiliary'' spaces $\mc V$ in $\mc V\otimes\mc V) $. In addition, when the auxiliary spaces are loop spaces $\mc V=V\otimes\C[[z,z^{-1}]]$, these labels encapsulate an additional dependence as formal series in positive and negative powers of the complex variables $z_1$ and $z_2$, becoming the so-called spectral parameters when (\ref{DQR}) is represented, e.g.\ in evaluation form. Denoting by $\N^*_n$ the set $\{1,\hdots,n\}$, for any $n\in\N^*=\N\setminus\{0\}$, both $R$ and $T$  depend in addition on a f\/inite family $(\l_i)_{i\in\N^*_n}$ of $c$-number complex ``dynamical'' parameters understood as coordinates on the dual algebra $\mf h^* $ of a $n$-dimensional  complex Lie algebra $\mf h$. The term ``dynamical'' comes from the identif\/ication of these parameters in the classical limit as being the position variables in  the context of classical Calogero--Moser or Ruijsenaar--Schneider models. We shall consider here only the case of a $n$-dimensional Abelian algebra $\mf h$. Non-Abelian cases were introduced in~\cite{Xu} and extensively considered e.g. in~\cite{ DoMu, EtVar_2, EtVar_3, EtVar_1}.

Following \cite{EtVar_1}, in addition with the choosing of a basis $(h_i)_{i\in\N^*_n}$ of $\mf h$ and its dual basis $(h^i)_{i\in\N^*_n}$, being the natural basis of $\mf h^\ast$, we assume that the f\/inite vector  space $V$ is a $n$-dimensional diagonalizable module of $\mf h$, hereafter refereed as a Etingof-module of $\mf h$. That is: $V$ is a $n$-dimensional vector space with the weight decomposition $V=\bigoplus\limits_{\mu\in\mf h^*}V[\mu]$, where the weight spaces~$V[\mu]$ are irreducible modules of~$\mf h$, hence are one-dimensional. The operator $R$ is therefore represented by an $n^2\times  n^2$  matrix.

This allows to understand the notation $T_a(\l+\g h_b)$, for any distinct labels $a$ and $b$:  $\l$ is a~vector in $\mf h^* $ and $h_b$ denotes the canonical element of $\mf h \otimes \mf h^*$ with a~natural action of $\mf h$ on any given vector of~$V$. As a matter of fact, for example $a=1$ and $b=2$, this yields the usual vector shift by $\g h_2$ def\/ined, for any $v_1,v_2\in V$ as
  \[
T_1(\l+\g h_2)v_1\otimes v_2=T_1(\l+\g\mu_2)v_1\otimes v_2,
  \]
where $\mu_2$ is the weight of the vector~$v_2$.

The shift, denoted $\g h_q$, is similarly def\/ined as resulting from the action on $h_b$ of $\phi \otimes\1$, where $\phi$: $\mf h\longrightarrow\mf A$ is an algebra morphism, $\1$ being the identity operator in the space $V$. If (\ref{DQR}) is acted upon by $\1\otimes\1\otimes\rho_H$, where $\rho_H$ is a representation of the quantum algebra $\mf A$ on a Hilbert space $H$ assumed also to be a diagonalizable module of $\mf h$, then $\rho_H(h_q)$ acts naturally on $H$ (in particular on a basis of common eigenvectors of $\mf h$ assuming the axiom of choice) yielding also a~shift vector in $\mf h^* $.

Requiring now that the $R$-matrix obey the so-called zero-weight condition under adjoint action of any element $h\in\mf h$
  \[
\left[h_1+h_2,R_{12}\right]=0
  \]
allows to establish that the associativity condition on the quantum algebra (\ref{DQR}) implies as a~consistency condition the so-called dynamical quantum Yang--Baxter algebra for $R$
  \begin{gather}
R_{12}(\l+\g h_3)R_{13}(\l-\g h_2)R_{23}(\l+\g h_1)=R_{23}(\l-\g h_1)R_{13}(\l+\g h_2)R_{12}(\l-\g h_3 ). \label{DQYBE}
  \end{gather}
Using the zero-weight condition allows to rewrite (\ref{DQYBE}) in an alternative way which we shall consider from now on;
  \begin{gather}
R_{12}(\l+2\g h_3)R_{13}(\l)R_{23}(\l+2\g h_1)=R_{23}(\l)R_{13}(\l+2\g h_2)R_{12}(\l), \label{DQYBE2} \tag{DQYBE}
  \end{gather}
where the re-def\/inition $R_{ab}\longrightarrow R'_{ab}=\mathrm{Ad}\, \exp\g( h\cdot\dd_a+h\cdot\dd_b)R_{ab}$ is performed, $h\cdot\dd$ denoting the dif\/ferential operator $\sum\limits_{i=1}^n h_i\partial_{\l_i}$. Due to the zero-weight condition on the $R$-matrix, the action of this operator yields another $c$-number matrix in $\End(\mc V\otimes\mc V)$ instead of the expected dif\/ference operator-valued matrix. Note that it may happen that the matrix $R$ be of dynamical zero-weight, i.e.~$[h\cdot\dd_a+h\cdot\dd_b,R_{ab}]=0$, in which case $R'=R$.

Early examples of solutions in this non-af\/f\/ine case have been brought to light under the hypothesis that $R$ obeys in addition a so-called Hecke condition \cite{Isa}. The classif\/ication of Hecke type solutions in the non-af\/f\/ine case has been succeeded for a long time starting with the pioneering works of Etingof et al.~\cite{EtSchif, EtVar_1}. It restricts the eigenvalues of the permuted $R$-matrix $\check R=PR$, $P$ being the permutation operator of vector spaces $V\otimes\1$ and $\1\otimes V$, to take only the value $\varrho$ on each one-dimensional vector space $V_{ii}=\C v_i\otimes v_i$, for any index $i\in\N^*_n$, and the two distinct values $\varrho$ and $-\k$ on each two-dimensional vector space $V_{ij}=\C v_i\otimes v_j\oplus\C v_j\otimes v_i$, for any pair of distinct indices $(i,j)\in(\N^*_n)^2$, $(v_i)_{i\in\N^*_n}$ being a basis of the space $V$.

The less constraining, so-called ``weak Hecke'' condition, not explored in \cite{EtVar_1}, consists in assuming only that the eigenvalue condition without assumption on the structure of eigenspaces. In other words, one only assumes the existence of two $c$-numbers $\varrho$ and $\kappa$, with $\varrho\neq-\kappa$, such that
  \begin{displaymath}
(\check R-\varrho)(\check R+\k)=0.
  \end{displaymath}
We shall not assume a priori any Hecke or weak Hecke condition in our discussion. However, an important remark is in order here. The weak Hecke condition is understood as a quantization of the skew-symmetry condition on the classical dynamical $r$-matrices $r_{12}=-r_{21}$ \cite{EtVar_1}. It must be pointed out here that the classical limit of \ref{DQYBE2} is only identif\/ied with  the consistent associativity condition for the ``sole'' skew-symmetric part $a_{12}-a_{21}$ of a classical $r$-matrix parametrizing the linear Poisson bracket structure of a Lax matrix for a given classical integrable system
  \[
\{l_1,l_2\}=[a_{12},l_1]-[a_{21},l_2].
  \]
Only when the initial $r$-matrix is skew-symmetric do we then have a direct connection between classical and quantum dynamical Yang--Baxter equation. Dropping the weak Hecke condition in the quantum case therefore severs this link from classical to quantum Yang--Baxter equation and may thus modify the understanding of (\ref{DQYBE}) as a deformation by a parameter $\hbar$ of a classical structure. Nevertheless it does not destroy any of the characteristic quantum structures: co\-pro\-duct, coactions, fusion of $T$-matrices and quantum trace formulas yielding quantum commuting Hamiltonians, and as such one is perfectly justif\/ied in  considering a generalized classif\/ication of a priori non-weak Hecke solutions in the context of building new quantum integrable systems of spin-chain or $N$-body type.

The issue of classifying non-af\/f\/ine $R$-matrices, solutions of \ref{DQYBE2}, when the (weak) Hecke condition is dropped, already appears in the literature~\cite{Ju}, but in the very particular case of $\mc Gl_2(\C)$ and for trigonometric behavior only. A further set of solutions, in addition to the expected set of Hecke-type solutions, is obtained. In the context of the six-vertex model, these solutions are interpreted as free-fermion-type solutions, and show a weak Hecke-type, but non-Hecke-type, behavior $R_{12,21}=f(\l_1+\l_2)$, where $f$ is a trigonometric function.

We therefore propose here a complete classif\/ication of invertible $R$-matrices solving \ref{DQYBE2} for $V=\C^n$. We remind that we choose $\mf h$ to be the Cartan algebra of $\mc Gl_n(\C)$ with basis vectors $h_i=e^{(n)}_{ii}\in\mc M_n(\C)$  in the standard $n\times  n$ matrix notation. This f\/ixes in turn the normalization of the coordinate $\l$ up to an overall multiplicator set so as to eliminate the prefactor $2 \g$. This classif\/ication is proposed within the following framework.
  \begin{enumerate}[label={\bf\roman*}.]\itemsep=0pt
\item We consider non-spectral parameter dependent $R$-matrices. They are generally called ``constant'' in the literature on quantum $R$-matrices but this denomination will never be used here in this sense since it may lead to ambiguities with respect to the  presence in our matrices of ``dynamical'' parameters. This implies that a priori no elliptic dependence of the solutions in the dynamical variables is expected: at least in the Hecke case all dynamical elliptic quantum $R$-matrices are until now af\/f\/ine solutions.
\item We assume the matrix $R$ to be invertible. Non-invertible $R$-matrices are expected to correspond to an inadequate choice of auxiliary space $V$ (e.g.~reducible). It precludes even the proof of commutation of the traces of monodromy matrices, at least by using the dynamical quantum group structure, hence such $R$-matrices present in our view a lesser interest.
\item We assume that the elements of the matrix $R$ have suf\/f\/icient regularity properties as functions of their dynamical variables, so that we are able to solve any equation of the form $A(\l)B(\l)=0$ as $A(\l)=0$ or $B(\l)=0$ on the whole domain of variation $\C^n$ of $\l$ except of course possible isolated singularities. In other words, we eliminate the possibility of ``domain-wise zero'' functions with no overlapping non-zero values. This may of course exclude potentially signif\/icant solutions but considerably simplif\/ies the (already quite lengthy) discussion of solutions to~\ref{DQYBE2}.
\item Finally we shall hereafter consider as ``(pseudo)-constant'' all functions of the variable $\l$ with an integer periodicity, consistent with the chosen normalization of the basis $(h_i)_{i\in\N^*_n}$. Indeed such functions may not be distinguished from constants in the equations which we shall treat.
  \end{enumerate}

After having given some preliminary results in Sections~\ref{sec_2} and~\ref{sec_3} presents key procedures allowing to def\/ine an underlying partition of the indices $\N^*_n$ into $r$ subsets together with an associated ``reduced'' $\D$-incidence matrix $\mc M_R\in\mc M_r(\{0,1\})$ derived from the $\D$-incidence matrix~$\mc M$. The giving of this partition and the
associated matrix $\mc M_R$ essentially determines the general structure of the
$R$-matrix in terms of constituting blocks.

In Section~\ref{sec_4}, we shall establish the complete forms of all such blocks by solving system (\ref{syst}). The Hecke-type solutions will appear as a very particular solution\footnote{For more details, see Subsection~\ref{subsec_5_5}.}.

Section~\ref{sec_5} then presents the form of a general solution of \ref{DQYBE2}, and addresses the issue of the moduli structure of the set of solutions. The building blocks of any solution are in particular identif\/ied as weak Hecke type solutions or scaling thereof. The continuity of solutions in the moduli space are also studied in details.

Finally we brief\/ly conclude on the open problems and outlooks.

\section{Preparatory material} \label{sec_2}

The following parametrization is adopted for the $R$-matrix
  \[
R=\sum_{i,j=1}^n\D_{ij}e^{(n)}_{ij}\otimes e^{(n)}_{ji}+\sum_{i\neq j=1}^nd_{ij}e^{(n)}_{ii}\otimes e^{(n)}_{jj}.
  \]
A key fact of our resolution is that since the $R$-matrix is assumed to be invertible, its determinant is non zero. Let $n\geq2$. Since the matrix $R$ satisf\/ies the zero weight-condition, for any $i,j\in\N^*_n$, the vector spaces $\C e^{(n)}_{ii}\otimes e^{(n)}_{ii}$ and $\C e^{(n)}_{ij}\otimes e^{(n)}_{ji}\oplus\C e^{(n)}_{ij}\otimes e^{(n)}_{ji}$ are stable. Then its determinant is given by the factorized form
  \begin{gather}
\det(R)=\prod\limits_{i=1}^n \D_{ii} \prod\limits_{j=i+1}^n\{d_{ij}d_{ji}-\D_{ij}\D_{ji}\}. \tag{det} \label{det}
  \end{gather}
This implies that all $\D_{ii}$ are non-zero, and that $\D_{ij}\D_{ji}\neq0$, if $d_{ij}d_{ji}=0$, and vice versa.

Using this parametrization, we now obtain the equations obeyed by the coef\/f\/icients of the $R$-matrix from projecting \ref{DQYBE2} on the basis $(e^{(n)}_{ij}\otimes e^{(n)}_{kl}\otimes e^{(n)}_{mp})_{i,j,k,l,m,p\in\N^*_n}$ of $n^2\times n^2\times n^2$ matrices. Only f\/ifteen terms are left due to the zero-weight condition. Occurrence of a shift by~$2\g$ (normalized to~$1$) of the $i$-th component of the dynamical vector $\l$ will be denoted ``$(i)$''. Distinct labels $i$, $j$ and $k$ mean distinct indices. The equations then read
  \begin{alignat}{3}
& \D_{ii}\D_{ii}(i)\{\D_{ii}(i)-\D_{ii}\}=0 \qquad && (G_0), & \nonumber\\
& d_{ij}d_{ij}(i)\{\D_{ii}(j)-\D_{ii}\}=0 \qquad && (F_1),& \nonumber\\
& d_{ji}d_{ji}(i)\{\D_{ii}(j)-\D_{ii}\}=0 \qquad && (F_2),& \nonumber\\
& d_{ij}\{\D_{ii}(j)\D_{ij}(i)-\D_{ii}(j)\D_{ij}-\D_{ji}\D_{ij}(i)\}=0 \qquad && (F_3),& \nonumber\\
& d_{ji}\{\D_{ii}(j)\D_{ij}(i)-\D_{ii}(j)\D_{ij}-\D_{ji}\D_{ij}(i)\}=0 \qquad && (F_4),& \nonumber\\
& d_{ij}(i)\{\D_{ii}\D_{ji}(i)-\D_{ii}\D_{ji} +\D_{ji}\D_{ij}(i)\}=0 \qquad && (F_5),& \nonumber\\
& d_{ji}(i)\{\D_{ii}\D_{ji}(i)-\D_{ii}\D_{ji}+\D_{ji}\D_{ij}(i)\}=0 \qquad && (F_6),& \nonumber\\
& \D_{ii}^2(j)\D_{ij}-(d_{ij}d_{ji})\D_{ij}(i)-\D_{ii}(j)\D_{ij}^2=0 \qquad && (F_7),& \nonumber\\
& \D_{ii}^2\D_{ji}(i)-(d_{ij}d_{ji})(i)\D_{ij}-\D_{ii}\D_{ji}^2(i)=0 \qquad && (F_8), & \tag{$S$} \label{syst}\\
& \D_{ii}d_{ij}(i)d_{ji}(i)-\D_{ii}(j)d_{ij}d_{ji}+\D_{ij}(i)\D_{ji}\{\D_{ij}(i)-\D_{ji}\}=0 \qquad && (F_9),& \nonumber\\
& d_{ij}(k)d_{jk}(i)d_{ik}-d_{ij}d_{jk}d_{ik}(j)=0 \qquad && (E_1),& \nonumber\\
& d_{jk}d_{ik}(j)\{\D_{ij}(k)-\D_{ij}\}=0 \qquad && (E_2),& \nonumber\\
& d_{ij}(k)d_{ik}\{\D_{jk}(i)-\D_{jk}\}=0 \qquad && (E_3),& \nonumber\\
& d_{ij}(k)\{\D_{ij}(k)\D_{jk}+\D_{ji}(k)\D_{ik}-\D_{ik}\D_{jk}\}=0 \qquad && (E_4),& \nonumber\\
& d_{jk}\{\D_{ij}(k)\D_{jk}+\D_{ik}(j)\D_{kj}-\D_{ij}(k)\D_{ik}(j)\}=0 \qquad && (E_5),& \nonumber\\
& d_{ij}(k)d_{ji}(k)\D_{ik}-d_{jk}d_{kj}\D_{ik}(j)+\D_{ij}(k)\D_{jk}\{\D_{ij}(k)-\D_{jk}\}=0 \qquad && (E_6). & \nonumber
    \end{alignat}

Treating together coef\/f\/icients of $e^{(n)}_{ij}\otimes e^{(n)}_{ji}$ and $e^{(n)}_{ii}\otimes e^{(n)}_{ii}$ as $\D$-coef\/f\/icients is consistent since both tensor products may be understood as representing some universal objects $e \otimes e^*$, components of a universal $R$-matrix $\mc R$ in some abstract algebraic setting. The $d$-coef\/f\/icients of $e^{(n)}_{ii} \otimes e^{(n)}_{jj}$ are in this sense more representation dependent objects and we shall see indeed that they exhibit some gauge freedom in their explicit expression.

More generally, in order to eliminate what may appear as spurious solutions we immediately recall three easy ways of obtaining ``new'' solutions to \ref{DQYBE2} from previously obtained solutions.

 Let $(\a^i)_{i\in\N^*_n}$ be a family of functions of the variable $\l$. Def\/ine the dynamical diagonal operator $F_{12}=\ee^{\a_1(h_2)}\ee^{\a_2}$, where $\a$ is the $\l$-dependent vector $\a=\sum\limits_{i=1}^n\a^i h_i\in\mf h$.

\begin{proposition}[dynamical diagonal twist covariance]\label{subprop_1a}\label{prop_1}
If the matrix $R$ is a solution of \ref{DQYBE2}, then the twist action $R'=F_{12}RF_{21}^{-1}$ is also a solution of~\ref{DQYBE2}.

Denoting $\b_i=\ee^{\a^i}$, this is the origin of a particular, hereafter denoted, ``twist-gauge'' ar\-bitra\-riness on the $d$-coefficients, defined as\footnote{For more details, see Propositions~\ref{subprop_9a},~\ref{subprop_9b} and~\ref{subprop_9c}. \label{footnote_1}}
\[
d_{ij}\rightarrow d'_{ij}=\frac{\b_i(j)}{\b_i}\frac{\b_j}{\b_j(i)}d_{ij}, \qquad \forall \, i,j\in\N^*_n.
\]
\end{proposition}
    \begin{proof}For any distinct labels $a$, $b$ and $c$, the operator $\ee^{\pm\a_c}$ commutes with any operator with labels $a$ and/or $b$ and shifted in the space of index $c$, such as $R_{ab}(h_c)$, $\ee^{\pm\a_a(h_c)}$ or $\ee^{\pm\a_a(h_b+h_c)}$. Moreover, the zero-weight condition implies that $\ee^{\pm\a_a(h_b+h_c)}$ also commute with $R_{bc}$. By directly plugging $R'$ into the l.h.s.\ of \ref{DQYBE2} and using \ref{DQYBE2} for $R$, we can write
      \begin{gather*}
R'_{12}(h_3)R'_{13}R'_{23}(h_1)
=\ee^{\a_1(h_2+h_3)}\ee^{\a_2(h_3)}R_{12}(h_3)\ee^{-\a_2(h_1+h_3)}\ee^{\a_3}R_{13}\ee^{-\a_3(h_1)}\\
\hphantom{R'_{12}(h_3)R'_{13}R'_{23}(h_1)=}{}
\times \ee^{-\a_1}\ee^{\a_2(h_1+h_3)}
\ee^{\a_3(h_1)}R_{23}(h_1)
\ee^{-\a_3(h_1+h_2)}\ee^{-\a_2(h_1)}
\\
\hphantom{R'_{12}(h_3)R'_{13}R'_{23}(h_1)}{}
= \ee^{\a_1(h_2+h_3)}\ee^{\a_2(h_3)}\ee^{\a_3}R_{12}(h_3)R_{13}R_{23}(h_1)\ee^{-\a_3(h_1+h_2)}\ee^{-\a_2(h_1)}\ee^{-\a_1}
\\
\hphantom{R'_{12}(h_3)R'_{13}R'_{23}(h_1)}{}
= \ee^{\a_1(h_2+h_3)}\ee^{\a_2(h_3)}\ee^{\a_3}R_{23}R_{13}(h_2)R_{12}\ee^{-\a_3(h_1+h_2)}\ee^{-\a_2(h_1)}\ee^{-\a_1}
\\
\hphantom{R'_{12}(h_3)R'_{13}R'_{23}(h_1)}{}
= \ee^{\a_2(h_3)}\ee^{\a_3}R_{23}\ee^{-\a_3(h_2)}R'_{13}(h_2)\ee^{\a_1(h_2)}R_{12}\ee^{-\a_2(h_1)}\ee^{-\a_1}
\\
\hphantom{R'_{12}(h_3)R'_{13}R'_{23}(h_1)}{}
= R'_{23}R'_{13}(h_2)R'_{12},
      \end{gather*}
where the equality $\ee^{-\a_a(h_b)}\ee^{\a_a(h_b)}=\1\otimes\1\otimes\1$ is used when needed. It is then immediate to check that
\[
R'=\sum_{i,j=1}^n\D_{ij}e^{(n)}_{ij}\otimes e^{(n)}_{ji}+\sum_{i\neq j=1}^nd'_{ij}e^{(n)}_{ii}\otimes e^{(n)}_{jj},
\]
where the $d$-coef\/f\/icients of $R'$ are given as in the proposition.
    \end{proof}

  \begin{corollary}
Let $(\a^{ij})_{i,j\in\N^*_n}\in\C^{n^2}$ be a family of constants, denoting $\b_{ij}=\ee^{\a^{ij}-\a^{ji}}$,
 there exists a non-dynamical gauge arbitrariness on the $d$-coefficients as$^{\text{\rm\scriptsize \ref{footnote_1}}}$
  \[
d_{ij}\rightarrow d'_{ij}=\b_{ij}d_{ij}, \qquad \forall\,  i,j\in\N^*_n.
  \]
  \end{corollary}

    \begin{proof}
Introducing the family $(\a^i)_{i\in\N^*_n}$ of functions of the variable $\l$, def\/ined as $\a^i=\sum\limits_{k=1}^n\a^{ik}\l_k$, for any $i\in\N^*_n$, it is straightforward to verify that $\b_{ij}=\frac{\b_i(j)}{\b_i}\frac{\b_j}{\b_j(i)}$, for any $i,j\in\N^*_n$.
    \end{proof}

\begin{remark}
The dynamical twist operator $F$ can be identif\/ied as the evaluation representation of a dynamical coboundary operator.
\end{remark}

 Let $R^{aa}$ and $R^{bb}$ be two $R$-matrices, solutions of \ref{DQYBE2} respectively represented on auxiliary spaces $V_a$ and $V_b$, being Etingof-modules of the underlying dynamical Abelian algebras~$\mf h_a$ and~$\mf h_b$. Then $V_a\oplus V_b$ is an Etingof-module for $\mf h_a+\mf h_b$.

Let $g_{ab}$ and $g_{ba}$ be two non-zero constants, $\1^{ab}$ and $\1^{ba}$ respectively the identity operator in the subspaces $V_a\otimes V_b$ and $V_b\otimes V_a$. Def\/ine the new object
\[
R^{ab,ab}=R^{aa}+g_{ab}\1^{ab}+g_{ba}\1^{ba}+R^{bb}\in\End((V_a\oplus V_b)\otimes(V_a\oplus V_b)),
\]
where the sum ``$+$'' should be understood as a sum of the canonical injections of each component operator into $\End((V_a\oplus V_b)\otimes(V_a\oplus V_b))$.
\begin{proposition}[decoupled $R$-matrices]\label{subprop_1b}
The matrix $R^{ab,ab}$ is an invertible solution of \ref{DQYBE2} represented on auxiliary space $V_a\oplus V_b$ with underlying dynamical Abelian algebra $\mf h_a+\mf h_b$.
  \end{proposition}

    \begin{proof}
Obvious by left-right projecting \ref{DQYBE2} onto the eight subspaces of $(V_a\oplus V_b)^{\otimes 3}$ yielding a priori sixty-four equations. The new $R$-matrix is diagonal in these subspaces hence only eight equations survive.

Among them, the only non-trivial equations are the \ref{DQYBE2} for $R^{aa}$ and $R^{bb}$, lying respectively in $\End(V_a^{\otimes 3})$ and $\End(V_b^{\otimes 3})$, up to the canonical injection into $\End((V_a\oplus V_b)^{\otimes 3})$, since $R^{aa,bb}$ depends only on coordinates in $\mf h_{a,b}^*$, and by def\/inition of the canonical injection $\mf h_{a,b}$ acts as the operator $0$ on $V_{b,a}$. The six other equations are trivial because in addition they contain two factors $\1$ out of three.
    \end{proof}

Iterating $m$ times this procedure will naturally produce $R$-matrices combining $m$ ``sub''-$R$-matrices, hereafter denoted ``irreductible components'', with $m(m-1)$ identity matrices. To this end, it is not necessary to assume that the quantities $g_{ab}$ and $g_{ba}$ factorizing the identity operators $\1^{ab}$ and $\1^{ba}$ linking the matrices $R^{aa}$ and $R^{bb}$ should be constants\footnote{For a more indepth characterization of the quantities $g_{ab}$, see Proposition~\ref{prop_10}.}. Since, as above, the canonical injection $\mf  h_c$ acts as $0$ on $V_{b,a}$, for any third distinct label $c$, it is suf\/f\/icient to assume~$g_{ab}$ and~$g_{ba}$ to be non-zero 1-periodic functions in coordinates in $\mf h^*_a$ and $\mf h^*_b$, the dependence on any coordinate in $\mf h^*_c$ remaining free.

Finally, a third construction of new solutions to system (\ref{syst}) from already known ones now stems from the form itself of (\ref{syst}).

\par Let $R$ be a matrix, solution of $\mc Gl_n(\C)$-\ref{DQYBE2}, with Cartan algebra $\mf h^{(n)}$ having basis vectors $h^{(n)}_i=e^{(n)}_{ii}$, for any $i\in\N^*_n$, and  $\I=\{i_a,\; a\in\N^*_m\}\subseteq\N^*_n$ an ordered subset of $m$ indices. We introduce the matrices $e^\I_{ij}=e^{(m)}_{\s^\I(i)\s^\I(j)}\in\mc M_m(\C)$, for any $i,j\in\I$, and def\/ine the bijection $\s^\I$: $\I\longrightarrow\N^*_m$ as $\s^\I(i_a)=a$, for any $a\in\N^*_m$.

\begin{proposition}[contracted $R$-matrices]\label{subprop_1c}
The contracted matrix $R^\I=\sum\limits_{i,j,k,l\in\I} R_{ij,kl}e^\I_{ij}\otimes e^\I_{kl}$ of the matrix $R$ to the subset $\I$ is a solution of $\mc Gl_m(\C)$-\ref{DQYBE2}, with dynamical algebra $\mf h^{(m)}$ having basis vectors $h^{(m)}_a=e^{(m)}_{aa}$, for any $a\in\N^*_m$.
  \end{proposition}

    \begin{proof}
Obvious by direct examination of the indices structure of the set of equations (\ref{syst}). No sum over free indices occur, due to the zero-weight condition. Both lhs and rhs of all equations in (\ref{syst}) can therefore be consistently restricted to any subset of indices.
    \end{proof}

\begin{remark}
Formally the matrix $e^\I_{ij}$ consists in the matrix $e^{(n)}_{ij}$, from which the lines and columns, whose label does not belong to the subset $\I$, are removed.
\end{remark}

 We shall completely solve system (\ref{syst}) within the four conditions specif\/ied above, all the while setting aside in the course of the discussions all forms of solutions corresponding to the three constructions explicited in Propositions~\ref{subprop_1a}, \ref{subprop_1b} and~\ref{subprop_1c},
and a last one explicited later in Propositions~\ref{subprop_9a},~\ref{subprop_9b} and~\ref{subprop_9c}. A key ingredient for this procedure will be the $\D$-incidence matrix $\mc M\in\mc M_n(\{0,1\})$ of coef\/f\/icients def\/ined as $m_{ij}=0$ if and only if $\D_{ij}=0$.

\section[The $\D$-incidence matrix and equivalence classes]{The $\Delta$-incidence matrix and equivalence classes} \label{sec_3}

We shall f\/irst of all consider several consistency conditions on the cancelation of $d$-coef\/f\/icients and $\D$-coef\/f\/icients, which will then lead to the def\/inition of the partition of indices indicated above.

\subsection[$d$-indices]{$\boldsymbol{d}$-indices}

Two properties are established.
\begin{proposition}[symmetry]\label{subprop_2a}
Let $i,j\in\N^*_n$ such that $d_{ij}=0$. Then, $d_{ji}=0$.
  \end{proposition}

      \begin{proof}
If $d_{ij}=0$, $\D_{ij}\D_{ji}\neq0$. From $(G_0)$ one gets $\D_{ii}(i)=\D_{ii}$ and $\D_{jj}(j)=\D_{jj}$.

From $(F_7)$, one gets $\D_{ij}=\frac{\D_{ii}^2(j)}{\D_{ii}}$. This implies now that $ \D_{ij}(i)=\D_{ij}$. $(F_4)$ then becomes $d_{ji}\D_{ji}\D_{ij}(i)=0$,
hence $d_{ji}=0$.
    \end{proof}

\begin{proposition}[transitivity]\label{subprop_2b}
Let $i,j,k\in\N^*_n$ such that $d_{ij}=0$ and $d_{jk}=0$. Then, $d_{ik}=0$.
  \end{proposition}

    \begin{proof}
From $d_{ij}=0$ and $(F_9)$, one now gets $\D_{ij}=\D_{ji}$. From $\D_{ij}(i)=\D_{ij}$ now follows that $\D_{ji}(i)=\D_{ji}$, hence $\D_{ij}(j)=\D_{ij}$.

From $(F_8)$, $\D_{ij}=\D_{ji}=\D_{ii}=\D_{jj}=\D$, where the function $\D$ is independent of variables~$\l_i$ and~$\l_j$. Similarly, one also has $\D_{jk}=\D_{kj}=\D_{kk}=\D_{jj}=\D$, independently of variables~$\l_k$ and~$\l_j$.

Writing now $(E_6)$ with indices $jki$ and $(E_5)$ with indices $jik$ yields
\[
d_{ik}d_{ki}=\D_{ki}\{\D-\D_{ki}\}=\D_{ki}\{\D-\D_{ik}\} \qquad \textrm{and} \qquad d_{ik}\{\D-\D_{ik}-\D_{ki}\}=0.
\]
From which we deduce, if $d_{ik}\neq0$, that $\D=\D_{ik}+\D_{ki}$. Then $d_{ik}d_{ki}=\D_{ki}\{\D-\D_{ki}\}=\D_{ki}\D_{ik}$, and $\det(R)=0$. Hence, one must have $d_{ik}=0$.
    \end{proof}

\begin{corollary}
Adding the axiom $i\mc D i$, for any $i\in\N^*_n$, the relation defined by
\[
i\mc Dj \quad \Leftrightarrow \quad d_{ij}=0
\]
is an equivalence relation on the set of indices $\N^*_n$.
  \end{corollary}

\begin{remark}
The $\mc D$-class generated by any index $i\in\N^*_n$ will be denoted
\[
\I(i)=\{j\in\N^*_n \; \big| \; j\mc Di\},
\]
and we will introduce the additional subset
\[
\I_0=\{i\in\N^*_n \; \big| \; \I(i)=\{i\}\}
\]
of so-called ``free'' indices.

For any subset $\I$ of the set of indices $\N^*_n$ and any $m\in\N^*$, let us also def\/ine the set $\I^{(m,\nD)}=\{(i_a)_{a\in\N^*_m}\in(\N^*_n)^m\; |\; a\neq b\Rightarrow i_a\nD i_b\}$. In the following, we will actually consider only the case $m\in\{2,3\}$. An element of $\I^{(2,\nD)}$ (resp.~$\I^{(3,\nD)}$) will be refereed as a $\nD$-pair (resp.~$\nD$-triplet) of indices.
\end{remark}

\subsection[$\D$-indices]{$\boldsymbol{\D}$-indices}

We establish a key property regarding the propagation of the vanishing of $\D$-coef\/f\/icients.
\begin{proposition}\label{subprop_3a}
Let $i,j\in\N^*_n$ such that $\D_{ij}=0$. Then, $\D_{ik}\D_{kj}=0$, for any $k\in\N^*_n$.
  \end{proposition}

\begin{proposition}[contraposition]\label{subprop_3b}
Let $i,j\in\N^*_n$. Equivalently, if there exist $k\in\N^*_n$ such that $\D_{ik}\D_{kj}\neq 0$, then $\D_{ij}\neq 0$.
\end{proposition}

    \begin{proof}
If $\D_{ij}=0$ then $d_{ij}d_{ji}\neq0$. It follows from Proposition~\ref{subprop_2b} that $d_{ik}\neq 0$ or $d_{kj}\neq 0$, for all $k\neq i,j$. Assume that $d_{ik}\neq0$ hence $d_{ki}\neq 0$. $(E_4)$ with indices $ikj$ reads
\[
d_{ik}(j)[\D_{ik}(i)\D_{kj}+\D_{ij}\{\D_{ki}(j)-\D_{kj}\}]=0,
\]
hence $\D_{ik}=0$ or $\D_{kj}=0$.

If instead $d_{kj}\neq 0$ hence $d_{jk}\neq 0$. $(E_5)$ with indices $ijk$ directly yields $\D_{ik}(j)\D_{kj}=0$ with the same conclusion.
    \end{proof}

\begin{proposition}\label{subprop_3c}
The relation defined by
\[
i\D j \quad \Leftrightarrow \quad \D_{ij}\D_{ji}\neq0
\]
is an equivalence relation on the set of indices $\N^*_n$.

Moreover, any $\mc D$-class is included in a single $\D$-class.
\end{proposition}

    \begin{proof}
Ref\/lexivity and symmetry are obvious. Transitivity follows immediately from Proposition~\ref{subprop_3b}. If $i\D k\Leftrightarrow\D_{ik}\D_{ki}\neq0$ and $k\D j\Leftrightarrow\D_{kj}\D_{jk}\neq0$, hence $\D_{ik}\D_{kj}\neq0$ and $\D_{jk}\D_{ki}\neq0$. Then, $\D_{ij}\neq0$ and $\D_{ji}\neq0$, i.e.~$i\D j$.

The second part of the proposition follows immediately from~(\ref{det}).
    \end{proof}

  \begin{corollary}
Denote $\{\J_p,\; p\in\N^*_r\}$ the set of $r$ $\D$-classes, which partitions the set of indices~$\N^*_n$.

For any $p\in\N^*_r$, there exist $l_p\in\N$, a so-called ``free'' subset $\I^{(p)}_0=\J_p\cap\I_0$ of free indices $($possibly empty$)$, and $l_p$ $\mc D$-classes generated by non-free indices $($possibly none$)$, denoted $\I^{(p)}_l$ with $l\in\N^*_{l_p}$, such that $\J_p=\bigcup\limits_{l=0}^{l_p}\I^{(p)}_l$ is a partition. Finally, $i\mc Dj$, if and only if $\exists\, l\in\N^*_{l_p}\; |\; i,j\in\I^{(p)}_l$.
  \end{corollary}

\subsection[(Reduced) $\D$-incidence matrix]{(Reduced) $\boldsymbol{\D}$-incidence matrix}

The $\D$-incidence matrix $\mc M=\sum\limits_{i,j=1}^nm_{ij}e^{(n)}_{ij}\in\mc M_n(\{0,1\})$ is def\/ined as follows
  \begin{displaymath}
m_{ij}=1 \quad \Leftrightarrow \quad \D_{ij}\neq0 \qquad \textrm{and} \qquad m_{ij}=0 \quad \Leftrightarrow \quad \D_{ij}=0.
  \end{displaymath}
Let us now use the $\D$-class partition and Propositions~\ref{subprop_3a}, \ref{subprop_3c} and~\ref{subprop_3c} to better characterize the form of the $\D$-incidence matrix $\mc M$ of a solution of \ref{DQYBE2}. The key object here will be the so-called reduced $\D$-incidence matrix $\mc M_R$.
\begin{proposition}\label{subprop_4a}
Let $\I$, $\J$ two distinct $\D$-classes such that $\exists \, (i,j)\in\I\times\J\; \big|\; \D_{ij}\neq0$. Then, for any pair of indices $(i,j)\in\I\times\J$, $\D_{ij}\neq0$.
  \end{proposition}

    \begin{proof}
Let $i'\in\I$ and $j'\in\J$. Applying Proposition~\ref{subprop_3b} to $\D_{i'i}\D_{ij}\neq0$, we deduce $\D_{i'j}\neq0$. Then $\D_{i'j'}\neq 0$, since $\D_{i'j}\D_{jj'}\neq0$.
    \end{proof}

\begin{remark}
In the proof of this proposition, note here that nothing forbids $i'=i$ and/or $j'=j$. To facilitate their writing and reading, this convention will be also used in Proposition~\ref{prop_8}, Lemmas~\ref{sublem_4a} and~\ref{sublem_4b}, as well as Theorems~\ref{subtheo_3a},~\ref{subtheo_3b} and~\ref{theo_4}.
\end{remark}

  \begin{corollary}
Let $\I$, $\J$ two distinct $\D$-classes. Then either all connecting $\D$-coefficients in~$\D_{ij}$, with $(i,j)\in\I\times\J\}$, are zero or all are non-zero.
  \end{corollary}

 This justif\/ies that the property of vanishing of $\D$-coef\/f\/icients shall be from now on denoted with overall $\D$-class indices as $\D_{\I\J}=0$ or $\D_{\I\J}\neq0$. This leads now to introduce a reduced $\D$-incidence matrix $\mc M_R=\sum\limits_{p,p'=1}^rm^R_{pp'}e^{(r)}_{pp'}\in\mc M_r(\{0,1\})$, def\/ined as
\[
m^R_{pp'}=1 \quad \Leftrightarrow \quad \D_{\J_p\J_{p'}}\neq0 \qquad \textrm{and} \qquad m^R_{pp'}=0 \quad \Leftrightarrow \quad \D_{\J_p\J_{p'}}=0.
\]

\begin{proposition}  \label{subprop_4b}
The relation defined by
\[
\I\succeq\J \quad \Leftrightarrow \quad \D_{\I\J}\neq0
\]
is a partial order on the set of $\D$-classes.
\end{proposition}

    \begin{proof}
If $\I\succeq\J$ and $\J\succeq\I$, then $\D_{\I\J}\neq0$ and $\D_{\J\I}\neq0$. Hence, for all $(i,j)\in\I\times\J$, $\D_{ij}\D_{ji}\neq0$, i.e.~$\I=\J$.

If $\I\succeq\J$ and $\J\succeq\K$, then $\D_{\I\J}\neq0$ and $\D_{\J\K}\neq0$. Hence, from Proposition~\ref{subprop_3b}, $\D_{ik}\neq0$, for all $(i,k)\in\I\times\K$, i.e.~$\I\succeq\K$.
    \end{proof}

\looseness=-1
Two $\D$-classes $\I$ and $\J$ shall be refereed hereafter as ``comparable'', if and only if $\I\succeq\J$ or $\J\succeq\I$, which will be denoted $\I\comp\J$. This order on $\D$-classes is of course not total, because there may exist $\D$-classes which are not comparable, i.e.\ such that $\D_{\I\J}=\D_{\J\I}=0$, being denoted $\I\ncomp\J$.

The order $\succeq$ is to be used to give a canonical form to the matrix $\mc M_R$ in two steps, and more particularly the strict order $\succ$ deduced from $\succeq$ by restriction to distinct $\D$-classes. Unless otherwise stated, in the following, the subsets $\I$, $\J$ and $\K$ are three distinct $\D$-classes.

  \begin{proposition}[triangularity]\label{prop_5}\sloppy
The reduced $\D$-incidence matrix $\mc M_R$ is triangularisable in~$\mc M_r(\{0,1\})$.
  \end{proposition}

    \begin{proof}
The strict order $\succ$ def\/ines a natural oriented graph on the set of $\D$-classes.
Triangularity property of the order implies that no cycle exists in this graph.
To any $\D$-class $\I$ one can then associate all linear subgraphs ending on $\I$ as $\J_{p_1}\succ\J_{p_2}\succ\cdots\succ\J_{p_k}\succ\I$. There exist only a~f\/inite number of such graphs (possibly none) due to the non-cyclicity property. One can thus associate to the $\D$-class $\I$ the largest value of
$k$ introduced above, denoted by $k(\I)$.

We now label $\D$-classes according to increasing values of $k(\I)$, with the additional convention that $\D$-classes of same value of $k(\I)$ are labeled successively and arbitrarily. The labels  are denoted as $l(\I)\in\N^*_r$ in increasing value, and we have the crucial following lemma.
      \begin{lemma} \label{lem_1}
If $l(\I)<l(\J)$, then $\D_{\J\I}=0$.
  \end{lemma}

        \begin{proof}
By contraposition, if $\D_{\J\I}\neq0$ and $\I\neq\J$, then $\J\succ\I$. Hence $k(\I)\geq k(\J)+1>k(\J)$, which is impossible if $l(\I)<l(\J)$, by def\/inition of the labeling by increasing values of $k(\I)$.
        \end{proof}

Let us now introduce the permutation $\s$: $p\longmapsto l(\J_p)\in\mf S_r$, its associated permutation matrix $P_\s=\sum\limits_{p=1}^re^{(r)}_{\s(p)p}\in\mc Gl_r(\{0,1\})$, of inverse $P^{-1}_\s=P_{\s^{-1}}$, and the permuted reduced $\D$-incidence matrix $\mc M^\s_R=P_\s \mc M_RP^{-1}_\s$. It is straightforward to check that $m^{R,\s}_{pp'}=m^R_{\s^{-1}(p)\s^{-1}(p')}$. From Lemma~\ref{lem_1}, we deduce that, if $p=\s(q)<\s(q')=p'$, then $\D_{\J_q\J_{q'}}=0$, and $m^{R,\s}_{pp'}=m^R_{qq'}=0$, i.e.\ the matrix $\mc M^\s_R$ is upper-triangular.
    \end{proof}

  \begin{corollary}
Denoting $\J^\s_p=\J_{\s(p)}$, if $p<p'$, then either $\J^\s_p\ncomp\J^\s_{p'}$ or $\J^\s_p\succ\J^\s_{p'}$.
  \end{corollary}

The characterization of a canonical form for the matrix $\mc M_R$ can now be further precise.
\begin{proposition}\label{subprop_6a}
If $\I\ncomp\J$ and $\I\succ\K$, then $\J\ncomp\K$.
 \end{proposition}

    \begin{proof}
By assumption, remark that $\D_{\I\J}=\D_{\J\I}=\D_{\K\I}=0$ and $\D_{\I\K}\neq0$. Let $(i,j,k)\in\I\times\J\times\K$.

Since $\D_{ij}=0$ and $\D_{ki}=0$, from~(\ref{det}), $d_{ij}d_{ji}\neq0$ and $d_{ik}d_{ki}\neq0$, but $\D_{ik}\neq0$.

When written with indices $ijk$, $(E_4)$ reduces to $d_{ij}(k)\D_{ik}\D_{jk}=0$, hence $\D_{jk}=0$.

When written with indices $ikj$, $(E_4)$ reduces to $d_{ik}(j)\D_{kj}\D_{ik}(j)=0$, hence $\D_{kj}=0$.
    \end{proof}

\begin{proposition}\label{subprop_6b}
If $\I\ncomp\J$ and $\K\succ\I$, then $\J\ncomp\K$.
 \end{proposition}

    \begin{proof}
Identical to the proof of Proposition~\ref{subprop_6a} using $(E_5)$, written with indices $jik$ and $ikj$.
    \end{proof}

\begin{proposition}\label{subprop_6c}
If $\I\ncomp\J$ $($resp.\ $\I\comp\J)$ and $\I\comp\K$, then $\J\ncomp\K$ $($resp. $\J\comp\K)$.
 \end{proposition}

  \begin{proof}
From Propositions~\ref{subprop_6a} and~\ref{subprop_6b}, if $\I\ncomp\J$ and $\I\comp\K$, then $\J\ncomp\K$.

If $\I\comp\J$, and assuming that $\J\ncomp\K$, there is a contradiction with $\I\comp\K$, then $\J\comp\K$.
  \end{proof}

  \begin{corollary}\label{prop_6.cor}
Let $p<p',p''$. Hence,
  \begin{enumerate}[label={\bf\roman*}.]\itemsep=0pt
\item if $\J^\s_p\ncomp\J^\s_{p'}$ and $\J^\s_p\succ\J^\s_{p''}$, then  $\J^\s_p\ncomp\J^\s_{p''}$;
\item if $\J^\s_p\succ\J^\s_{p'}$ and $\J^\s_p\succ\J^\s_{p''}$, with $p'<p''$, then  $\J^\s_{p'}\succ\J^\s_{p''}$.
  \end{enumerate}
  \end{corollary}

  \begin{proposition}[block upper-triangularity]\label{prop_7}
The reduced $\D$-incidence matrix $\mc M_R$ is similar to a block upper-triangular matrix in $\mc M_r(\{0,1\})$.

That is: there exists a permutation $\pi\in\mf S_r$ and a partition of the set $\N^*_r$ in $s$ subsets $\PP_q=\{p_q+1,\dots,p_{q+1}\}$ $($with the convention that $p_1=0$ and $p_{s+1}=r)$, of respective cardinality $r_q$, such that
\[
m^{R,\pi}_{pp'}=1 \quad \Leftrightarrow \quad \exists\,  q\in\N^*_s\; \big|\; (p,p')\in\PP_q^{(2,<)}.
\]
i.e.\ the matrix $\mc M^\pi_R=\sum\limits_{p,p'=1}^rm^{R,\pi}_{pp'}e^{(r)}_{pp'}$ is graphically represented by blocks as
\[
\mc M^\pi_R=\!\left(\!\!
\raisebox{0.5\depth}{
\xymatrixcolsep{1ex}%
\xymatrixrowsep{1ex}%
\xymatrix{\relax
\mc T_{r_1}						& \mc O_{r_1r_2} \ar@{.}[rrrr] \ar@{.}[dddrrrr]	& 	& 	& 			& \mc O_{r_1r_s} \ar@{.}[ddd]
\\
\mc O_{r_2r_1} \ar@{.}[ddd] \ar@{.}[dddrrrr]		& \mc T_{r_2} \ar@{.}[ddrrr]			& 	& 	& 			&
\\
							& 						& 	& 	& 			&
\\
							& 						& 	& 	& \mc T_{r_{s-1}}	& \mc O_{r_{s-1}r_s}
\\
\mc O_{r_sr_1} \ar@{.}[rrrr]				& 						& 	& 	& \mc O_{r_sr_{s-1}}	& \mc T_{r_s}}}
    \right)\!\!\in\mathcal M_r(\{0,1\}),
\]
where the type $\mc T$, $\mc O$ block matrices are def\/ined by
\[
\mc T_{r'}=\!\left(\!\!
\raisebox{0.5\depth}{
\xymatrixcolsep{1ex}%
\xymatrixrowsep{1ex}%
\xymatrix{\relax
1 \ar@{.}[rrr] \ar@{.}[dddrrr]	& 	& 	& 1 \ar@{.}[ddd]
\\
0 \ar@{.}[dd] \ar@{.}[ddrr]	& 	& 	&
\\
				& 	& 	&
\\
0 \ar@{.}[rr] 			& 	& 0	& 1}}
    \right)\!\!\in\mathcal M_{r'}(\{0,1\})
\]
and
\[
\mc O_{r'r''}=\!\left(\!\!
\raisebox{0.5\depth}{
\xymatrixcolsep{1ex}%
\xymatrixrowsep{1ex}%
\xymatrix{\relax
0 \ar@{.}[rr] \ar@{.}[dd] \ar@{.}[ddrr]	& 	& 0 \ar@{.}[dd]
\\
					& 	&
\\
0 \ar@{.}[rr] 				& 	& 0}}
    \right)\!\!\in\mathcal M_{r',r''}(\{0\}),\qquad \mc O_{r'}=\mc O_{r'r'}\!\!\in\mathcal M_{r'}(\{0\}).
\]
\end{proposition}

\begin{remark}\looseness=-1
For any set $\I$ of integers and any $m\in\N^*$, by analogy with the def\/inition
of the set $\I^{(m,\nD)}$, we adopt the notations $\I^{(m,<)}=\{(i_a)_{a\in\N^*_m}\in\I^m\; |\; a<b\Rightarrow i_a<i_b\}$
and $\I^{(m,\nD,<)}=\{(i_a)_{a\in\N^*_m}\in\I^{(m,\nD)}\; |\; a<b\Rightarrow i_a<i_b\}$.
For example, a pair of labels $q,q'\in\N^*_s$ such that $q<q'$ (resp.\ a $\nD$-pair of
indices $(i,j)\in\I^2$ such that $i<j$) belongs to the set $\N_s^{*(2,<)}$ (resp.\  $\I^{(2,\nD,<)}$).
\end{remark}

    \begin{proof}
The proof relies on a recursion procedure on the value of the size $r$ of the matrix $\mc M_R$. The proposition being trivial for $r\in\{1,2\}$, let us assume that $r\geq3$.

{\bf 1.~Re-ordering from line~1.}
Starting from the matrix $\mc M^\s_R$, whose existence is guaranteed by Proposition~\ref{prop_5}, its upper-triangularity is used following
Corollary~\ref{prop_6.cor}. 

Remember that label ordering and class-ordering run contrary to each other.

Note $p^{(1)}_1=1$. Since $\J^\s_1$ is always comparable to itself, the set of $\D$-classes comparable to $\J^\s_1$ is not empty, and we will denote $r_1\in\N^*_r$ its cardinality. If $r_1=1$, i.e.\ if $\J^\s_1$ is not comparable to any other $\D$-class, line $1$ of matrix $\mc M_R^\s$ consists of an one-label block $m^{R,\s}_{11}=1$, and the process stops.

Assuming that $r_1\in\{2,\dots,r\}$, consider the subset $\{p\in\{2,\dots,r\}\; \big|\; \J^\s_1\succ\J^\s_p\}\neq\varnothing$. This set is naturally totally ordered. Let us then denote its elements as $p^{(1)}_q$ by increasing value, where $q\in\{2,\dots,r_1\}$. Then, by convention, $(p^{(1)}_q,p^{(1)}_{q'})\in\N_{p^{(1)}_{r_1}}^{*(2,<)}$, if and only if $(q,q')\in\N_{r_1}^{*(2,<)}$.

Moreover, by construction, we have that $\J^\s_1\succ\J^\s_{p^{(1)}_q}$ and $\J^\s_1\succ\J^\s_{p^{(1)}_{q'}}$, for any $q,q'\in\{2,\dots,r_1\}$. Then, from Corollary~\ref{prop_6.cor},  $\J^\s_{p^{(1)}_q}\succ\J^\s_{p^{(1)}_{q'}}$, if and only if $(q,q')\in\N_{r_1}^{*(2,<)}$, i.e.\ $\J^\s_1\succ\J^\s_{p^{(1)}_2}\succ\cdots\succ\J^\s_{p^{(1)}_{r_1}}$, where no sign $\succ$ can be reversed. In particular, the $\D$-classes in $\big\{\J^\s_{p^{(1)}_q},\; q\in\{2,\dots,r_1\}\big\}$ are comparable one-to-one. Since $\J^\s_1$ is only comparable to the $\D$-classes in $\big\{\J^\s_{p^{(1)}_q},\; q\in\{2,\dots,r_1\}\big\}$, no other $\D$-class is comparable to any $\D$-class $\J^\s_{p^{(1)}_q}$, with $q\in\{2,\dots,r_1\}$. This implies that
\[
m^{R,\s}_{p^{(1)}_qp^{(1)}_{q'}}=1\quad \Leftrightarrow\quad (q,q')\in\N_{r_1}^{*(2,<)},
\]
and that
\[
m^{R,\s}_{p^{(1)}_qp}=m^{R,\s}_{pp^{(1)}_q}=0,\qquad \forall \, q\in\N^*_{r_1} \qquad \textrm{and} \qquad \forall \, p\in\N^*_r\setminus\{p^{(1)}_q, \; q\in\N^*_{r_1}\}.
\]

Let $\pi_1\in\mf S_r$ be the unique permutation such that
  \[
\pi_1(p^{(1)}_q)=q,\qquad \forall\; q\in\N^*_{r_1},
  \]
and that $\pi_1$ is increasing on $\N^*_r\setminus\{p^{(1)}_q,\; q\in\N^*_{r_1}\}$. We apply the same reasoning as in the end of the proof of Proposition~\ref{prop_5}. The coef\/f\/icients of the permuted matrix $\mc M^{\pi_1\circ\s}_R$ now satisfy the following equalities
\begin{gather*}
m^{R,\pi_1\circ\s}_{pp'}=1\quad \Leftrightarrow \quad (p,p')\in\N_{r_1}^{*(2,<)},
\qquad {\rm and}\\
m^{R,\pi_1\circ\s}_{pp'}=m^{R,\pi_1\circ\s}_{p'p}=0,\qquad \forall\, p\in\N^*_{r_1} \qquad \textrm{and} \qquad \forall\, p'\in\{r_1+1,\dots,r\}.
\end{gather*}
Furthermore, the increasing property of the permutation $\pi_1$ transfers the upper-triangularity of the matrix $\mc M_R^\s$ to the matrix $\mc M^{\pi_1\circ\s}_R$, which can f\/inally be graphically represented by blocks as
  \begin{displaymath}
\mc M^{\pi_1\circ\s}_R=\left(
      \begin{BMAT}(e){c.c}{c.c}
\mc T_{r_1} 	& ~\mc O_{r_1r'}
\\
\mc O_{r'r_1} 	& \mc M'_R
      \end{BMAT}
    \right)\!.
  \end{displaymath}

{\bf 2.~Recursion on $\boldsymbol{r}$.}
Let assume that the statement is true for any reduced $\D$-incidence matrix of size $r'\in\N^*_{r-1}$ associated with a solution of \ref{DQYBE2}. Using the previously def\/ined re-ordering procedure on the f\/irst line of a matrix $\mc M^\s_R\in\mc M_r(\{0,1\})$, there exists a upper-triangular reduced $\D$-incidence matrix $\mc M'_R\in\mc M_{r'}(\{0,1\})$ of size $r'=r-r_1<r$, which is moreover associated with a solution of \ref{DQYBE2} from Proposition~\ref{subprop_1c}. The recursion hypothesis can now be applied to the f\/irst line of the matrix $\mc M'_R$ describing the order of the $r'$ remaining $\D$-classes.

{\bf 3.~Recursive construction of $\boldsymbol{\pi}$ and $\boldsymbol{\{\PP_q,\; q\in\N^*_s\}}$.}
Since the number of $\D$-classes is f\/inite, the process described above comes to an end after a f\/inite number $s\in\N^*_r$ of iterations.

The $q^\mathrm{th}$ iteration insures the existence of an integer $r_q\in\N^*_s$ and a permutation $\pi_q\in\mf S_r$, built by recursion. Def\/ining $p_m=\sum\limits_{m'=1}^{m-1}r_{m'}\in\N^*_r$, for any $m\in\N^*_q$, the integer $r_q$ is the cardinality of the set of $\D$-classes comparable to the $\D$-class $\J^{\pi_{p-1}\circ\cdots\circ\pi_1\circ\s}_{p_{q-1}+1}$, being the f\/irst remaining $\D$-class after $q-1$ iterations. Introducing the totally ordered set $\{p^{(q)}_{q'},\; q'\in\N^*_{r_q}\}$ of indices of such $\D$-classes and putting $p_q=p_{q-1}+r_q$, the permutation $\pi_q$ re-orders the indices as follows
\[
\pi_q(p)=p,\qquad \forall\, p\in\N^*_{p_q} \qquad \textrm{and} \qquad \pi_q(p^{(q)}_{q'})=p_{q-1}+q',\qquad \forall\, q'\in\N^*_{r_q},
\]
$\pi_q$ being increasing on $\{p_q+1,\dots,r\}\setminus\{p^{(q)}_{q'},\; q'\in\N^*_{r_q}\}$. Finally, the permutation $\pi=\pi_s\circ\cdots\circ\pi_1\circ\s\in\mf S_r$ leads to the expected permuted matrix $\mc M^\pi_R$, and the partition $\N^*_r=\bigcup\limits_{q=1}^s\PP_q$ stands by construction.
    \end{proof}

\subsection{Classif\/ication}

For convenience, we will now identify in the following the reduced $\D$-incidence matrix $\mc M_R$ and its associated block upper-triangular matrix $\mc M^\pi_R$, as well as the $\D$-classes in $\{\J_p,\; p\in\N^*_r\}$ and the re-ordered $\D$-classes in $\{\J^\pi_p,\; p\in\N^*_r\}$. Let us conclude this section by fully describing the $\D$-incidence structure of a general $R$-matrix to complete the classif\/ication of solutions of \ref{DQYBE2}, together with the required steps to end the resolution of system (\ref{syst}).

  \begin{theorem}[$\D$-incidence matrices]\label{theo_1}
Let $n\geq2$. Then, any $R$-matrix, solution of \ref{DQYBE2}, is characterized in particular by
    \begin{itemize}\itemsep=0pt
\item an ordered partition of the indices $\N^*_n$ into $r$ $\D$-classes $\J_p$ of respective cardinality $n_p$,
\item an ordered partition of the indices $\N^*_r$ into $s$ subsets $\PP_q=\{p_q+1,\dots,p_{q+1}\}$, of respective cardinality $r_q$ $($with the convention that $p_1=0$ and $p_{s+1}=r)$,
\item an ordered partition of each $\D$-class $\J_p$ into a ``free'' subset $\I^{(p)}_0=\J_p\cap\I_0$ $($possibly empty$)$ of cardinality $n^{(p)}_0$, and $l_p$ $\mc D$-classes $\I^{(p)}_l$ generated by non-free indices $($possibly none$)$ of respective cardinality $n^{(p)}_l$;
    \end{itemize}
such that the following union is an ordered partition of the set of indices $\N^*_n$
\[
\N^*_n=\bigcup_{q=1}^s\K_q=\bigcup_{p=1}^r\J_p=\bigcup_{p=1}^r \bigcup_{l=0}^{l_p}\I^{(p)}_l,
\]
denoting $\K_q=\bigcup\limits_{p\in\PP_q}\J_p$ the set of $\D$-classes, of cardinality $N_q=\sum\limits_{p\in\PP_q}n_p\in\N^*_n$, associated to each subset $\PP_q$.

Re-expanding its reduced $\D$-incidence matrix $\mc M_R$, the $R$-matrix has a $\D$-incidence matrix $\mc M=\sum\limits_{p,p'\in\N^*_r \atop p\leq p'}m^R_{pp'}e^{(r)}_{pp'}\otimes \mc E_{n_pn_{p'}}$, which can be graphically represented as
\[
\mc M=\!\left(\!\!
\raisebox{0.5\depth}{
\xymatrixcolsep{1ex}%
\xymatrixrowsep{1ex}%
\xymatrix{\relax
\mc T^{(1)}								& \mc O^{(1,2)} \ar@{.}[rrrr] \ar@{.}[dddrrrr]	& 	& 		& 		& \mc O^{(1,s)} \ar@{.}[ddd]
\\
\mc O^{(2,1)} \ar@{.}[ddd] \ar@{.}[dddrrrr]				& \mc T^{(2)} \ar@{.}[ddrrr]			&	& 		& 		&
\\
									& 						& 	& 		& 		&
\\
									& 						& 	& 		& \mc T^{(s-1)}	& \mc O^{(s-1,s)}
\\
\mc O^{(s,1)} \ar@{.}[rrrr]						& 						& 	& 		& \mc O^{(s,s-1)}	& \mc T^{(s)}}}
    \right)\!\!\in\mathcal M_n(\{0,1\}),
\]
where the matrices $\mc T^{(q)}=\sum\limits_{p,p'\in\PP_q \atop p\leq p'}e^{(r_q)}_{pp'}\otimes \mc E_{n_pn_{p'}}$ are graphically represented as
\[
\mc T^{(q)}=\!\left(\!\!
\raisebox{0.5\depth}{
\xymatrixcolsep{1ex}%
\xymatrixrowsep{1ex}%
\xymatrix{\relax
\mc E_{n_{p_q+1}} \ar@{.}[rrr] \ar@{.}[dddrrr]		& 	& 				& \mc E_{n_{p_q+1}n_{p_{q+1}}} \ar@{.}[ddd]
\\
\mc O_{n_{p_q+2}n_{p_q+1}} \ar@{.}[dd] \ar@{.}[ddrr]	& 	& 				&
\\
							& 	& 				&
\\
\mc O_{n_{p_{q+1}}n_{p_q+1}} \ar@{.}[rr] 		& 	& \mc O_{n_{p_{q+1}-1}n_{p_{q+1}}}	& \mc E_{n_{p_{q+1}}}
}}
    \right)\!\!\in\mathcal M_{N_q}(\{0,1\}),
\]
with $\mc O^{(q,q')}=\mc O_{N_qN_{q'}}$, and where the type $\mc E$ matrices are defined like the type $\mc O$ matrices except that~$0$ is replaced by~$1$, i.e.\
\[
\mc E_{r'r''}\in\mc M_{r'r''}(\{1\}), \qquad \mc E_{r'}=\mc E_{r'r'}\in\mc M_{r'}(\{1\}).
\]
  \end{theorem}

\begin{remark} By ordered partition, we mean that the indices appear in the partition in the canonical order of integers. For example, a $\nD$-pair of indices $(i,j)\in\K_q\times\K_{q'}$ with $(q,q')\in\N_s^{*(2,<)}$ satisf\/ies by construction $i<j$, i.e.\ is an ordered pair $(i,j)\in\N_n^{*(2,\nD,<)}$.
\end{remark}

    \begin{proof}
Theorem~\ref{theo_1} is almost entirely a direct consequence of Proposition~\ref{prop_7}, the only unproved point being the re-ordering of each subset $\J_p=\bigcup\limits_{l=0}^{l_p}\I^{(p)}_l$ as an ordered partition, for any $p\in\PP_q$, with $q\in\N^*_s$. To this end, for any $l\in\N_{l_p}$, we f\/irst denote the elements of the subset $\I^{(p)}_l$, when not empty, by increasing values as $p^{(l)}_i$, with $i\in\N^*_{n^{(p)}_l}$.

If $l_p=0$, i.e.\ if $n^{(p)}_0\geq1$, then $\J_p=\I^{(p)}_0$ is a single free subset, and is already ordered. If $n^{(p)}_0\geq1$ and $l_p\geq1$, we def\/ine the permutation $\s_p\in\mf S_n$, whose support is a subset of $\J_p$, as
\[
\s_p(p^{(0)}_i)=i+p_q, \qquad \forall\, i\in\N^*_{n^{(p)}_0}
\]
and
\[
\s_p(p^{(l)}_i)=i+\sum_{l'=0}^{l-1}n^{(p)}_{l'}+p_q, \qquad \forall\, i\in\N^*_{n^{(p)}_l} \qquad \textrm{and}\qquad \forall\, l\in\N^*_{l_p}.
\]
If $n^{(p)}_0=0$, i.e.\ if $l_p\geq1$, then $\J_p=\bigcup\limits_{l=1}^{l_p}\I^{(p)}_l$ does not contain free indices, and we def\/ine the permutation $\s_p\in\mf S_n$ just as above, but omitting the f\/irst part of this def\/inition.

Therefore, the $\D$-class $\J_p$ can be written as the following ordered partition
\[
\J^\s_p=\s_p(\I^{(p)}_0)\cup\s_q(\J_p\setminus\I^{(p)}_0)=\bigcup_{l=0}^{l_p}\s_p\big(\I^{(p)}_l\big),
\]
where the exponent ``${}^\s$'' indicates that the permutation $\s_p$ is applied. Moreover, since the supports of the permutations $\{\s_p\}_{p\in\N^*_r}$ are disjoint and since the set $\{\J_p,\; p\in\N^*_r\}$ has a natural order from Proposition~\ref{prop_7}, the permutation $\s=\s_1\circ\cdots\circ\s_r\in\mf S_n$ re-orders as expected each element of the set of indices $\N^*_n$, i.e.\ $(\N^*_n)^\s=\bigcup\limits_{p=1}^r\J^\s_p$ is an ordered partition. Finally, for convenience, as earlier, we drop the exponent ``${}^\s$'', and identify the subsets $\{\J_p,\; p\in\N^*_r\}$ and the set of indices $\N^*_n$ with the re-ordered ones $\{\J^\s_p,\; p\in\N^*_r\}$ and $(\N^*_n)^\s$.
    \end{proof}

  \begin{corollary}\label{theo_1.cor}
In addition with the family of diagonal elements $(\D_{ii})_{i\in\N^*_n}$, the associated non-zero $R$-matrix elements to be determined are the coefficients
    \begin{enumerate}[label={\bf\roman*}$)$]\itemsep=0pt
\item $\D_{ij}$ for all pairs of indices $(i,j)$, $i$ and $j$ belonging to the same $\mc D$-class $\I^{(p)}_l$.
\item $\D_{ij}$, $\D_{ji}$, $d_{ij}$ and $d_{ji}$, for all $\nD$-pairs of indices $(i,j)$, $i$ and $j$ belonging to the same $\D$-class~$\J_p$. This covers the cases of indices $i$ and $j$
      \begin{itemize}\itemsep=0pt
\item both in the free subset $\I^{(p)}_0\neq\varnothing$, accordingly the corresponding contracted $R$-matrix will be refereed as ``full'' since all zero-weight elements are a priori non-zero;
\item in the free subset $\I^{(p)}_0\neq\varnothing$ and in a $\mc D$-class $\I^{(p)}_l$;
\item in two distinct $\mc D$-classes~$\I^{(p)}_l$ and~$\I^{(p)}_{l'}$, with $l<l'$.
      \end{itemize}
\item $\D_{ij}$, $d_{ij}$ and $d_{ji}$, for all $\nD$-pairs of indices $(i,j)$, $i$ and $j$ belonging to two distinct $\D$-classes~$\J_p$ and~$\J_{p'}$ of the same subset $\K_q$, i.e.\ with $(p,p')\in\PP_q^{(2,<)}$. This covers the cases of indi\-ces~$i$ and~$j$
      \begin{itemize}\itemsep=0pt
\item in the two free subsets $\I^{(p)}_0\neq\varnothing$ and $\I^{(p')}_0\neq\varnothing$;
\item in the free subset $\I^{(p)}_0\neq\varnothing$ and in a $\mc D$-class $\I^{(p')}_l$, as well as the non-equivalent symmetric case of any pair of indices in a $\mc D$-class~$\I^{(p)}_l$ and the free subset~$\I^{(p')}_0\neq\varnothing$;
\item in two $\mc D$-classes~$\I^{(p)}_l$ and $\I^{(p')}_{l'}$.
      \end{itemize}
\item $d_{ij}$ and $d_{ji}$, for all $\nD$-pairs of indices $(i,j)$, $i$ and $j$ belonging to two distinct subsets $\K_q$ and $\K_{q'}$, with $(q,q')\in\N_s^{*(2,<)}$.
    \end{enumerate}
  \end{corollary}

    \begin{proof}
This is a simple study of cases, when the indices $i$ and $j$ belong respectively to any possible subsets $\I^{(p)}_l$ and $\I^{(p')}_{l'}$, with $p,p'\in\N^*_r\; |\; p\leq p'$. Cases {\bf iii} and {\bf iv} are respectively reduced to $(p,p')\in\PP_q^{(2,<)}$ and $(q,q')\in\N_s^{*(2,<)}$ thanks to Proposition~\ref{prop_7} on the upper-triangularity of the reduced $\D$-incidence matrix $\mc M_R$.
    \end{proof}

\section{Resolution} \label{sec_4}

The resolution of the system (\ref{syst}) needs the introduction of the functions ``sum'' $S_{ij}$ and ``determinant'' $\Si_{ij}$ def\/ined, for any pair of indices $(i,j)\in(\N^*_n)^2$, as
  \[
S_{ij}=S_{ji}=\D_{ij}+\D_{ji}\qquad \textrm{and} \qquad \Si_{ij}=\Si_{ji}=
    \begin{vmatrix}
d_{ij}		& \D_{ij}
\\
\D_{ji}		& d_{ji}
    \end{vmatrix}\neq0.
  \]

\subsection{Preliminaries}
We will begin these f\/irst considerations on the resolution of system (\ref{syst}) by
solving cases {\bf i} and~{\bf iv} of
Corollary~\ref{theo_1.cor}. To this end, let $q\in\N^*_s$ and consider the subset $\K_q$.

\begin{proposition}[inside a $\mc D$-class]\label{prop_8}
Let $i\in\K_q\setminus(\K_q\cap\I_0)$. Then, there exists a non-zero constant $\D_{\I(i)}$ such that
the solution of system {\rm (\ref{syst})} restricted to the subset $\I(i)$ is given by
\[
\D_{jj'}=\D_{\I(i)},\qquad \forall\, j,j'\in\I(i).
\]
\end{proposition}

    \begin{proof}
This corresponds to the case {\bf i} of Corollary~\ref{theo_1.cor}. From the proof of Proposition~\ref{subprop_2a}, there exists a function $\D_{\I(i)}$ independent of the variable $\l_k$,
for any $k\in\I(i)$, such that $\D_{jj}=\D_{jj'}=\D_{j'j}=\D_{\I(i)}$, for any $j,j'\in\I(i)$, with $j\neq j'$. It remains to prove that $\D_{\I(i)}$ is a~1-periodic function in the variable $\l_k$, for any $k\in\N^*_n\setminus\I(i)$. If $n=2$, the proof of the proposition ends here.

Assuming that $n\geq3$, it is possible to suppose without loss of generality that $\I(i)\subsetneq\N^*_n$, the case of equality having been already treated. Let then $k\in\N^*_n\setminus\I(i)$. By construction, $(j,k)\in\N_n^{*(2,\nD)}$ and $(j',k)\in\N_n^{*(2,\nD)}$, then $(E_2)$ with indices $jj'k$ implies that $\D_{\I(i)}(k)=\D_{jj'}(k)=\D_{jj'}=\D_{\I(i)}$.

Reciprocally, it is straightforward to check that this is indeed a solution of system (\ref{syst}) restricted to the $\mc D$-class $\I(i)$. The set of solutions of system (\ref{syst}) restricted to a $\mc D$-class $\I(i)$ is exactly parametrized by the constant $\D_{\I(i)}$.
    \end{proof}

If the subset $\K_q$ is reduced to a single $\mc D$-class, the resolution ends here.

We must now consider that the subset $\K_q$ is not reduced to a single $\mc D$-class. In particular, there exists a $\nD$-pair of indices $(i,j)\in\K_q^{(2,\nD,<)}$. This also suggests to extend the notation $\D_{\I(i)}$ to $\D_{ii}$, even if $i\in\I_0$.

  \begin{corollary}\label{prop_8.cor}
For any pair of indices $(i,j)\in\K_q^{(2,\nD)}$ satisfying one of the cases {\bf ii}$)$--{\bf iv}$)$ of Corollary~{\rm \ref{theo_1.cor}}, $(F_1)$--$(F_9)$ of system {\rm (\ref{syst})} is equivalent to
  \begin{alignat}{4}
& (G_0),\;(F_1)\Leftrightarrow(F_2)\,\! && \ \D_{\I(i)}(i)=\D_{\I(i)}(j)=\D_{\I(i)}\neq0\  && (G'_0),&\nonumber\\
& \hphantom{(G_0),}\;(F_3)\Leftrightarrow(F_4) && \ \D_{\I(i)}\{\D_{ij}(i)-\D_{ij}\}-\D_{ji}\D_{ij}(i)=0\  && (F'_1),&\nonumber\\
& \hphantom{(G_0),}\;(F_5)\Leftrightarrow(F_6) && \ \D_{\I(i)}\{\D_{ji}(i)-\D_{ji}\}+\D_{ji}\D_{ij}(i)=0\  && (F'_2),&\nonumber\\
& \hphantom{(G_0),}\;(F_7) && \ \D_{\I(i)}\D_{ij}\{\D_{ii}-\D_{ij}\}-d_{ij}d_{ji}\D_{ij}(i)=0\  && (F'_3), &\!\!\!\!\!\!\!\! \label{syst'}\tag{$S'$}\\
& \hphantom{(G_0),}\;(F_8) && \ \D_{\I(i)}\D_{ji}(i)\{\D_{ii}-\D_{ji}(i)\}-d_{ij}(i)d_{ji}(i)\D_{ij}=0\  && (F'_4),&\nonumber\\
& \hphantom{(G_0),}\;(F_9) && \ \D_{\I(i)}\{d_{ij}(i)d_{ji}(i)-d_{ij}d_{ji}\}+\D_{ij}(i)\D_{ji}\{\D_{ij}(i)-\D_{ji}\}=0\;\; \! && (F'_5).&\nonumber
  \end{alignat}
  \end{corollary}

For later purpose, we will now introduce several lemmas, which restrain a priori the dependences of the $\D$-coef\/f\/icients (Lemmas~\ref{sublem_2a} and~\ref{sublem_2b}) and $d$-coef\/f\/icients (Lemmas~\ref{sublem_3a} and \ref{sublem_3b}) on the variable $\l_k$, for any $k\in\N^*_n$, as well as their acceptable form.
\begin{lemma}\label{sublem_2a}\label{lem_2}
Let $i\in\K_q\cap\I_0$. Then, $\D_{\I(i)}\neq0$ is a constant.
\end{lemma}
    \begin{proof}
Since $i\in\I_0$, for any $j\in\N^*_n\setminus\{i\}$, system (\ref{syst'}) applies to the pair of indices $(i,j)\in(\N^*_n)^2$ (cf.\ Corollary~\ref{prop_8.cor}). $(G'_0)$ then implies that $\D_{\I(i)}$ is constant.
    \end{proof}

\begin{lemma}\label{sublem_2b}
Let $(i,j)\in\K_q^{(2,\nD,<)}$ be a $\nD$-pair of indices. Then, $\D_{ij}$ and $\D_{ji}$ are $1$-periodic functions in the variable $\l_k$, for any $k\in\N^*_n\setminus(\I(i)\cup\I(j))$.
\end{lemma}
    \begin{proof}
If $n=2$, i.e.\ $\N^*_n\setminus(\I(i)\cup\I(j))=\varnothing$, the lemma is empty.

Assuming that $n\geq3$, let $k\in\N^*_n\setminus(\I(i)\cup\I(j))$. Since $(i,j,k)\in\N_n^{*(3,\nD)}$, $(E_2)$ with indices $ijk$ and with indices $jik$ implies that $\D_{ij}(k)=\D_{ij}$ and $\D_{ji}(k)=\D_{ji}$.
    \end{proof}

\subsection{Decoupling procedure}
This section is dedicated to the specif\/ic characterization of the decoupled $R$-matrices, as def\/ined in Proposition~\ref{subprop_1c}, the main result being that any $R$-matrix, solution of~\ref{DQYBE2}, characterized by a block-upper triangular matrix reduced $\D$-incidence matrix $\mc M_R$ with two or more triangular blocks, is in fact decoupled, up to a particular transformation explicited in the following. For the moment, let us focus on two fundamental lemmas, which describe the form of the non-zero $d$-coef\/f\/icients.
\begin{lemma}\label{sublem_3a}
Let $s\geq2$. Then, for any $(q,q')\in\N_s^{*(2,<)}$, there exists a non-zero constant $\Si_{qq'}$ and a family of non-zero functions $(g_{ij})_{\!(i,j)\in\K_q\times\K_{q'}}$ $($with the property that $g_{ij}g_{ji}=1)$, such that, for any $(i,j)\in\K_q\times\K_{q'}$
\[
\Si_{ij}=\Si_{qq'}, \qquad d_{ij}=\sqrt{\Si_{qq'}}g_{ij} \qquad \textrm{and}\qquad d_{ji}=\sqrt{\Si_{qq'}}g_{ji}.
\]
\end{lemma}

    \begin{proof}
Let $(i,j)\in\K_q\times\K_{q'}$, from Corollary~\ref{prop_8.cor}, system (\ref{syst'}) stands for the pair of indices $(i,j)$. Hence, from $(F'_5)$, we get that $\Si_{ij}(i)=d_{ij}(i)d_{ji}(i)=d_{ij}d_{ji}=\Si_{ij}$($=\Si_{ji}(j)$ by symmetry), i.e.\ the function $\Si_{ij}$ is 1-periodic in variables $\l_i$ and $\l_j$. If $n=2$, or if $n\geq2$ and $N_q=N_{q'}=1$, the pair $(i,j)$ is the only such pair of indices to consider.

Assuming that $N_q\geq2$ and $N_{q'}\geq1$, let $k\in\K_q\setminus\{i\}$. It follows that $\D_{ij}=\D_{ji}=\D_{kj}=\D_{jk}=0$ and $i\nD j\nD k$. From Proposition~\ref{prop_8} (when $k\mc Di$) or Lemma~\ref{sublem_2b} (when $k\nD i$), the function $\D_{ik}$ is a 1-periodic function in the variable $\l_j$. This implies, when used in $(E_6)$, $\Si_{ij}(k)=\Si_{kj}=\Si_{kj}(k)=\Si_{ij}$,\footnote{Theorem~\ref{theo_2} enunciates a similar result to this part of the reasoning for any pair of indices $(i,j)\in\K_q^{(2,\nD,<)}$.} which is the expected result if $N_{q'}=1$, as far as the determinant is concerned. If $N_q=1$ and $N_{q'}\geq2$, the symmetrical result is obtained by exchanging the indices $i$ and $j$, as well as the labels~$q$ and~$q'$.

Assuming that $N_q\geq2$ and $N_{q'}\geq2$, both previous results apply, so that the function~$\Si_{ij}$ does not depend on the index $i\in\K_q$ nor on the index $j\in\K_{q'}$, but only on the subsets~$\K_q$ and~$\K_{q'}$. There exists then a non-zero function, denoted $\Si_{qq'}$ by language abuse, which is 1-periodic in the variable $\l_k$, for any $k\in\K_q\cup\K_{q'}$, such that $\Si_{ij}=\Si_{qq'}$, for any $(i,j)\in\K_q\times\K_{q'}$. If $s=2$, i.e.\ $\N^*_n=\K_q\cup\K_{q'}$, the subsets $\K_q$ and $\K_{q'}$ are the only subsets to consider.

Assuming that $s\geq3$, let $q''\in\N^*_s\setminus\{q,q'\}$ and $k\in\K_{q''}$. Since $(i,j,k)\in\N_n^{*(3,\nD)}$, $(E_1)$ is non-trivial when written with indices $kji$ and $kji$. This yields that
  \begin{gather*}
d_{ki}(j)d_{ij}(k)d_{kj}d_{ji}(k)=d_{ki}d_{ij}d_{kj}(i)d_{ji}(k)=d_{ij}d_{kj}d_{ji}d_{ki}(j) \quad \Rightarrow \quad d_{ij}(k)d_{ji}(k)=d_{ij}d_{ji},
  \end{gather*}
implying that $\Si_{qq'}$ is also 1-periodic in the variable $\l_k$, for any $k\in\N^*_n\setminus(\K_q\cup\K_{q'})$, hence is constant. There exist then two non-zero functions $g_{ij}$ and $g_{ji}$, such that the functions $d_{ij}=\sqrt{\Si_{qq'}}g_{ij}$ and $d_{ji}=\sqrt{\Si_{qq'}}g_{ji}$ are the general solution of this equation, with the condition $g_{ij}g_{ji}=1$.
    \end{proof}

\begin{lemma}\label{sublem_3b}
Let $(i,j)\in\K_q^{(2,\nD,<)}$ be a $\nD$-pair of indices. Then, there exist two non-zero functions $d^0_{ij}$ and $d^0_{ji}$ of the variable $\l_k$, for any $k\in\I(i)\cup\I(j)$, and $1$-periodic in other variable, and two non-zero functions $g_{ij}$ and $g_{ji}$ $($with the property that $g_{ij}g_{ji}=1)$, such that
  \[
d_{ij}=g_{ij}d^0_{ij} \qquad \textrm{and} \qquad d_{ji}=g_{ji}d^0_{ji}.
  \]
\end{lemma}

\begin{proof}
From Corollary~\ref{prop_8.cor}, system (\ref{syst'}) stands for any pair of indices $(i,j)\in\K_q^{(2,\nD,<)}$.
$(F'_1)+(F'_2)$ implies that $S_{ij}(i)=S_{ij}$(=$S_{ij}(j)$ by symmetry), i.e.\ the function $S_{ij}$ is 1-periodic in va\-riab\-les $\l_i$ and $\l_j$. When inserted in $(F'_2)$, this yields
  \begin{gather*}
\D_{ji}\D_{ij}(i)=-\D_{\I(i)}\{\D_{ji}(i)-\D_{ji}\}=\D_{\I(i)}\{\D_{ij}(i)-\D_{ij}\} \\ \Leftrightarrow \quad \D_{\I(i)}\D_{ij}=\D_{ij}(i)\{\D_{\I(i)}-\D_{ji}\}.
  \end{gather*}
Hence, from $(F'_3)$, and since $\D_{ij}\neq0$, we deduce that
\[
\D_{ij}(i)[\{\D_{\I(i)}-\D_{ij}\}\{\D_{\I(i)}-\D_{ji}\}-d_{ij}d_{ji}]=0 \quad \Leftrightarrow \quad \Si_{ij}=\D_{\I(i)}\{\D_{\I(i)}-S_{ij}\},
\]
which implies that the function $\Si_{ij}$ is also 1-periodic in variables $\l_i$ and $\l_j$, because $\D_{\I(i)}$ is constant from Proposition~\ref{prop_8} (when $i\in\K_q\setminus(\K_q\cap\I_0)$) or Lemma~\ref{sublem_2a} (when $i\in\K_q\cap\I_0$). Moreover, by exchanging the indices $i$ and $j$, we get that
  \begin{gather}
\Si_{ij}=\D_{\I(i)}\{\D_{\I(i)}-S_{ij}\}=\D_{\I(j)}\{\D_{\I(j)}-S_{ij}\}, \qquad \forall\, (i,j)\in\K_q^{(2,\nD,<)}. \label{Si_ij}
  \end{gather}

Assuming that $\N^*_n\neq\I(i)\cup\I(j)$, let $k\in\N^*_n\setminus(\I(i)\cup\I(j))$. Otherwise, the lemma is trivial. From Proposition~\ref{prop_8} and Lemma~\ref{sublem_2b}, which express that the functions $\D_{ij}$ and $\D_{ji}$ are 1-periodic in the variable $\l_k$, (\ref{Si_ij}) implies that the functions $S_{ij}$ and $\Si_{ij}$ are 1-periodic in the variables $\l_i$, $\l_j$ and $\l_k$. Then, the function $d^0_{ij}=B_{ij}-\D_{ij}$ is a particular solution of
\[
d_{ij}d_{ji}=\Si_{ij}+\D_{ij}\D_{ji},
\]
being 1-periodic in the variable $\l_k$, for any $k\in\N^*_n\setminus(\I(i)\cup\I(j))$, where the quantity $B_{ij}=\frac{S_{ij}+\sqrt{S^2_{ij}+4\Si_{ij}}}2$ is a root of the polynomial $P_{ij}(X)=X^2-S_{ij}X-\Si_{ij}$. Hence there exists a~non-zero function $g_{ij}$ such that $d_{ij}=g_{ij}d^0_{ij}$ is the general solution of this equation, with the condition $g_{ij}g_{ji}=1$.
    \end{proof}

\begin{remark} Theorems~\ref{subtheo_3a},~\ref{subtheo_3b} and~\ref{theo_4} will provide explicit expressions for the non-zero functions $d^0_{ij}$ and $d^0_{ji}$, with $(i,j)\in\K_q^{(2,\nD,<)}$ and $q\in\N^*_s$, which will appear as the multiplicative invariant part of the $d$-coef\/f\/icients.

For $(i,j)\in\K_q^{(2,\nD,<)}$, with $q\in\N^*_s$, there exists a second realization of the functions~$d^0_{ij}$ and~$d^0_{ji}$, which also determine the functions $g_{ji}$, given by
\begin{gather*}
d^{0\prime}_{ij}=d^{0\prime}_{ji}=\sqrt{d^0_{ij}d^0_{ji}} \qquad \textrm{and} \qquad g'_{ij}=g_{ij}g^0_{ij}, \\ \textrm{with} \qquad g^0_{ij}=\sqrt{\frac{d^0_{ij}}{d^0_{ji}}}, \qquad \forall\, (i,j)\in\K_q^{(2,\nD,<)}.
\end{gather*}
Formally, this is the parametrization used in Lemma~\ref{sublem_3a}. In particular, both previous realizations of the functions $g^0_{ij}$ are also 1-periodic in the variable $\l_k$, for any $k\in\N^*_n\setminus(\I(i)\cup\I(j))$.

Moreover, let us point out that, extending the notation $d^0_{ij}$ to any pair of indices $(i,j)\in\K_q\times\K_{q'}$, with $(q,q')\in\N_s^{*(2,<)}$, i.e.\ if we set $d^0_{ij}=\sqrt{\Si_{qq'}}$ in this case, the family of non-zero functions $(d^0_{ij})_{(i,j)\in\N_n^{*(2,\!\!\nD\!\!)}}$ introduced by Lemmas~\ref{sublem_3a} and~\ref{sublem_3b} trivially satisf\/ies $(E_1)$.
\end{remark}

Since we have introduced all the needed tools, we can now separately study $(E_1)$ in details. This particular treatment is justif\/ied by the fact that this equation, which is the only equation where three $d$-coef\/f\/icients appear, is decoupled from other equations of system~(\ref{syst}). It only constrains the functions $g_{ij}$, with $(i,j)\in\N_n^{*(2,\nD,<)}$.

To this end, it is assumed that $n\geq3$, let $i,j,k\in\N^*_n$. Since $d$-coef\/f\/icients are concerned, it is possible to consider that the triplet $(i,j,k)$ is a $\nD$-triplet, i.e.\ $(i,j,k)\in\N_n^{*(3,\nD)}$. Otherwise, there exists $\{i',j'\}\subseteq\{i,j,k\}$ such that $i'\mc Dj'$, and $(E_1)$ becomes trivial.

We f\/irst establish that \ref{DQYBE2} shows another type of covariance, of which the twist cova\-rian\-ce is an example (cf.\ Proposition~\ref{subprop_1a}). This new symmetry of \ref{DQYBE2} is of great importance for characterizing the decoupled $R$-matrices. Let us now give the following def\/initions by analogy with~\cite{EtSchif}.

\begin{definition}[multiplicative 2-forms]\label{def_1}
Let $\I$ be a subset of the set of indices $\N^*_n$.
    \begin{enumerate}[label={\bf\roman*}.]\itemsep=0pt
\item A family of non-zero functions $(\a_{ij})_{(i,j)\in\I^{(2,\!\!\nD\!\!)}}$ (resp.\ $(\a_{ij})_{(i,j)\in\I^2}$) of the variable $\l$, such that $\a_{ij}\a_{ji}=1$, for any $(i,j)\in\I^{(2,\nD)}$ (resp.\ $(i,j)\in\I^2$), is called a $\nD$-multiplicative 2-form (resp. multiplicative 2-form).
\item A $\nD$-multiplicative 2-form (resp. multiplicative 2-form) $(\a_{ij})_{(i,j)\in\I^{(2,\!\!\nD\!\!)}}$ (resp.\ $(\a_{ij})_{(i,j)\in\I^2}$) is said to be $\nD$-closed (resp.\ closed), if it satisf\/ies the cyclic relation
\[
\frac{\a_{ij}(k)}{\a_{ij}}\frac{\a_{jk}(i)}{\a_{jk}}\frac{\a_{ki}(j)}{\a_{ki}}=1,
\qquad \forall\, (i,j,k)\in\I^{(3,\nD)} \qquad \textrm{(resp.} \quad \forall\, (i,j,k)\in\I^3\textrm).
\]
\item A $\nD$-multiplicative 2-form (resp.\ multiplicative 2-form) $(\a_{ij})_{(i,j)\in\I^{(2,\!\!\nD\!\!)}}$ (resp. $(\a_{ij})_{(i,j)\in\I^2}$) is said to be $\nD$-exact (resp.\ exact), if there exists a family of non-zero functions $(\a_i)_{i\in\I}$ of the variable $\l$, such that
\[
\a_{ij}=\frac{\a_i(j)}{\a_i}\frac{\a_j}{\a_j(i)}, \qquad \forall\, (i,j)\in\I^{(2,\nD)} \qquad \textrm{(resp.} \quad \forall\, (i,j)\in\I^2\textrm).
\]
    \end{enumerate}
  \end{definition}

\begin{proposition}\label{subprop_9a}\label{prop_9}
Let $(\a_{ij})_{(i,j)\in\I^{(2,\!\!\nD\!\!)}}$ $($resp.\ $(\a_{ij})_{(i,j)\in\I^2})$ be a $\nD$-closed $\nD$-multiplicative $2$-form $($resp.\ closed multiplicative $2$-form$)$. If the matrix $R$ is a solution of \ref{DQYBE2}, then the matrix
\[
R'=\sum_{i,j=1}^n\D_{ij}e^{(n)}_{ij}\otimes e^{(n)}_{ji}+\sum_{i\neq j=1}^n\a_{ij}d_{ij}e^{(n)}_{ii}\otimes e^{(n)}_{jj}.
\]
is also a solution of \ref{DQYBE2}.
\end{proposition}

    \begin{proof}
This is directly seen on system (\ref{syst}) and by remarking that the transformation def\/ined above respects the $\mc D$-classes, and then the ordered partition of the set of indices $\N^*_n$.

$(F_1)$--$(F_6)$ and $(E_2)$--$(E_5)$ are factorized by $d$-coef\/f\/icients, then either they are trivially verif\/ied (if $i\mc Dj$, when, for example, $(F_1)$--$(F_6)$ are considered with indices $ij$) or the $d$-coef\/f\/icients can be simplif\/ied (if $i\nD j$, for the same example).

$(F_7)$, $(F_8)$ and $(E_6)$ depend on $d$-coef\/f\/icients only through the product $d_{ij}d_{ji}$, which is clearly invariant under the previous transformation, since $(\a_{ij})_{(i,j)\in\I^{(2,\!\!\nD\!\!)}}$ (resp.\ $(\a_{ij})_{(i,j)\in\I^2}$) is a $\nD$-multiplicative 2-form (resp. multiplicative 2-form). The same kind of argument applies to $(E_1)$, which is also invariant, since $(\a_{ij})_{(i,j)\in\I^{(2,\!\!\nD\!\!)}}$ (resp. $(\a_{ij})_{(i,j)\in\I^2}$) is in addition assumed to be $\nD$-closed (resp.\ closed).
    \end{proof}

\begin{corollary}\label{subprop_9a.cor}
Let $\I$ be a subset of the set of indices $\N^*_n$ of cardinality $m$, and $(\a_{ij})_{(i,j)\in\I^{(2,\!\!\nD\!\!)}}$ $($resp.\ $(\a_{ij})_{(i,j)\in\I^2})$ be a $\nD$-closed $\nD$-multiplicative $2$-form $($resp.\ closed multiplicative $2$-form$)$. Following Proposition~{\rm \ref{subprop_1c}}, the previous proposition implies that the contracted matrix $(R')^\I$ of the matrix $R'$ to the subset $\I$ is  a solution of $\mc Gl_m(\C)$-\ref{DQYBE2}.
  \end{corollary}

\begin{proposition}\label{subprop_9b}
The family of non-zero functions $(g_{ij})_{\!(i,j)\in\N_n^{*(2,\!\!\nD\!\!,<)}}$, introduced in Lemmas~{\rm \ref{sublem_3a}} and~{\rm \ref{sublem_3b}}, is a $\nD$-closed $\nD$-multiplicative $2$-form.
\end{proposition}

    \begin{proof}
This is an obvious corollary of the remark of Lemmas~\ref{sublem_3a} and~\ref{sublem_3b}, since the family of non-zero functions $(d^0_{ij})_{(i,j)\in\N_n^{*(2,\!\!\nD\!\!,<)}}$ introduced by Lemmas~\ref{sublem_3a} and~\ref{sublem_3b} trivially satisf\/ies $(E_1)$. Then, using $g_{ij}g_{ji}=1$, $(E_1)$ with indices $ijk$ is simply the cyclic relation $\frac{g_{ij}(k)}{g_{ij}}\frac{g_{jk}(i)}{g_{jk}}\frac{g_{ki}(j)}{g_{ki}}=1$.
    \end{proof}

  \begin{corollary}
The $\nD$-closed $\nD$-multiplicative $2$-form $(g_{ij})_{\!(i,j)\in\N_n^{*(2,\!\!\nD\!\!,<)}}$ can be factorized out by $\nD$-multiplicative covariance, as described by Proposition~{\rm \ref{subprop_9a}}.
  \end{corollary}

\begin{proposition}\label{subprop_9c}
Any ($\nD$-)exact ($\nD$-)multiplicative $2$-form is $(\nD${\rm -)}closed.
\end{proposition}

\begin{remark}
Under the assumption that, for any $(i,j)\in\N_n^{*(2,<)}$, the non-zero function $\a_{ij}$ is a holomorphic function of the variable~$\l$ in a simply connected domain of $\C^n$, there exists a~multiplicative analog of the Poincar\'e lemma for dif\/ferential forms, the so-called multiplicative Poincar\'e lemma. It  enunciates that the reciprocal of Proposition~\ref{subprop_9c} is also true, that is: a~multiplicative 2-form $(\a_{ij})_{(i,j)\in\N_n^{*(2,<)}}$ is exact, if and only if it is closed~\cite{EtSchif}. This directly implies that the multiplicative covariance of Proposition~\ref{subprop_9a} coincides under this assumption with the twist covariance of Proposition~\ref{subprop_1a}.

\looseness=-1
In particular, if $\I_0=\N^*_n$, i.e.\ if the set of indices $\N^*_n$ only contain free indices, a $\nD$-closed $\nD$-multiplicative 2-form is a closed multiplicative 2-form, and then is exact, which is the case e.g.\ for (weak) Hecke-type solutions of \ref{DQYBE2}\footnote{For more details, see Subsection~\ref{subsec_5_5}.}. In this case, by analogy with dif\/ferential forms, the closed multiplicative 2-form $(\a_{ij})_{(i,j)\in\N_n^{*(2,<)}}$ will be refereed as a gauge 2-form, since it can universally be factorized out thanks to Proposition~\ref{subprop_1a}, in the sense that it is representation-independent.

Considering the general problem, we do not succeed to solve whether or not any $\nD$-closed $\nD$-multiplicative 2-form is $\nD$-exact. However, as we will see in the proof of the following proposition, it does not really matter in practice, since the notion of $\nD$-multiplicative covariance is actually the minimal main tool allowing to achieve the characterization of the decoupled $R$-matrices.

\looseness=-1
The issue which therefore remains is to get a general classif\/ication of $\nD$-closed $\nD$-multiplica\-ti\-ve 2-forms, when the $\mc D$-classes have a non-trivial structure. Note that if the set of indices~$\N^*_n$ is split into {\it two} $\mc D$-classes any $\nD$-multiplicative 2-form is $\nD$-closed, since no cyclic relation exists.
\end{remark}

\begin{proposition}\label{prop_10}
Any $R$-matrix, solution of \ref{DQYBE2}, characterized by a block-upper triangular reduced $\D$-incidence matrix $\mc M_R$ with two or more triangular blocks, is $\nD$-multiplicatively reducible to a decoupled $R$-matrix, and vice versa.
\end{proposition}

    \begin{proof}
This results from successive implementations of Proposition~\ref{subprop_1b}.

Assuming $s\geq2$, let $(q,q')\in\N_s^{*(2,<)}$ and consider the solutions of system (\ref{syst}) restricted to the subsets $\K_q$ and $\K_{q'}$. According to Proposition~\ref{subprop_1c}, matrix elements of the $R$-matrix with both indices either in the subset $\K_q$ or in the subset $\K_{q'}$ realize a contraction-type solution of a~lower-dimensional, more precisely of a $N_q$-dimensional or $N_{q'}$-dimensional~\ref{DQYBE2}. Due to the block-upper triangularity, the only remaining non-zero matrix elements are the $d$-coef\/f\/i\-cients~$d_{ij}$, with $(i,j)\in\K_q\times\K_{q'}$.

Lemma~\ref{sublem_3a} now solves this issue, for any pair of labels $(q,q')\in\N_s^{*(2,<)}$. Indeed, Lemmas~\ref{sublem_3a} and \ref{sublem_3b}, and Proposition~\ref{subprop_9b} prove the existence of a $\nD$-closed $\nD$-multiplicative 2-form $(g_{ij})_{\!(i,j)\in\N_n^{*(2,\!\!\nD\!\!,<)}}$, such that
\[
d_{ij}=\sqrt{\Si_{qq'}}g_{ij} \qquad \textrm{and} \qquad d_{ji}=\sqrt{\Si_{qq'}}g_{ji}, \qquad \forall\, (i,j)\in\K_q\times\K_{q'},
\]
where $\Si_{qq'}$ is a non-zero constant, or
\[
d_{ij}=g_{ij}d^0_{ij} \qquad \textrm{and} \qquad d_{ji}=g_{ji}d^0_{ji}, \qquad \forall\, (i,j)\in\K_q^{(2,\nD,<)},
\]
where the functions $d^0_{ij}$ and $d^0_{ji}$ are 1-periodic in the variable $\l_k$, for any $k\in\N^*_n\setminus\K_q$. Moreover, from Lemmas~\ref{sublem_2a} and \ref{sublem_2b}, the $\D$-coef\/f\/icients $\D_{ii'}$, with $i,i'\in\K_q$, are also 1-periodic in the variable~$\l_k$, for any $k\in\N^*_n\setminus\K_q$, with $q\in\N^*_s$.

From Proposition~\ref{subprop_9b} and its corollary, the $\nD$-multiplicative 2-form $(g_{ij})_{\!(i,j)\in\N_n^{*(2,\!\!\nD\!\!,<)}}$ is $\nD$-closed, and then can precisely be factorized out by the $\nD$-multiplicative covariance. This brings, on the one hand, the $d$-coef\/f\/icients $d_{ij}$, with $(i,j)\in\K_q\times\K_{q'}$ and $(q,q')\in\N_s^{*(2,<)}$, to be equal to an overall block-pair dependent constant $\sqrt{\Si_{qq'}}$, and, on the other hand, the $d$-coef\/f\/icient $d_{ij}$ to be equal to $d^0_{ij}$, for any $(i,j)\in\K_q^{(2,\nD)}$.

To summarize, any solution of \ref{DQYBE2} def\/ined by its block-upper triangular reduced $\D$-incidence matrix $\mc M_R$ is necessarily $\nD$-multiplicatively covariant to a multiply decoupled $R$-matrix obtained from successive applications of Proposition~\ref{subprop_1b}. But this proposition shows that such decoupled $R$-matrices are also solutions of \ref{DQYBE2}. The reciprocal is obvious.
    \end{proof}

  \begin{corollary}
It is therefore relevant to focus our discussion of solutions of system {\rm (\ref{syst})} to the cases {\rm {\bf ii}} and~{\rm {\bf iii}} of Corollary~{\rm \ref{theo_1.cor}}, where the indices $i,j\in\K_q$, with $q\in\N^*_s$ and $N_q\geq2$.
  \end{corollary}

\subsection{Sum and determinant}
In this section, as stated in the following fundamental result, the
functions $S_{ij}$ and $\Si_{ij}$ are shown to be actually constant
independent of indices $i$ and $j$, as soon as the pair of indices $(i,j)$
is a~$\nD$-pair, depending then only on the subset $\K_q$, i.e.\ only on the label $q$.
Moreover, they actually parametrize the set of solutions of system (\ref{syst}) restricted to
the subset $\K_q$, to be specif\/ied later.

\begin{theorem}[inside a set $\K_q$ of $\D$-classes]\label{theo_2}
There exist a constant $S_q$ and a non-zero constant~$\Si_q$ such that
\[
S_{ij}=S_q \qquad \textrm{and} \qquad \Si_{ij}=\Si_q, \qquad \forall\,  (i,j)\in\K_q^{(2,\nD,<)};
\]
refereed as the ``sum'' and the ``determinant'' in the subset~$\K_q$.

Moreover, denoting $D_q=\sqrt{S_q^2+4\Si_q}$ the ``discriminant'' in the subset $\K_q$, there exists a~fa\-mi\-ly of $\sum\limits_{p\in\PP_q}(n^{(p)}_0+l_p)$ signs $(\e_\I)_{\I\in\{\I(i),\; i\in\K_q\}}$ such that
    \begin{gather}
\D_{\I(i)}=\frac{S_q+\e_{\I(i)}D_q}2, \qquad \forall \, i\in\K_q. \label{D_I(i)_s}
    \end{gather}
\end{theorem}

    \begin{proof}
If $n=2$, i.e.\ if $\N^*_n=\K_q=\{i,j\}$, the theorem is a direct corollary of the proof of Lemma~\ref{sublem_3b} and (\ref{Si_ij}).

Assuming that $n\geq3$, the proof of Lemma~\ref{sublem_3b} and (\ref{Si_ij}) shows that the functions~$S_{ij}$ and~$\Si_{ij}$ are 1-periodic in the variables $\l_i$, $\l_j$ and $\l_k$, with $k\in\N^*_n\setminus(\I(i)\cup\I(j))$.  If $N_q=2$, i.e.\ if $\K_q=\{i,j\}$, the proof of the theorem ends.

The proof goes now in three steps.

{\bf 1.~Periodicity.} Assuming that $N_q\geq3$, it becomes possible to introduce a~third index $k\in\K_q\setminus\{i,j\}$. Two symmetrical possibilities $k\nD i$ or $k\nD j$ are to be considered. Indeed, ad absurdum, $k\mc Di$ and $k\mc Dj$ leads to the contradiction $i\mc Dj$.
        \begin{itemize}\itemsep=0pt
\item If $k\nD j$, any $d$-coef\/f\/icient involving one of the indices $i$, $k$ and the index $j$ is non-zero. Moreover, since $i,k\in\K_q$, $\D_{ik}\neq0$ or $\D_{ki}\neq0$. Without loss of generality, it is possible to assume that $\D_{ik}\neq0$, the case $\D_{ki}\neq0$ being treated similarly by exchanging the indices~$i$ and~$k$. Hence $(E_4)$ and $(E_5)$ both with indices $ijk$ give
          \begin{gather}
\D_{ij}(k)\D_{jk}+\D_{ik}\D_{ji}(k)=\D_{ik}\D_{jk}, \tag{$E'_4$}
          \end{gather}
and
          \begin{gather}
\D_{ij}(k)\D_{jk}+\D_{ik}(j)\D_{kj}=\D_{ik}(j)\D_{ij}(k).\tag{$E'_5$}
          \end{gather}
From Proposition~\ref{prop_8} (when $k\mc Di$) and from Lemma~\ref{sublem_2b} (when $k\nD i$), the function $\D_{ik}$ is 1-periodic in the variable $\l_j$. Then, by the substraction $(E'_4)$--$(E'_5)$, we get that
\[
\D_{ik}\{S_{ij}(k)-S_{kj}\}=0.
\]
From which we deduce that $S_{ij}(k)=S_{kj}$. However, we have seen that the function $S_{kj}$ is 1-periodic in the variable $\l_k$, since $k\nD j$. The function $S_{ij}$ is thus also 1-periodic in the variable $\l_k$, for any $k\in\K_q\setminus(\{i\}\cup\I(j))$. Moreover, we obtain that
\[
S_{kj}=S_{ij}
\]
and
\[
\Si_{kj}=\D_{\I(j)}\{\D_{\I(j)}-S_{kj}\}=\D_{\I(j)}\{\D_{\I(j)}-S_{ij}\}=\Si_{ij},\qquad \forall\, (k,j)\in(\K_q\setminus\{i\})^{(2,\nD)}.
\]

\item If $k\nD i$, the previous reasoning is symmetrically done, exchanging the indices $i$ and $j$. This yields that the function $S_{ij}$ is 1-periodic in the variable $\l_k$, for any $k\in\K_q\setminus(\I(i)\cup\{j\})$, and
\[
S_{ik}=S_{ij} \qquad \textrm{and} \qquad \Si_{ik}=\Si_{ij}, \qquad \forall\, (i,k)\in(\K_q\setminus\{j\})^{(2,\nD)}.
\]
        \end{itemize}
If $\I(j)=\{j\}$, or if $\I(i)=\{i\}$, the functions $S_{ij}$ and $\Si_{ij}$ are thus in particular respectively 1-periodic in the variable $\l_k$, for any $k\in\I(i)\setminus\{i\}$ or $k\in\I(j)\setminus\{j\}$, from application of the f\/irst or the second previous point, and then are constant from above.

Assuming that $|\I(i)|\geq2$ and $|\I(j)|\geq2$, both previous points apply, implying that the functions $S_{ij}$ and $\Si_{ij}$ are actually 1-periodic in variable $\l_k$, with $k\in(\I(i)\cup\I(j))\setminus\{i,j\}$, and then are constant in this case too. This ends the proof of the periodicity property expressed in the theorem.

 {\bf 2.~Existence of $\boldsymbol{S_q}$ and $\boldsymbol{\Si_q}$.}
Always under the assumption that $N_q\geq3$, Step~1 has been seen to justify the existence of a $\nD$-pair of indices $(i',j')\in\K_q^{(2,\nD)}$, distinct from the $\nD$-pair $(i,j)$. Since these two $\nD$-pairs are distinct, it is always possible to impose that $i'\in\K_q\setminus\{i,j\}$. This suggests to rather adopt the notation $(k,j')$, where $k\in\K_q\setminus\{i,j\}$.
        \begin{itemize}\itemsep=0pt
\item If $j'\in\{i,j\}$, then $k\nD j'\nD i'$, where we def\/ine the index $i'\in\{i,j\}$ so that $\{i',j'\}=\{i,j\}$. From Step~1, if $j'=j$, the case $j'=i$ being treated similarly by exchanging the indices $i$ and $j$, we directly deduce that $S_{kj'}=S_{j'i'}=S_{ij}$ and $\Si_{kj'}=\Si_{ij}$.

If $k\mc Di'$, the pairs of indices $(k,j')$ and $(j',i')$ are the only $\nD$-pairs in $\{i',j',k\}$.

On the contrary, if $k\nD i'$, i.e.\ if $(i',j',k)\in\K_q^{(3,\nD)}$, the pair of indices $(k,i')$ has to be also considered. The result we have just obtained applies to the indices $k\nD i'\nD j'$, leading to the second needed set of equations $S_{ki'}=S_{ij}$ and $\Si_{ki'}=\Si_{ij}$.

In particular, if $N_q=3$, i.e.\ if $\K_q=\{i,j,k\}$, the existence of the constant $S_q$ is proved.
\item Assuming that $N_q\geq4$, since the f\/irst point of this reasoning already dealt with the case $j'\in\{i,j\}$, we can consider here without loss of generality the case $j'\notin\{i,j\}$.

However, since once more either $k\nD i$ or $k\nD j$, and either $j'\nD i$ or $j'\nD j$, there exists $j_1,j_2\in\{i,j\}$, such that $k\nD j_1$ and $j'\nD j_2$. Def\/ining the index $i_1\in\{i,j\}$ so that $\{i_1,j_1\}=\{i,j\}$, then $j_2\in\{i_1,j_1\}$.

If $j_2=j_1$, then $k\nD j'\nD j_1\nD i_1$, and the f\/irst point of this reasoning applies successively to the subsets $\{k,j',j_1\}$ and $\{j',j_1,i_1\}$, implying that
\[
S_{kj'}=S_{j'j_1}=S_{j_1i_1}=S_{ij} \quad \Rightarrow \quad \Si_{kj'}=\Si_{ij}.
\]

If $j_2=i_1$, then $k\nD j'\nD j_2\nD j_1$, and the f\/irst point of this reasoning applies successively to the subsets $\{k,j',j_2\}$ and $\{j',j_2,j_1\}$, implying that
\[
S_{kj'}=S_{j'j_2}=S_{j_2j_1}=S_{ij} \quad \Rightarrow \quad \Si_{kj'}=\Si_{ij}.
\]
        \end{itemize}
This implies that there exists two constants $S_q=S_{ij}$ and $\Si_q=\Si_{ij}$, such that
\[
S_{i'j'}=S_q \qquad \textrm{and} \qquad \Si_{i'j'}=\Si_q, \qquad \forall\, (i',j')\in\K_q^{(2,\nD)},
\]
ending the proof of the f\/irst part of the theorem.

{\bf 3.~Existence of $\boldsymbol{(\e_\I)_{\I\in\{\I(i),\; i\in\K_q\}}}$.}
The previous two steps now imply, from~(\ref{Si_ij}), that the family of constants $(\D_\I)_{\I\in\{\I(i),\; i\in\K_q\}}$, $S_q$ and $\Si_q$ satisfy the following quadratic equation
        \begin{gather}
\D_{\I(i)}^2-S_q\D_{\I(i)}-\Si_q=0. \label{D_I(i)_e} \tag{$\ref{D_I(i)_s}'$}
        \end{gather}
From which we deduce the existence of a sign $\e_{\I(i)}\in\{\pm\}$ for each of the $\sum\limits_{p\in\PP_q}(n^{(p)}_0+l_p)$ $\mc D$-classes $\I(i)\subseteq\K_q$.
    \end{proof}

\begin{remark}
 We have to insist on the fact that (\ref{D_I(i)_e}) does not impose that the constant $\D_{\I(i)}$ and the sign $\e_{\I(i)}$ to be independent from the $\mc D$-class $\I(i)$. Considering a $\nD$-pair of indices $(i,j)\in\K_q^{(2,\nD)}$, $\D_{\I(i)}$ and $\D_{\I(j)}$ are solutions of (\ref{D_I(i)_e}), which is equivalent to $\e_{\I(i)}=\pm\e_{\I(j)}$, and does not indeed constrain the family of signs $(\e_\I)_{\I\in\{\I(i),\; i\in\K_q\}}$. This remark will be crucial later to distinguish between Hecke, weak Hecke and non-Hecke type solutions\footnote{For more details, see Subsection~\ref{subsec_5_5}, and particularly Propositions~\ref{subprop_15a} and~\ref{subprop_15b}.}.
 \end{remark}

\subsection[Inside a $\D$-class]{Inside a $\boldsymbol{\D}$-class}

This section will present in details the explicit resolution of system (\ref{syst})
restricted to any $\D$-class~$\J_p$, with $p\in\PP_q$, of the subset $\K_q$.
Case {\bf i} of Corollary~\ref{theo_1.cor} being already solved in Theo\-rem~\ref{theo_2}
thanks to Proposition~\ref{prop_8}, we have to focus on case~{\bf ii}, in which any pair of indices $(i,j)$ under study is a $\nD$-pair. In general, the solution will be parametrized by the values of the sum $S_q$ and the constant $T_q=\frac{D_q-S_q}{D_q+S_q}\in\C^*=\C\setminus\{0\}$, in addition with the family of signs $(\e_\I)_{\I\in\{\I(i),\; i\in\K_q\}}$. More precisely we will see that three cases are to be distinguished.

\begin{remark}
The quantity $T_q$ is well def\/ined and is non-zero, for any constants $S_q$
and $\Si_q\neq0$ (since, by construction, $D_q=\sqrt{S_q^2+4\Si_q}\neq\pm S_q$).
Moreover, $S_q=0$ if and only if $T_q=1$.

When $S_q\neq0$, there exists $t_q=\frac{D_q}{S_q}\in\C\setminus\{\pm1\}$ such
that $T_q=-\frac{1-t_q}{1+t_q}=-\frac{1-|t_q|^2-2\i\Im(t_q)}{|1+t_q|^2}\in\C^*$, where the limit $|t_q|\rightarrow\infty$, or equivalently the limit $S_q\rightarrow0$, exists. We then deduce that $T_q\in\R^*_-$, if and only if $t_q\in\; ]{-}1,1[$.
\end{remark}

  For later purpose, when $S_q\neq0$ (when $T_q\neq1$), we also introduce the non-zero constants $A_q$ and $B_q$, viewed as functions of $T_q$, def\/ined as
\[
A_q=
           \begin{cases}
\log(T_q)		& \textrm{if}~T_q\notin\R^*_-,
\\
\log(-T_q)-\i\pi	& \textrm{if}~T_q\in\R^*_-,
      \end{cases} \qquad \textrm{and}\qquad B_q=\frac{S_q+D_q}2=\frac{S_q}{1-\ee^{A_q}},
\]
where the principal value of the function $\log$: $\C\setminus\R_-\longrightarrow\C$ is used when needed. When, $S_q=0$, the constant $B_q$ can also be def\/ined, and is equal to $\sqrt{\Si_q}$.

Before beginning the resolution, we need to introduce the following technical lemma.

\begin{lemma}[multiplicative shift] \label{sublem_4a}
Let $A\in\C$, $i\in\K_q$ and a family of non-zero functions $(\b_j)_{j\in\I(i)}$ of the variable $\l_k$, for any $k\in\I(i)$, and $1$-periodic in any other variable, such that, for any $j\in\I(i)$ and $j'\in\I(i)\setminus\{j\}$
    \begin{gather}
\b_j(j)=\ee^{A\e_{\I(i)}}\b_j \qquad \textrm{and} \qquad \b_j(j')=\ee^{A\e_{\I(i)}}\b_{j'}. \label{f_i_e}
    \end{gather}
Then, there exists a non-zero constant $f_{\I(i)}$ such that, for any $j\in\I(i)$
    \begin{gather}
\b_j=\b_{\I(i)}=\ee^{A\e_{\I(i)}\L_{\I(i)}}f_{\I(i)}, \label{f_i_s} \tag{$\ref{f_i_e}'$}
    \end{gather}
where we define the variable $\L_{\I(i)}=\sum\limits_{k\in\I(i)}\l_k$.
\end{lemma}

    \begin{proof}
The case $A=0$ being trivial, we will focus on the case $A\in\C^*$.

If $|\I(i)|=1$, i.e.\ $\I(i)=\{i\}$, (\ref{f_i_e}) reduces to $\b_i(i)=\ee^{A\e_{\I(i)}}\b_i$. Hence there exists a non-zero function $f_i$, such that $\b_i=\ee^{A\e_{\I(i)}\l_i}f_i$, and the proof of the lemma ends.

Assuming that $|\I(i)|\geq2$, from (\ref{f_i_e}), for any $j\in\I(i)\setminus\{i\}$, we deduce that $\b_i(j)=\ee^{A\e_{\I(i)}}\b_j=\b_j(j)$, i.e.\ $\b_i=\b_j$. Hence there exists a non-zero function $\b_{\I(i)}$ of the variable $\l_k$, for any $k\in\I(i)$, and 1-periodic in any other variable, such that
\[
\b_{\I(i)}=\b_j, \qquad \forall\, j\in\I(i).
\]
From (\ref{f_i_e}), the function $\b_{\I(i)}$ satisf\/ies
    \begin{gather}
\b_{\I(i)}(j)=\ee^{A\e_{\I(i)}}\b_{\I(i)},~~\forall\, j\in\I(i). \label{b_I(i)_e} \tag{$\ref{f_i_e}''$}
    \end{gather}
We now def\/ine the function $f_{\I(i)}$ of the variable $\l_k$, for any $k\in\I(i)$, and 1-periodic in any other variable as $f_{\I(i)}=\ee^{-A\e_{\I(i)}\b_{\I(i)}\L_{\I(i)}}\b_{\I(i)}$. From (\ref{b_I(i)_e}), we directly deduce that $f_{\I(i)}$ is now periodic in the variable $\l_k$, for any $k\in\I(i)$, and then is constant.
    \end{proof}

This result possesses an obvious linear limit.
\begin{lemma}[additive shift]\label{sublem_4b}
Let $i\in\K_q$ and a family of functions $(\b_j)_{j\in\I(i)}$ of the va\-riab\-le~$\l_k$, for any $k\in\I(i)$, and $1$-periodic in any other variable, such that, for any $j\in\I(i)$ and $j'\in\I(i)\setminus\{j\}$
    \begin{gather}
\b_j(j)=\b_j+\e_{\I(i)} \qquad \textrm{and} \qquad \b_j(j')=\b_{j'}+\e_{\I(i)}. \label{f^0_i_e}
    \end{gather}
Then, there exists a constant $f_{\I(i)}$ such that, for any $j\in\I(i)$
    \begin{gather}
\b_j=\b_{\I(i)}=\e_{\I(i)}\L_{\I(i)}+f_{\I(i)}. \label{f^0_i_s} \tag{$\ref{f^0_i_e}'$}
    \end{gather}
\end{lemma}

    \begin{proof}
Let $a\in\C$ and introduce, for any $j\in\I(i)$, the function of the variable $\l_k$, for any $k\in\I(i)$
\[
\b^a_j=\ee^{a\b_j}.
\]
By construction, the family of functions $(\b^a_j)_{j\in\I(i)}$ satisf\/ies the assumptions of Lemma~\ref{sublem_4a}. Hence, there exists a non-zero constant $f^a_{\I(i)}$ such that
\[
\b^a_j=\b^a_{\I(i)}=\ee^{a\e_{\I(i)}\L_{\I(i)}}f^a_{\I(i)}, \qquad \forall\, j\in\I(i).
\]
However, for any $j\in\I(i)$, the function $a\longmapsto \b^a_j$ is holomorphic on $\C$, and then as well as the functions $a\longmapsto \b^a_{\I(i)}$ and $a\longmapsto f^a_{\I(i)}$. These three functions admit a Taylor expansion in the neighboorhood of 0. In particular, there exists a constant $f_{\I(i)}=\frac\dd{\dd a}f^a_{\I(i)}\big|_{a=0}$ such that
\begin{gather*}
\b^a_j=f^0_{\I(i)}+a\b_j+o(a)=f^0_{\I(i)}+a(\e_{\I(i)}\L_{\I(i)}+f_{\I(i)})+o(a) \\
 \Rightarrow \quad \b_j=\e_{\I(i)}\L_{\I(i)}+f_{\I(i)}=\b_{\I(i)}.\tag*{\qed}
\end{gather*}
  \renewcommand{\qed}{}
    \end{proof}

We now enunciate the fundamental result of the resolution in any $\D$-class $\J_p$ as well as in Theorems~\ref{theo_4} and~\ref{theo_5}, as justif\/ied by the special treatment of Propositions~\ref{subprop_9a},~\ref{subprop_9b} and~\ref{subprop_9c}.

\begin{theorem}[trigonometric behavior]\label{subtheo_3a}
Assuming that $S_q\neq0$, i.e.\ $T_q\in\C^*\setminus\{1\}$, there exist a family of $n^{(p)}_0+l_p$ non-zero constants $(f_\I)_{\I\in\{\I(i),\; i\in\J_p\}}$ $($with the convention that $f_{\I(\min\J_p)}=1)$, and a $\nD$-multiplicative $2$-form $(g_{ij})_{\!(i,j)\in\J_p^{(2,\!\!\nD\!\!,<)}}$, such that the solution of system {\rm (\ref{syst})} restricted to the $\D$-class $\J_p$ is given by the following expressions
  \begin{gather*}
\D_{\I(i)}=\frac{S_q}{1-\ee^{A_q\e_\I(i)}}, \qquad \forall\,  i\in\J_p;
  \\
\D_{ij}((\l_k)_{k\in\I(i)\cup\I(j)})=\frac{S_q}{1-\ee^{A_q(\e_{\I(i)}\L_{\I(i)}-\e_{\I(j)}\L_{\I(j)})}
\frac{f_{\I(i)}}{f_{\I(j)}}}=\D_{\I(i)\I(j)}(\L_{\I(i)},\L_{\I(j)})
  \end{gather*}
and
\[
d_{ij}=g_{ij}\{B_q-\D_{\I(i)\I(j)}\},\qquad \forall\, (i,j)\in\J_p^{(2,\nD)}.
\]
\end{theorem}
    \begin{proof}

{\bf 1.~Diagonal $\boldsymbol{\D}$-coef\/f\/icients.}
From (\ref{D_I(i)_s}), if we adopt for a time the notations $A^\pm_q=\log(\pm T_q)$, where the exponent ``${}^\pm $'' means respectively that $T_q\notin\R^*_-$ or $T_q\in\R^*_-$, we deduce that
  \begin{gather*}
1-\frac{S_q}{\D_{\I(i)}}=\frac{\e_{\I(i)}D_q-S_q}{\e_{\I(i)}D_q+S_q}=T^{\e_{\I(i)}}_q=\pm\ee^{A^\pm_q\e_{\I(i)}}=\ee^{A_q\e_{\I(i)}}\\
\Leftrightarrow \quad \D_{{\I(i)}}=\frac{S_q}{1\mp\ee^{A^\pm_q\e_{\I(i)}}}=\frac{S_q}{1-\ee^{A_q\e_{\I(i)}}}.
  \end{gather*}
For the rest of the article, we will omit to make the explicit split between the cases $T_q\notin\R^*_-$ and $T_q\in\R^*_-$, unless otherwise stated. If $n_p=1$ or if the $\D$-class $\J_p$ is reduced to a single $\mc D$-class (cf.\ Proposition~\ref{prop_8}), the proof of the theorem ends here (cf.\ case {\bf i} of Corollary~\ref{theo_1.cor}).

{\bf 2.~Of\/f-diagonal $\boldsymbol{\D}$-coef\/f\/icients.}
Assuming that $n_p\geq2$ and that $\J_p$ is not reduced to a~$\mc D$-class, there exists a $\nD$-pair of indices $(i,j)\in\J_p^{(2,\nD)}$. Let $(i',j')\in\J_p^{(2,\nD)}$ be a $\nD$-pair of indices such that $(i',j')\in\I(i)\times\I(j)$. From Lemmas~\ref{sublem_2a} and \ref{sublem_2b}, only the dependence in the variable $\l_k$, for any $k\in\I(i)\cup\I(j)$, of the function $\D_{i'j'}$ remains to be determined. To this end, $(F'_1)\Leftrightarrow(F'_2)$ with indices $i'j'$ is re-written, since $\D_{\I(i)}\D_{\I(j)}\D_{\I(i)\I(j)}\D_{\I(j)\I(i)}\neq0$, as
  \[
\frac{\D_{\I(i)}}{\D_{i'j'}(i')}-1=\!\left\{1-\frac{S_q}{\D_{\I(i)}}\right\}\!\frac{\D_{\I(i)}}{\D_{i'j'}}
=\ee^{A_q\e_{\I(i)}}\frac{\D_{\I(i)}}{\D_{i'j'}} \quad \Leftrightarrow \quad \frac1{\D_{i'j'}(i)}=\frac{\ee^{A_q\e_{\I(i)}}}{\D_{i'j'}}+\frac1{\D_{\I(i)}}.
  \]
Denoting $\b_{i'j'}=\frac{S_q}{\D_{i'j'}}-1=\frac{\D_{j'i'}}{\D_{i'j'}}\neq0$, we deduce that
  \[
\b_{i'j'}(i')=\ee^{A_q\e_{\I(i)}}\frac{S_q}{\D_{i'j'}}+\frac{S_q}{\D_{\I(i)}}-1=\ee^{A_q\e_{\I(i)}}\b_{i'j'},
  \]
and by symmetry
  \[
\b_{i'j'}(j')=\ee^{-A_q\e_{\I(j)}}\b_{i'j'}.
  \]
        \begin{itemize}\itemsep=0pt
\item If $|\I(j)|=1$, i.e.\ if $\I(j)=\{j\}$, the only $\nD$-pairs to consider are $(i',j)$, for any $i'\in\I(i)$. From above, the function $\b_{i'j}$ satisf\/ies
  \[
\b_{i'j}(i')=\ee^{A_q\e_{\I(i)}}\b_{i'j} \qquad \textrm{and} \qquad \b_{i'j}(j)=\ee^{-A_q\e_{\I(j)}}\b_{i'j}.
  \]
\item If $|\I(j)|\geq2$, i.e.\ if $\I(j)$ is a $\mc D$-class, let $k\in\I(j)$ and $k'\in\I(j)\setminus\{k\}$. Since $k'\nD i'\nD k$, for any $i'\in\I(i)$, we have
  \[
\b_{i'k}(k)=\ee^{-A_q\e_{\I(j)}}\b_{i'k} \qquad \textrm{and} \qquad \b_{i'k'}(k')=\ee^{-A_q\e_{\I(j)}}\b_{i'k'}.
  \]
Moreover, $(E'_4)$ can be used with indices $i'kk'$, and yields
          \begin{gather}
\b_{i'k}(k')=\ee^{-A_q\e_{\I(j)}}\b_{i'k'}. \label{h_ij}
          \end{gather}
        \end{itemize}
In both cases, Lemma~\ref{sublem_4a} is now applied to the family of functions $(\b_{i'k})_{k\in\I(j)}$ of the variable~$\l_k$, for any $k\in\I(i)\cup\I(j)$. Hence, there exists a non-zero function $\b_{i'\I(j)}$ of the variable~$\l_k$, for any $k\in\I(i)$, such that, for any $i'\in\I(i)$
           \begin{gather}
\b_{i'j'}=\ee^{-A_q\e_{\I(j)}\L_{\I(j)}}\b_{i'\I(j)} \qquad \textrm{and} \qquad \b_{i'\I(j)}(i')=\ee^{A_q\e_{\I(i)}}\b_{i'\I(j)}, \qquad \forall\,  j'\in\I(j). \label{h_iI(j)}
           \end{gather}
        \begin{itemize}\itemsep=0pt
\item If $|\I(i)|=1$, i.e.\ if $\I(i)=\{i\}$,  the only $\nD$-pairs to consider are $(i,j')$, for any $j'\in\I(j)$. From above, the function $\b_{i\I(j)}$ satisf\/ies
  \[
\b_{i\I(j)}(i)=\ee^{A_q\e_{\I(i)}}\b_{i\I(j)}.
  \]
\item If $|\I(i)|\geq2$, let $k\in\I(i)$ and $k'\in\I(i)\setminus\{k\}$. The previous reasoning ensures the existence of a non-zero function $\b_{k\I(j)}$ of the variable $\l_k$, for any $k\in\I(i)$, which satisf\/ies (\ref{h_iI(j)}). Moreover, the exchange of the indices $i$ and $j$, as well as the indices $i'$ and $j'$ in (\ref{h_ij}) yields
  \[
\b_{k\I(j)}(k')=\ee^{A_q\e_\I(i)}\b_{k'\I(j)},
  \]
where the symmetry relation $\b_{i'j'}=\frac1{\b_{j'i'}}$ is used, for any $(i',j')\in\J_p^{(2,\nD)}\; |\; (i',j')\in\I(i)\times\I(j)$.
        \end{itemize}
In both cases, Lemma~\ref{sublem_4a} applies once more to the family of functions $(\b_{k\I(j)})_{k\in\I(i)}$ of the variable~$\l_k$, for any $k\in\I(i)$, ensuring the existence of the non-zero constant $f_{\I(i)\I(j)}$ such that
  \[
\b_{i'j'}(\L_{\I(i)},\L_{\I(j)})=\ee^{A_q(\e_{\I(i)}\L_{\I(i)}-\e_{\I(j)}\L_{\I(j)})}f_{\I(i)\I(j)}, \qquad \forall \, (i',j')\in\I(i)\times\I(j).
  \]
Finally, this implies that
  \begin{gather*}
\D_{ij}((\l_k)_{k\in\I(i)\cup\I(j)})=\frac{S_q}{1-\ee^{A_q(\e_{\I(i)}\L_{\I(i)}-\e_{\I(j)}\L_{\I(j)})}f_{\I(i)\I(j)}}\\
\hphantom{\D_{ij}((\l_k)_{k\in\I(i)\cup\I(j)})}{}
=\D_{\I(i)\I(j)}(\L_{\I(i)},\L_{\I(j)}), \qquad \forall\, (i,j)\in\J_p^{(2,\nD)};
  \end{gather*}
where, as in the second point above, by symmetry, $f_{\I(i)\I(j)}f_{\I(j)\I(i)}=1$.

{\bf 3.~Existence of the functions $\boldsymbol{(f_\I)_{\I\in\{\I(i),\; i\in\J_p\}}}$.}
If $\J_p=\I(i)\cup\I(j)$, assuming that $i<j$, it is suf\/f\/icient to set $f_{\I(i)}=1$ and $f_{\I(j)}=\frac1{f_{\I(i)\I(j)}}$.

Assuming that $\J_p\neq\I(i)\cup\I(j)$, let $k\in\J_p\setminus(\I(i)\cup\I(j))$, i.e.\ $(i,j,k)\in\J_p^{(3,\nD)}$. $(E'_4)$ can be used with indices $ijk$, and yields, by linear independence of the functions $\L_{\I(i)}\longmapsto\ee^{A_q\e_{\I(i)}\L_{\I(i)}}$, $\L_{\I(j)}\longmapsto\ee^{A_q\e_{\I(j)}\L_{\I(j)}}$ and $\L_{\I(k)}\longmapsto\ee^{A_q\e_{\I(k)}\L_{\I(k)}}$
  \[
\frac1{\D_{\I(j)\I(i)}\D_{\I(i)\I(k)}}+\frac1{\D_{\I(i)\I(j)}\D_{\I(j)\I(k)}}=\frac1{\D_{\I(i)\I(j)}\D_{\I(i)\I(k)}}
\quad \Leftrightarrow \quad f_{\I(i)\I(k)}f_{\I(k)\I(j)}=f_{\I(i)\I(j)}.
  \]
This second set of equations reduces the number of independent constants to the choice of a~family of $2(l_p+n^{(p)}_0)$ non-zero constants $(f_{\I\J},f_{\J\I})_{\I,\J\in\{\I(i),\; i\in\J_p\}}$, for any f\/ixed $\mc D$-class~$\I$, for example~$\I(i_{\min})$, where $i_{\min}=\min\J_p$. The f\/irst set of equations $f_{\I(i)\I(j)}f_{\I(j)\I(i)}=1$ reduces this number by half. The family of $l_p+n^{(p)}_0$ non-zero constants $(f_\I)_{\I\in\{\I(i),\; i\in\J_p\}}$ (f\/ixed by $f_{\I(i_{\min})}=1$ and $f_{\I(i)}=\frac1{f_{\I(i)\I(i_{\min})}}$, for any $i\in\J_p\setminus\{i_{\min}\}$) satisf\/ies the expected properties.

{\bf 4.~$\boldsymbol{d}$-coef\/f\/icients.}
Setting $B_q=\frac{S_q}{1-\ee^{A_q}}=B_{ij}$, for any $\nD$-pair $(i,j)\in\J_p^{(2,\nD,<)}$, Lemma~\ref{sublem_3b} and Proposition~\ref{subprop_9b} insure the existence of a $\nD$-closed $\nD$-multiplicative 2-form $(g_{ij})_{\!(i,j)\in\J_p^{(2,\!\!\nD\!\!,<)}}$ such that $d_{ij}=g_{ij}d^0_{\I(i)\I(j)}$, where $d^0_{\I(i)\I(j)}=B_q-\D_{\I(i)\I(j)}$.

Reciprocally, it is straightforward to check that the family of constants $(\D_\I)_{\I\in\{\I(i),\; i\in\J_p\}}$ and the family of functions $(\D_{\I(i)\I(j)},\D_{\I(j)\I(i)},d_{ij},d_{ji})_{(i,j)\in\J^{(2,\!\!\nD\!\!)}_p}$ are indeed solutions of system (\ref{syst}) restricted to the $\D$-class $\J_p$. Note in particular that, as mentioned in the remark of Lemmas~\ref{sublem_3a} and~\ref{sublem_3b}, the family of functions $(d_{ij})_{(i,j)\in\J_p^{(2,\!\!\nD\!\!,<)}}$ obeys $(E_1)$. The set of solutions of system (\ref{syst}) restricted to any $\D$-class $\J_p$ is exactly parametrized by the giving of the constants $S_q$, $\Si_q$ and $(\e_\I,f_\I)_{\I\in\{\I(i),\; i\in\J_p\}}$ and the $\nD$-multiplicative 2-form $(g_{ij})_{\!(i,j)\in\J_p^{(2,\!\!\nD\!\!,<)}}$.
    \end{proof}

\begin{remark} Because we use the principal value of the logarithm function, if $T_q\in\C^*\setminus\{1\}\cup\R^*_-$ then $\Im(A_q)=\arg(T_q)\in\; ]{-}\pi,\pi[$, and if $T_q\in\R^*_-$, then $\Im(A_q)=-\pi$. Therefore, if $T_q\in\C^*\setminus\{1\}\cup\R^*_-$, the function $\L_{\I(i)}\longmapsto\ee^{A_q\e_{\I(i)}\L_{\I(i)}}$ can be periodic of any period strictly greater than~2, but cannot be 2-periodic (since $A_q\neq0$, the period is greater than~$2$). This happens if and only if $T_q\in\R^*_-$, in which case can arise 2-periodic trigonometric functions such as $\L_{\I(i)}\longmapsto\ee^{\i\pi\e_{\I(i)}\L_{\I(i)}}$.

Moreover, Theorem~\ref{subtheo_3a} justif\/ies the choosing of the quantity $A_q$, through the choosing of a~particular complex logarithm. The expressions obtained for the solutions are indeed independent of this choice. Strictly speaking, $\Im(A_q)$ may be a priori def\/ined up to $2\pi$. However, for any $k\in\Z$, the function $\L_{\I(i)}\longmapsto\ee^{2i\pi k\e_{\I(i)}\L_{\I(i)}}$ is 1-periodic in any variable. Remembering that ``constant quantity'' means in fact ``1-periodic function in any variable'', it can be re-absorbed in each constant of the family $(f_\I)_{\I\in\{\I(i),\; i\in\J_p\}}$ by multiplying each one by the function $\L_{\I(i)}\longmapsto\ee^{-2\i\pi k(\e_{\I(i)}\L_{\I(i)}-\e_{\I(\min\J_p)}\L_{\I(\min\J_p)})}$. In particular, this preserves the convention $f_{\I(\min\J_p)}=1$.\footnote{Such manipulation will be also used in the proof of Propositions~\ref{subprop_11b} and~\ref{prop_12}.} Then, the universal covering of $\C^*$ by the Riemann surface $\mathbb S=\{(z,\t)\in\C^*\times\R \; |\; \t-\arg(z)\in2\pi\Z\}$, associated with the logarithm function $\log_\mathbb S$: $(z,\t)\in\mathbb S\longmapsto\log|z|+\i\t$, allows to naturally continuously extend expressions of Theorem~\ref{subtheo_3a}, viewed as functions of $T_q$, to the surface $\mathbb S$. This can be done as above by multiplying each constant $f_{\I(i)}$, viewed as a function of $T_q\in\mathbb S$, by the function $T_q\in\mathbb S\longmapsto\ee^{\i(\t-\arg(T_q))(\e_{\I(i)}\L_{\I(i)}-\e_{\I(\min\J_p)}\L_{\I(\min\J_p)})}$, which is 1-periodic in any variable, for any $T_q\in\mathbb S$.
 \end{remark}

\begin{theorem}[rational behavior]\label{subtheo_3b}
Assuming that $S_q=0$, i.e.\ $T_q=1$, there exist a family of $n^{(p)}_0+l_p$ constants $(f_\I)_{\I\in\{\I(i),\; i\in\J_p\}}$ $($with the convention that $f_{\I(\min\J_p)}=0)$, and a $\nD$-multiplicative $2$-form $(g_{ij})_{\!(i,j)\in\J_p^{(2,\!\!\nD\!\!,<)}}$, such that the solution of system {\rm (\ref{syst})} restricted to the $\D$-class $\J_p$ is given by the following expressions
  \begin{gather*}
\D_{\I(i)}=\e_{\I(i)}\sqrt{\Si_q}, \qquad \forall\, i\in\J_p;
\\
\D_{ij}((\l_k)_{k\in\I(i)\cup\I(j)}) =-\D_{ji}(\{\l_k\}_{k\in\I(i)\cup\I(j)})=\D_{\I(i)\I(j)}(\L_{\I(i)},\L_{\I(j)})=-\D_{\I(j)\I(i)}(\L_{\I(j)},\L_{\I(i)})
\\
\hphantom{\D_{ij}((\l_k)_{k\in\I(i)\cup\I(j)})}{}
=\frac{\sqrt{\Si_q}}{\e_{\I(i)}\L_{\I(i)}-\e_{\I(j)}\L_{\I(j)}+f_{\I(i)}-f_{\I(j)}}
  \end{gather*}
and
\[
d_{ij}=g_{ij}\{B_q-\D_{\I(i)\I(j)}\}, \qquad \forall \, (i,j)\in\J_p^{(2,\nD)}.
\]
\end{theorem}

    \begin{proof}
The proof of Theorem~\ref{subtheo_3a} to compute the functions $\D_{\I(i)\I(j)}$ and $d_{ij}$, for any $i,j\in\J_p$ can be directly adapted here, since the family of constants $(\D_\I)_{\I\in\{\I(i),\; i\in\J_p\}}$ are obviously obtained from~(\ref{D_I(i)_s}). If $n_p=1$ or if the $\D$-class $\J_p$ is reduced to a $\mc D$-class (cf.\ Proposition~\ref{prop_8}), the proof of the theorem ends here (cf.\ case {\bf i} of Corollary~\ref{theo_1.cor}).

Assuming that $n_p\geq2$ and that $\J_p$ is not reduced to a $\mc D$-class, there exists a $\nD$-pair of indices $(i,j)\in\J_p^{(2,\nD)}$. Let $(i',j')\in\J_p^{(2,\nD)}$ be a $\nD$-pair of indices such that $(i',j')\in\I(i)\times\I(j)$. From Lemmas~\ref{sublem_2a} and~\ref{sublem_2b}, only the dependence in the variable $\l_k$, for any $k\in\I(i)\cup\I(j)$, of the function $\D_{i'j'}$ remains to be determined. To this end, $(F'_1)\Leftrightarrow(F'_2)$ with indices $i'j'$ is re-written, since $\D_{\I(i)}\D_{\I(j)}\D_{\I(i)\I(j)}\neq0$ and denoting $h_{ij}=\frac{\sqrt{\Si_q}}{\D_{ij}}=-\b_{ji}$, as
  \[
h_{ij}(i)=h_{ij}+\e_{\I(i)} \qquad \textrm{and} \qquad h_{ij}(j)=h_{ij}-\e_{\I(j)},
  \]
thanks to the equality $S_q=\D_{ij}+\D_{ji}=0$. Moreover, for any $k\in\I(j)$ and any $k'\in\I(j)\setminus\{j\}$, $(E'_4)$ can be used with indices $i'kk'$, and yields
  \[
\b_{i'k}(k')=\b_{i'k'}-\e_{\I(j)},
  \]
the exact correspondent of (\ref{h_ij}), to which we apply Lemma~\ref{sublem_4b}. We deduce the existence of a~constant $f_{\I(i)\I(j)}$, such that
  \[
\b_{i'j'}(\L_{\I(i)},\L_{\I(j)})=\e_{\I(i)}\L_{\I(i)}-\e_{\I(j)}\L_{\I(j)}+f_{\I(i)\I(j)}, \qquad \forall\, (i',j')\in\I(i)\times\I(j).
  \]
Finally, this implies that
  \begin{gather*}
\D_{ij}((\l_k)_{k\in\I(i)\cup\I(j)})=\frac{\sqrt{\Si_q}}{\e_{\I(i)}\L_{\I(i)}\!-\e_{\I(j)}\L_{\I(j)}\!+f_{\I(i)\I(j)}}
=\D_{ij}(\L_{\I(i)},\L_{\I(j)}),\! \qquad \forall\, (i,j)\in\J_p^{(2,\nD)},
  \end{gather*}
where, as above, by symmetry, $f_{\I(i)\I(j)}=-f_{\I(j)\I(i)}$. If $\J_p=\I(i)\cup\I(j)$, assuming that $i<j$, it is suf\/f\/icient to set $f_{\I(i)}=0$ and $f_{\I(j)}=-f_{\I(i)\I(j)}$.

Assuming that $\J_p\neq\I(i)\cup\I(j)$, let $k\in\J_p\setminus(\I(i)\cup\I(j))$, i.e.\ $(i,j,k)\in\J_p^{(2,\nD)}$. Hence $(E'_4)$ can be used with indices $ijk$, and yields, by linear independence of the functions $\L_{\I(i)}\longmapsto\L_{\I(i)}$, $\L_{\I(j)}\longmapsto\L_{\I(j)}$ and $\L_{\I(k)}\longmapsto\L_{\I(k)}$
  \begin{gather*}
\frac1{\D_{\I(j)\I(i)}\D_{\I(i)\I(k)}}+\frac1{\D_{\I(i)\I(j)}\D_{\I(j)\I(k)}}=\frac1{\D_{\I(i)\I(j)}\D_{\I(i)\I(k)}}
\quad \Leftrightarrow \quad f_{\I(i)\I(k)}+f_{\I(k)\I(j)}=f_{\I(i)\I(j)}.
  \end{gather*}
This second set of equations reduces the number of independent constants to the choice of a~family of $2(l_p+n^{(p)}_0-1)$ constants $(f_{\I\J},f_{\J\I})_{\I,\J\in\{\I(i),\; i\in\J_p\}}$, for any f\/ixed $\mc D$-class $\I$, for example $\I(i_{\min})$, where $i_{\min}=\min\J_p$. The family $(f_\I)_{\I\in\{\I(i),\; i\in\J_p\}}$ def\/ined as $f_{\I(i_{\min})}=0$ and $f_{\I(i)}=-f_{\I(i)\I(i_{\min})}$, for any $i\in\J_p\setminus\{i_{\min}\}$, yields the expected result, the rest of the proof being identical to the proof of Theorem~\ref{subtheo_3a}.
    \end{proof}

\subsection[Inside a subset $\K_q$]{Inside a subset $\boldsymbol{\K_q}$}

We now end the resolution of system (\ref{syst}) by solving case {\bf iii} of Corollary~\ref{theo_1.cor}. Let $q\in\N^*_s$, and consider a subset $\K_q=\bigcup\limits_{p\in\PP_q}\J_p$ such that $r_q\geq2$, the case $r_q=1$ being already treated in Theorems~\ref{subtheo_3a} and~\ref{subtheo_3b}. We now have to determine the cross-terms between two distinct $\D$-classes. This is given by the following theorem.

  \begin{theorem}[trigonometric behavior]\label{theo_4}
Let $q\in\N^*_s$ such that $r_q\geq2$. Then, there exist two non-zero constant $S_q$ and $\Si_q$, a family of signs $(\e_\I)_{\I\in\{\I(i),\; i\in\K_q\}}$, a family of non-zero constants $(f_\I)_{\I\in\{\I(i),\; i\in\K_q\}}$ $($with the convention that $f_{\I(\min\J_p)}=1$, for any $p\in\PP_q)$, and a $\nD$-multiplicative $2$-form $(g_{ij})_{\!(i,j)\in\K_q^{(2,\!\!\nD\!\!,<)}}$, such that the $R$-matrix, solution of system {\rm (\ref{syst})} restricted to the subset~$\K_q$, is given by
    \begin{gather*}
R^{(q)}=\sum_{p\in\PP_q}R^{(q)}_p +S_q\sum_{(p,p')\in\PP_q^{(2,<)}}\sum_{(i,j)\in\J_p\times\J_{p'}}e^\cdot_{ij}\otimes e^\cdot_{ji}
\\
\hphantom{R^{(q)}=}{}
+\sqrt{\Si_q}\sum_{(p,p')\in\PP_q^{(2,<)}}\sum_{(i,j)\in\J_p\times\J_{p'}}\{g_{ij}e^\cdot_{ii}\otimes e^\cdot_{jj}+g_{ji}e^\cdot_{jj}\otimes e^\cdot_{ii}\},
    \end{gather*}
where the family of matrices $(R^{(p)}_q,(R^{(q,p)}_{\I(i)})_{i\in\J_p})$ is defined, for any $p\in\PP_q$, as
  \begin{gather*}
R^{(q)}_p=\sum_{i\in\J_p}R^{(q,p)}_{\I(i)} +\sum_{(i,j)\in\J_p^{(2,\!\!\nD\!\!)}}\left\{\D_{\I(i)\I(j)}e^\cdot_{ij}\otimes e^\cdot_{ji}+d_{ij}e^\cdot_{ii}\otimes e^\cdot_{jj}\right\}
\\
\hphantom{R^{(q)}_p}{}
=\sum_{i\in\J_p}R^{(q,p)}_{\I(i)}  +\sum_{(i,j)\in\J_p^{(2,\!\!\nD\!\!)}}\big\{\D_{\I(i)\I(j)}(e^\cdot_{ij}\otimes e^\cdot_{ji}-g_{ij}e^\cdot_{ii}\otimes e^\cdot_{jj})\\
\hphantom{R^{(q)}_p=}{}
+\D_{\I(j)\I(i)}(e^\cdot_{ji}\otimes e^\cdot_{ij}-g_{ji}e^\cdot_{jj}\otimes e^\cdot_{ii})\big\}
  +B_q\sum_{(i,j)\in\J_p^{(2,\!\!\nD\!\!)}}\{g_{ij}e^\cdot_{ii}\otimes e^\cdot_{jj}+g_{ji}e^\cdot_{jj}\otimes e^\cdot_{ii}\},
  \end{gather*}
and
  \[
R^{(q,p)}_{\I(i)}=\frac{\D_{\I(i)}}{|\I(i)|}\sum_{j,j'\in\I(i)}e^\cdot_{jj'}\otimes e^\cdot_{j'j}, \qquad \forall\,  i\in\J_p.
  \]
  \end{theorem}

\begin{remark} In this notation, the exponent ``$~\!^\cdot\!~$'' has to be chosen to appropriately specify the size of the matrices $R^{(q)}_p$ or $R^{(q)}$. According to Proposition~\ref{subprop_1c}, there are three possibilities for the matrix $R^{(q)}_p$
      \begin{itemize} \itemsep=0pt
\item if ``$~\!\cdot=\J_p$'', then $R^{(q)}_p$ is a solution of the $\mc Gl_{n_p}(\C)$-\ref{DQYBE2}, which is the contraction to the $\D$-class $\J_p$ of a $R$-matrix, solution of the $\mc Gl_n(\C)$-\ref{DQYBE2} or the $\mc Gl_{N_q}(\C)$-\ref{DQYBE2};
\item if ``$~\!\cdot=\K_q$'', then $R^{(q)}_p$ is the restriction to the $\D$-class $\J_p$ of a solution of the $\mc Gl_{N_q}(\C)$-\ref{DQYBE2}, which is the contraction to the subset $\K_q$ of a $R$-matrix, solution of the $\mc Gl_n(\C)$-\ref{DQYBE2};
\item if ``$~\!\cdot=(n)$'', then $R^{(q)}_p$ is the restriction to the $\D$-class $\J_p$ of a $R$-matrix, solution of the $\mc Gl_n(\C)$-\ref{DQYBE2}.
      \end{itemize}
Similarly, there are two possibilities for the matrix $R^{(q)}$
      \begin{itemize} \itemsep=0pt
\item if ``$~\!\cdot=\K_q$'', then $R^{(q)}$ is a solution of the $\mc Gl_{N_q}(\C)$-\ref{DQYBE2}, which is the contraction to the subset $\K_q$ of a $R$-matrix, solution of the $\mc Gl_n(\C)$-\ref{DQYBE2};
\item if ``$~\!\cdot=(n)$'', then $R^{(q)}$ is the restriction to the subset $\K_q$ of a $R$-matrix, solution of the $\mc Gl_n(\C)$-\ref{DQYBE2}.
      \end{itemize}
\end{remark}

    \begin{proof}
The existence of the constants $S_q$ and $\Si_q$ is insured by Theorem~\ref{theo_2}.

Assuming that $r_q\geq2$, for any $\D$-class $\J_p$, with $p\in\PP_q$, of the subset $\K_q$, Theorems~\ref{subtheo_3a} and~\ref{subtheo_3b} give the expressions of the constants of the family $(\D_\I)_{\I\in\{\I(i),\; i\in\J_p\}}$ and the functions of the family $(\D_{\I(i)\I(j)},\D_{\I(j)\I(i)},d_{ij},d_{ji})_{(i,j)\in\J_p^{(2,\!\!\nD\!\!)}}$, depending on the signs of the family $(\e_\I)_{\I\in\{\I(i),\; i\in\J_p\}}$, the constants of the family $(f_\I)_{\I\in\{\I(i),\; i\in\J_p\}}$, and the $\nD$-multiplicative 2-form $(g_{ij})_{\!(i,j)\in\J_p^{(2,\!\!\nD\!\!,<)}}$. It remains to determine the ``crossed'' functions $(\D_{ij},d_{ij},d_{ji})_{(i,j)\in\J_p\times\J_{p'}|(p,p')\in\PP_q^{(2,<)}}$ of the solutions of system (\ref{syst}) restricted to the subset $\K_q$. We also need to specify the dependence of the $\nD$-multiplicative 2-form $(g_{ij})_{\!(i,j)\in\J_p^{(2,\!\!\nD\!\!,<)}}$ on the variable $\l_k$, for any $k\in\K_q\setminus\J_p$, which is done in Propositions~\ref{subprop_9a},~\ref{subprop_9b} and~\ref{subprop_9c}.

From Theorem~\ref{theo_2}, we have $\D_{ij}=S_q=\D_{\I(i)\I(j)}$, for any $(i,j)\in\J_p\times\J_{p'}\; \big|\; (p,p')\in\PP_q^{(2,<)}$. This implies also that the constant $S_q$ is non-zero (otherwise $\D_{ij}=S_q=0$, for any $(i,j)\in\J_p\times\J_{p'}\; \big|\; (p,p')\in\PP_q^{(2,<)}$), which yields a contradiction with the construction of the subset $\K_q$.

The $d$-coef\/f\/icients are deduced from the fact that $d^0_{ij}=d^0_{ji}=\sqrt{\Si_q}$ is a particular solution of $d_{ij}d_{ji}=\Si_q$, for any $(i,j)\in\J_p\times\J_{p'}\; \big|\; (p,p')\in\PP_q^{(2,<)}$, as in the proof of Theorem~\ref{subtheo_3a}. This yields the expected result, the rest of the proof being almost identical.

Reciprocally, it is straightforward to check that the family of functions $(\D_{\I(i)\I(j)},\D_{\I(j)\I(i)},d_{ij}$, $d_{ji})_{(i,j)\in\K_q^{(2,\!\!\nD\!\!)}}$ are indeed solutions of system (\ref{syst}) restricted to the subset~$\K_q$. The set of solutions of system~(\ref{syst}) restricted to the subset $\K_q$ is exactly parametrized by the giving of the constants~$S_q$, $\Si_q$ and $(\e_\I,f_\I)_{\I\in\{\I(i),\; i\in\K_q\}}$, and the $\nD$-closed $\nD$-multiplicative 2-form $(g_{ij})_{\!(i,j)\in\K_q^{(2,\!\!\nD\!\!,<)}}$. This concludes the proof of the theorem.
    \end{proof}

\begin{remark}
Let us f\/irst insist on the fact that the proof above has been seen to justify that assuming that $r_q\geq2$ implies that $S_q\neq0$, and then forbids the rational behavior. In particular, if the rational behavior is assumed, i.e.\ if $S_q=0$, then $r_q=1$, meaning that there exists a single label $p\in\N^*_r$ such that $\K_q=\J_p$.

A trigonometric $R$-matrix, solution of \ref{DQYBE2} restricted to the subset $\K_q$, shows similitudes with a decoupled $R$-matrix presented in Proposition~\ref{subprop_1b}. As in this case, the $d$-coef\/f\/icients $d_{ij}$ and $d_{ji}$, for any $(i,j)\in\J_p\times\J_{p'}$, between two distinct subsets $\J_p$ and $\J_{p'}$, with $(p,p')\in\PP_q^{(2,\nD)}$, are given by an overall $\K_q$-dependent constant, up to $\nD$-multiplicative covariance (cf.\ Propositions~\ref{subprop_9a},~\ref{subprop_9b} and~\ref{subprop_9c}). The coupling (with $\D$-coef\/f\/icients) between the subsets~$\J_p$ and~$\J_{p'}$ is minimal, in the sense that the coupling part $\sum\limits_{(p,p')\in\PP_q^{(2,<)}}\sum\limits_{(i,j)\in\J_p\times\J_{p'}}e^\cdot_{ij}\otimes e^\cdot_{ji}$ of the $R$-matrix is non-dynamical. It is interesting to note a similarity of upper-triangularity structure of this non-dynamical part of the dynamical $R$-matrix with the Yangian $R$-matrix $R_\mathrm{Yang}=\sum\limits_{i<j}e_{ij}\otimes e_{ji}$.
  structure also appears in the non-dynamical operators $\{R^{(q,p)}_{\I(i)}\}_{i\in\J_p}$, describing the coupling inside any $\mc D$-class.
  \end{remark}

\section{General solution and structure of the set of solutions} \label{sec_5}

Given an ordered partition of the set of indices $\N^*_n$ as described in Theorem~\ref{theo_1}, Proposition~\ref{prop_10} and Theorems~\ref{theo_2}, \ref{subtheo_3a}, \ref{subtheo_3b} and~\ref{theo_4} allow to directly write the general form of any solution of \ref{DQYBE2}, compatible with this partition, where the only parameters of the $R$-matrix not explicitly constructed is the $\nD$-closed $\nD$-multiplicative 2-form $(g_{ij})_{\!(i,j)\in\N_n^{*(2,\!\!\nD\!\!,<)}}$

More precisely, we recall that Theorem~\ref{theo_1} yields a f\/irst, set-theoretical, ``parametrization'' of the solutions of \ref{DQYBE2}, which is given in terms of an ordered partition of the indices set $\N^*_n$ into the $s$ ordered subsets $\{\K_q,\; q\in\N^*_s\}$, being unions of the $\D$-classes $\{\J_p,\; p\in\N^*_r\}$, and the ordered partition of each $\D$-class $\J_p$ into $l_p$ $\mc D$-classes $\{\I(i),\; i\in\J_p\}$, either reduced to a~single element (case of a $\mc D$-class $\I(i)$ reduced to a free index $i$) or non-trivial (case of $\mc D$-class~$\I(i)$ generated by a non-free index~$i$).

  \begin{theorem}[general $R$-matrices]\label{theo_5}
Let $n\geq2$ and an ordered partition of the set $\N^*_n$. Then, there exist a family of constants $(S_q)_{q\in\N^*_s}$ $($with $S_q\neq0$ if $r_q\geq2)$ two families of non-zero-constants $(\Si_q)_{q\in\N^*_s}$ and $(\Si_{qq'})_{(q,q')\in\N_s^{*(2,<)}}$, a family of signs $(\e_\I)_{\I\in\{\I(i),\; i\in\N^*_n\}}$, a family of non-zero constants $(f_\I)_{\I\in\{\I(i),\; i\in\N^*_n\}}$ $($with the convention that $f_{\I(\min\J_p)}=0$, for any $p\in\PP_q$, with $q\in\N^*_s\; |\; r_q=1$, and $f_{\I(\min\J_p)}=1$, for any $p\in\PP_q$, with $q\in\N^*_s\; |\; r_q\geq2)$, and a $\nD$-closed $\nD$-multiplicative $2$-form $(g_{ij})_{\!(i,j)\in\N_n^{*(2,\!\!\nD\!\!,<)}}$, such that the $R$-matrix, solution of~\ref{DQYBE2}, is given by
\[
R=\sum_{q=1}^sR^{(q)}+\sum_{(q,q')\in\N_s^{*(2,<)}}\sqrt{\Si_{qq'}}\sum_{(i,j)\in\K_q\times\K_{q'}}\big\{g_{ij}e^{(n)}_{ii}\otimes e^{(n)}_{jj}+g_{ji}e^{(n)}_{jj}\otimes e^{(n)}_{ii}\big\}.
\]
  \end{theorem}

Let us now characterize the structure of the moduli space of \ref{DQYBE2}.

Putting aside the delicate issue of general $\nD$-closed $\nD$-multiplicative 2-forms, we see from Proposition~\ref{prop_10} that the general solution of \ref{DQYBE2} is therefore built, up to the $\nD$-multiplicative covariance, in terms of solutions of \ref{DQYBE2} restricted to each subset $\K_q$ together with cross-terms $(\Si_{qq'})_{(q,q')\in\N_s^{*(2,<)}}$ between each pair of such subsets. This takes care of the interpretation of set-theoretical parameters $\{\K_q,\; q\in\N^*_s\}$ and $c$-number complex parameters $(\Si_{qq'})_{(q,q')\in\N_s^{*(2,<)}}$ as def\/ining irreducible components of decoupled $R$-matrices according to Proposition~\ref{subprop_1b}.

\subsection[Continuity properties of the solutions with respect to the constants $(S_q,\Si_q)_{q\in\N^*_s}$]{Continuity properties of the solutions with respect\\ to the constants $\boldsymbol{(S_q,\Si_q)_{q\in\N^*_s}}$} \label{subsec_5_1}

Consider the family of $c$-number complex parameters $(S_q,\Si_q)_{q\in\N^*_s}$. As seen in Theorems~\ref{subtheo_3a},~\ref{subtheo_3b} and~\ref{theo_4}, the solution of system (\ref{syst}) restricted to any subset $\K_q$ is essentially characterized by the constants $S_q$ and $\Si_q\neq0$, through the quantity $T_q\neq0$, except for the family of signs $(\e_\I)_{\I\in\{\I(i),\; i\in\K_q\}}$, the set of constants $(f_\I)_{\I\in\{\I(i),\; i\in\K_q\}}$ and the $\nD$-multiplicative 2-form $(g_{ij})_{\!(i,j)\in\K_q^{(2,\!\!\nD\!\!,<)}}$. We have particularly exhibited three cases to be distinguished
  \begin{enumerate}[label={\bf\roman*}.]\itemsep=0pt
\item Rational behavior: $S_q=0$ and $T_q=1$;
\item Trigonometric behavior (periodicity 2): $S_q\neq0$ and $T_q\in\R^*_-$;
\item Trigonometric behavior (arbitrary periodicity strickly greater than 2): $S_q\neq0$ and $T_q\in\C^*\setminus\{1\}\cup\R^*_-$.
  \end{enumerate}
This naturally rises the question whether these three types of solutions are distinct or if it is possible to connect them one to each other, typically in this situation by continuity arguments. Such connections exist and are described by the two following propositions.

Let us immediately point out that the f\/irst two cases are clearly incompatible, and thus essentially dif\/ferent, in the sense that exploring the neighboorhood of the dimensionless variable $\frac{S_q}{\sqrt{\Si_q}}=0$ imposes equivalently to explore the neighboorhood of $T_q=1$, which cannot be asymptotically reached by points in $\R^*_-$. In other words, for any f\/ixed $\Si_q\neq0$, the quantity $T_q$ viewed as a function of $\frac{S_q}{\sqrt{\Si_q}}$ def\/ined on $\C$ is not continuous in 0, and solutions parameterized by $S_q=0$ and $T_q=1$ cannot be approached by solutions with $S_q\neq0$ and $T_q\in\R^*_-$. The periodicity 2 of a trigonometric solution cannot then become inf\/inite as required by the rational behavior.

\begin{remark}
From Corollary~\ref{subprop_9a.cor}, the $\nD$-multiplicative 2-form $(g_{ij})_{\!(i,j)\in\K_q^{(2,\!\!\nD\!\!,<)}}$ can be factorized out independently in each subset~$\K_q$. In other words, the $\nD$-multiplicative 2-form $(g_{ij})_{\!(i,j)\in\K_q^{(2,\!\!\nD\!\!,<)}}$ are non-relevant moduli to consider inside any f\/ixed irreducible component. Hence, unless otherwise stated, we will assume in this section that $g_{ij}=1$, for any $(i,j)\in\K_q^{(2,\nD,<)}$.
\end{remark}

Let $q\in\N^*_s$ such that $r_q=1$, a solution (parametrized by the constants $S^0_q=0$, $\Si_q\neq0$ and $(\e_\I,f^0_\I)_{\I\in\{\I(i),\; i\in\K_q\}}$) of system (\ref{syst}) restricted to the subset $\K_q$, and a parameter $\xi\in\C^*$.

\begin{proposition}[from trigonometric to rational]\label{subprop_11a}\label{prop_11}
There exists a solution $($parametrized by the constants $S^\xi_q\neq0$, $\Si_q$ and $(\e_\I,f^\xi_\I)_{\I\in\{\I(i),\; i\in\K_q\}})$ of system {\rm (\ref{syst})} restricted to the subset $\K_q$, such that the solution with $S^0_q=0$ is the limit of the solution with $S^\xi_q\neq0$, when $\xi\rightarrow0$.

In particular, if $T^\xi_q\notin\R^*_-$, for any $\xi\in\C^*$, the piecewise solution $($parametrized among others by the constants $S_q^\xi$ and $(f^\xi_\I)_{\I\in\{\I(i),\; i\in\K_q\}})$ is a continuous function of~$\xi$ on~$\C$.
\end{proposition}

    \begin{proof}
Since $r_q=1$, there exists an unique $p\in\N^*_r$ such that $\K_q=\J_p$. The idea of the proof is to Taylor expand each considered quantity in the neighboorhood of $\xi=0$, starting from $S^\xi_q$. Such expansion exists as soon as the quantity under study is a suf\/f\/iciently regular function of $\xi\in\C$ (or at least in a neighboorhood of $\xi=0$).

Let for example $S^\xi_q$ a non-zero holomorphic function of $\xi$ on $\C^*$ (or at least in a neighboorhood of $\xi=0$), such that $S^\xi_q=S_q\xi+o(\xi)$, with $S_q=\frac\dd{\dd\xi}S^\xi_q\big|_{\xi=0}\neq0$. Then, we deduce the following expansions
\[
D^\xi_q=2\sqrt{\Si_q}\{1+o(\xi)\}\neq0 \qquad \textrm{or} \qquad T^\xi_q=1-\frac{S_q}{\sqrt{\Si_q}}\xi+o(\xi)\notin\R^*_-,
\]
implying that
\[
A^\xi_q=-\frac{S_q}{\sqrt{\Si_q}}\xi+o(\xi).
\]

Now, we reason similarly for the constants $(f^\xi_\I)_{\I\in\{\I(i),\; i\in\K_q\}}$. We assume that their Taylor expansion in the neighboorhood of $\xi=0$ exists and stands, for any $i\in\K_q$,
\[
f^\xi_{\I(i)}=1-f^0_{\I(i)}\frac{S_q}{\sqrt{\Si_q}}\xi+o(\xi).
\]
(in particular, $f^\xi_{\I(i)}=1$, for any $\xi\in\C^*$, implies $f^0_{\I(i)}=0$, and then we verify that $f^0_{\I(\min\K_q)}=1$, as required by Theorem~\ref{subtheo_3b}). Then, we deduce that, for any $i\in\K_q$
\[
\D^\xi_{\I(i)}=\frac{S^\xi_q}{1-\ee^{A^\xi_q\e_{\I(i)}}}\xrightarrow[\xi\rightarrow0]{}\e_{\I(i)}\sqrt{\Si_q}=\D^0_{\I(i)};
\]
and, for any $(i,j)\in\K_q^{(2,\nD)}$
      \begin{gather*}
\D^\xi_{\I(i)\I(j)}(\L_{\I(i)},\L_{\I(j)}) =\frac{S^\xi_q}{1-\ee^{A^\xi_q(\e_{\I(i)}\L_{\I(i)}-\e_{\I(j)}\L_{\I(j)})}\frac{f^\xi_{\I(i)}}{f^\xi_{\I(j)}}}
\\
\hphantom{\D^\xi_{\I(i)\I(j)}(\L_{\I(i)},\L_{\I(j)}) }{}
 \xrightarrow[\xi\rightarrow0]{}\frac{\sqrt{\Si_q}}{\e_{\I(i)}\L_{\I(i)}-\e_{\I(j)}\L_{\I(j)}+f^0_{\I(i)}-f^0_{\I(j)}}
 =\D^0_{\I(i)\I(j)}(\L_{\I(i)},\L_{\I(j)}).
      \end{gather*}

The limit for $d$-coef\/f\/icients is trivially deduced from above, for any $\nD$-pair $(i,j)\in\K_q^{(2,\nD)}$.
    \end{proof}

  \begin{corollary}
Let $s\geq2$ and $m\in\N^*_s$. This proposition can be extended to any set $\bigcup\limits_{a=1}^m\K_{q_a}$, where $r_{q_a}=1$, for any $a\in\N^*_m$, providing that there exists a non-zero constant $\Si_q$, such that $\Si_{q_a}=\Si_{q_aq_b}=\Si_q$, for any $(a,b)\subseteq\N_m^{*(2,<)}$.
  \end{corollary}

    \begin{proof}
The case $m=1$ already being treated, let assume that $m\geq2$. Since, from the remark of Theorem~\ref{theo_4}, there exists an unique $p_a\in\N^*_r$ such that $\K_{q_a}=\J_{p_a}$, the solution, parametrized (among others) by the constants $S^\xi_q\neq0$, $\Si_q$ and $(\e_\I,f^\xi_\I)_{\I\in\{\I(i),\; i\in\K_q\}}$, of system~(\ref{syst}) restricted to the subset $\K_q=\bigcup\limits_{a=1}^n\J_{p_a}$ has the expected properties. The limit for coef\/f\/icients in the family $(\D^\xi_{\I(i)\I(j)},d^\xi_{ij})_{(i,j)\in\\K_q^{(2,\nD)}}$ are given by Proposition~\ref{subprop_11a}. The limit for the other $\D$-coef\/f\/icients is not problematic, since $\D^\xi_{\I(i)\I(j)}=S_q\rightarrow 0=\D^0_{\I(i)\I(j)}$, for any $(i,j)\in\J_{p_a}\times\J_{p_b}$, with $(a,b)\in\N_m^{*(2,<)}$, and the limit for $d$-coef\/f\/icients is trivially deduced from above.
    \end{proof}

 Let $q\in\N^*_s$, a solution (parametrized by the constants $S^0_q\neq0$, $\Si^0_q$, $T^0_q=-\frac{1-t^0_q}{1+t^0_q}\in\R^*_-$, i.e.\ $t^0_q=\frac{D^0_q}{S^0_q}\in \; ]{-}1,1[$, and $(\e_\I,f^0_\I)_{\I\in\{\I(i),\; i\in\K_q\}}$) of system (\ref{syst}) restricted to the subset $\K_q$, and a~parameter $\t\in\C\setminus\R$.

  \begin{proposition}[from arbitrary periodicity to periodicity 2]\label{subprop_11b}
There exists a solution $($para\-met\-ri\-zed by the constants $S^\t_q\neq0$, $\Si^\t_q$, $T^\t_q=-\frac{1-t^\t_q}{1+t^\t_q}\notin\R^*_-$, i.e.\ $t^\t_q=\frac{D^\t_q}{S^\t_q}\notin \; ]{-}1,1[$, and $(\e_\I,f^\t_\I)_{\I\in\{\I(i),\; i\in\K_q\}})$ of system {\rm (\ref{syst})} restricted to the subset $\K_q$, such that the solution with $T^0_q\in\R^*_-$ is the limit of the solution with $T^\t_q\notin\R^*_-$, when $\t\rightarrow0$.

In particular, the piecewise solution $($parametrized among others by the constants $S^{\t}_q$, $\Si^{\t}_q$ and $(f^\t_\I)_{\I\in\{\I(i),\; i\in\K_q\}})$ is a continuous function of~$\t$ on a neighboorhood of~$0$ in $\C\setminus\R^*$.
  \end{proposition}

    \begin{proof}\sloppy
Let $t^0_q\in \; ]{-}1,1[$. The proof consists in showing that the constants $S^0_q$, $\Si^0_q$ and $(f^0_\I)_{\I\in\{\I(i),\; i\in\K_q\}}$ parametrize a solution, characterized by $t^0_q\in\; ]{-}1,1[$, which is the limit of a~solution, to be precised, parametrized by the constants $S^\t\neq0$, $\Si^\t_q$ and $(f^\t_\I)_{\I\in\{\I(i),\; i\in\K_q\}}$, and characterized by $t^\t_q\notin\; ]{-}1,1[$, when $\t\rightarrow0$.

Let for example $S^\t_q$ and $\Si^\t_q$ two non-zero holomorphic functions of $\t$ on a neighboorhood of $\t=0$ in $\C\setminus\R$, such that $S^\t_q=S^0_q\left\{1-\frac\t{2t^0_q}+o(\t)\right\}$ and $\Si^\t_q=\Si^0_q\left\{1+\frac{(D^0_q)^2+(S^0_q)^2}{4\Si^0_q}\frac\t{t^0_q}+o(\t)\right\}$. Then, using previous notation of the proof of Theorem~\ref{subtheo_3a}, i.e.\ $A^0_q=\log(-T^0_q)-\i\pi=\log\frac{1-\t^0_q}{1+\t^0_q}-\i\pi$, we Taylor expand the quantity
  \[
T^\t_q=\ee^{A^0_q} \left\{1-\frac{2\t}{1-(t^0_q)^2}+o(\t)\right\} \notin\R^*_-.
  \]
From which, introducing the Heaviside function $H$, we deduce that
  \[
A^\t_q=\log(T^\t_q)=A^0_q+2\i\pi H(\Im(\t))+o(1),
  \]
which does not converge in the general case when taking the limit $\t\rightarrow0$. However, it is possible to appropriately choose the functions $(f^\t_\I)_{\I\in\{\I(i),\; i\in\K_q\}}$, viewed as functions of $\t$, so that the constants $(\D^\t_\I)_{\I\in\{\I(i),\; i\in\K_q\}}$ and the functions $(\D^\t_{\I(i)\I(j)})_{(i,j)\in\K_q^{(2,\!\!\nD\!\!)}}$ converge to the constants $(\D^0_\I)_{\I\in\{\I(i),\; i\in\K_q\}}$ and the functions $(f^0_\I)_{\I\in\{\I(i),\; i\in\K_q\}}$ in this limit. More precisely, we have that, for any $i\in\K_q$
  \[
\D^\t_{\I(i)}=\frac{S^\t_q}{1-\ee^{A^\t_q\e_{\I(i)}}}=\frac{S^\t_q}{1-\ee^{A^0_q\e_{\I(i)}}\ee^{o(1)}}
\xrightarrow[\t\rightarrow0]{}\frac{S^0_q}{1-\ee^{A^0_q\e_{\I(i)}}}=\D^0_{\I(i)};
  \]
and, for any $(i,j)\in\J_p^{(2,\nD)}$, with $p\in\PP_q$,
  \begin{gather*}
\D^\t_{\I(i)\I(j)}(\L_{\I(i)},\L_{\I(j)}) =\frac{S^\t_q}{1-\ee^{A^\t_q(\e_{\I(i)}\L_{\I(i)}-\e_{\I(j)}\L_{\I(j)})}\frac{f^\t_{\I(i)}}{f^\t_{\I(j)}}}
\\
\hphantom{\D^\t_{\I(i)\I(j)}(\L_{\I(i)},\L_{\I(j)})}{}
=\frac{S^\t_q}{1-\ee^{A^0_q(\e_{\I(i)}\L_{\I(i)}-\e_{\I(j)}\L_{\I(j)})}\ee^{\left(2\i\pi H(\Im(\t))+o(1)\right)(\e_{\I(i)}\L_{\I(i)}-\e_{\I(j)}\L_{\I(j)})}\frac{f^\t_{\I(i)}}{f^\t_{\I(j)}}}.
  \end{gather*}
Remarking that the function $\L_{\I(i)}\longmapsto\ee^{-2\i\pi H(\Im(t_q-\t_q))\e_{\I(i)}\L_{\I(i)}}$ is 1-periodic in the variable $\L_{\I(i)}$, this prescribes a way to appropriately def\/ine the limit $t_q\rightarrow \t_q$. As in the remark following Theorem~\ref{subtheo_3a}, it is actually suf\/f\/icient to consider constants $(f^\t_\I)_{\I\in\{\I(i),\; i\in\K_q\}}$, from which it is possible to factorize the non-continuous part $\L_{\I(i)}\longmapsto\ee^{-2\i\pi H(\Im(\t))\e_{\I(i)}\L_{\I(i)}}$. Then, we will assume that there exist non-zero constants $(f_\I)_{\I\in\{\I(i),\; i\in\K_q\}}$, which verify, for any $i\in\J_p$, with $p\in\PP_q$,
\[
f^\t_{\I(i)}=\ee^{-2\i\pi H(\Im(\t))(\e_{\I(i)}\L_{\I(i)}-\e_{\I(\min\J_p)}\L_{\I(\min\J_p)})}f_{\I(i)},
\]
where the constant $f_{\I(i)}$ is a continuous function of $\t\in\C\setminus\R^*$ (or at least in a neighboorhood of~$0$ in $\C\setminus\R^*$), with the additional condition that $f_{\I(i)}(\t)\xrightarrow[\t\rightarrow0]{}f^0_{\I(i)}$, for any $i\in\K_q$. In particular, we verify that $f^\t_{\I(\min\J_p)}=f_{\I(\min\J_p)}=f^0_{\I(\min\J_p)}=1$, for any $p\in\PP_q$, as required by Theorem~\ref{subtheo_3a}. This choice implies that, for any $(i,j)\in\J_p^{(2,\nD)}$, with $p\in\PP_q$,
  \begin{gather*}
\D^\t_{\I(i)\I(j)}(\L_{\I(i)},\L_{\I(j)}) =\frac{S^\t_q}{1-\ee^{A^0_q(\e_{\I(i)}\L_{\I(i)}-\e_{\I(j)}\L_{\I(j)})}\ee^{o(1)}\frac{f^0_{\I(i)}}{f^0_{\I(j)}}}
\\
\hphantom{\D^\t_{\I(i)\I(j)}(\L_{\I(i)},\L_{\I(j)}) }{}
\xrightarrow[\t\rightarrow0]{}\frac{S^0_q}{1-\ee^{A^0_q(\e_{\I(i)}\L_{\I(i)}
-\e_{\I(j)}\L_{\I(j)})}\frac{f^0_{\I(i)}}{f^0_{\I(j)}}}=\D^0_{\I(i)\I(j)}(\L_{\I(i)},\L_{\I(j)}).
  \end{gather*}
If $r_q=1$, as far as $\D$-coef\/f\/icients are concerned, the proof of the theorem ends here.

Assuming that $r_q\geq2$, the limit for the other $\D$-coef\/f\/icients is not problematic, since $\D_{\I(i)\I(j)}=S_q\rightarrow S^\t_q=\D^\t_{\I(i)\I(j)}$, for any $(i,j)\in\J_p\times\J_{p'}$, with $(p,p')\in\PP_q^{(2,<)}$.

The limit for $d$-coef\/f\/icients is trivially deduced from above, for any $(i,j)\in\K_q^{(2,\nD)}$.
    \end{proof}

It is therefore suf\/f\/icient to consider the family of constants $(S_q,\Si_q)_{q\in\N^*_s}$ as general parameters characterized by the case {\bf iii} above, cases {\bf i} and {\bf ii} being independent limits thereof.

\subsection{Scaling of solutions} \label{subsec_5_2}

Let us now examine the interpretation of the set-theoretical parameter identif\/ied with the specif\/ication of the ordered partition of any subset $\K_q$ into $\D$-classes $\{\J_p,\; p\in\N^*_r\}$. There exists another kind of continuity property of the solutions, which relies on the freedom allowed by the choice of the constants $(f_\I)_{\I\in\{\I(i),\; i\in\K_q\}}$. This result brings to light the fact that the solution built on a subset $\K_q$ such that $r_q=1$, i.e.\ $\K_q$ is reduced to a single $\D$-class $\J_p$, is the elementary solution, in the sense that it can generate any other solution by a limit process of an adequate re-scaling of the constants $(f_\I)_{\I\in\{\I(i),\; i\in\K_q\}}$.

 Let $q\in\N^*_s$ such that $r_q\geq2$, a solution (parametrized by the constants $S_q\neq0$, $\Si_q$ and $(\e_\I,f_\I)_{\I\in\{\I(i),\; i\in\K_q\}}$) of system (\ref{syst}) restricted to the subset $\K_q$ and a parameter $\eta>0$.
  \begin{proposition} \label{prop_12}\sloppy
There exist a solution $($parametrized by the constants $S_q$, $\Si_q$ and $(\e_\I,f^\eta_\I)_{\I\in\{\I(i),\; i\in\K_q\}})$ of system {\rm (\ref{syst})} restricted to the subset $\K_q$, such that $r_q^\eta=1$ and that the solution with $(f_\I)_{\I\in\{\I(i),\; i\in\K_q\}}$ is $\nD$-multiplicatively reducible to the limit of the solution with $(f^\eta_\I)_{\I\in\{\I(i),\; i\in\K_q\}}$, up to a permutation of the indices in $\K_q$, when $\eta\rightarrow0^+$.
  \end{proposition}

    \begin{proof}
We f\/irst construct a permutation $\s_q$: $\K_q\longrightarrow\K_q$, such that, after re-ordering, the subset~$\K_q$ is an ordered partition as required by Theorem~\ref{theo_1} in the case of $r_q=1$. We accordingly bring all free indices to the beginning of the subset. Introducing the ordered partition of the subset $\K_q$ in free subsets and $\D$-classes
\[
\K_q=\bigcup_{p\in\PP_q}\J_p=\bigcup_{p\in\PP_q}~\bigcup_{l=0}^{l_p}\I^{(p)}_l,
\]
this can be done by def\/ining, for any $p\in\PP_q$, the permutation $\s_q\in\mf S_n$, whose support is a~sub\-set of $\K_q$, as
\[
\s_q(i)=i-\sum_{p'\in\PP_q \atop p'\leq p}n_{p'}+n_p+\sum_{p'\in\PP_q \atop p'\leq p}n^{(p')}_0-n^{(p)}_0, \qquad \forall\;  i\in\I^{(p)}_0;
\]
and
\[
\s_q(i)=i+\sum_{p'\in\PP_q \atop p'\geq p}n^{(p')}_0-n^{(p)}_0, \qquad \forall \,  i\in\J_p\setminus\I^{(p)}_0.
\]
Therefore, the subset $\K_q$ can be written as the following ordered partition
\[
\K^\s_q=\left(\bigcup_{p\in\PP_q}\s_q(\I^{(p)}_0)\right)\bigcup\left(\bigcup_{p\in\PP_q}\s_q(\J_p\setminus\I^{(p)}_0)\right) ,
\]
where we recall that the exponent ``${}^\s $'' indicates that the permutation $\s_q$ is applied. Moreover, the permutation $\s_q$ respects the relation $\nD$, meaning that the pair of indices $(i,j)\in\K_q^{(2,<)}$ is a~$\nD$-pair, if and only if the pair of indices $(\s_q(i),\s_q(j))\in(\K^\s_q)^2$ is a~$\nD$-pair.

Theorems~\ref{subtheo_3a} and~\ref{subtheo_3b}  insure the existence of a full solution (parametrized by the constants $S_q$, $\Si_q$ and $(\e_\I,f^\eta_\I)_{\I\in\{\I(i),\; i\in\K_q^\s\}}$, and the $\nD$-closed $\nD$-multiplicative 2-form $(g^\eta_{ij})_{\!(i,j)\in(\K^\s_q)^{(2,\!\!\nD\!\!,<)}}$) of system (\ref{syst}) restricted to the subset $\K_q$, where the constants $(f^\eta_\I)_{\I\in\{\I(i),\; i\in\K_q^\s\}}$ are def\/ined, for any $p\in\PP_q$, as
\[
f^\eta_{\I(\s_q(i))}=\eta^{-p+p_q+1}f_{\I(i)}, \qquad \forall\, i\in\J_p.
\]
In particular, we verify that $f^\eta_{\I(\min\K_q)}=f^\eta_{\I(\s_q(\min\K_q))}=f^\eta_{\I(\s_q(\min\J_{p_q+1}))}=f_{\I(\min\J_{p_q+1})}=1$, as required by Theorem~\ref{subtheo_3a}. Therefore, by construction, we have, for any $p\in\PP_q$
\[
\frac{f^\eta_{\I(\s_q(j))}}{f^\eta_{\I(\s_q(i))}}=\frac{f_{\I(j)}}{f_{\I(i)}}, \qquad \forall\, (i,j)\in\J_p^{(2,\nD)};
\]
and, for any $(p,p')\in\PP_q^{(2,<)}$,
\[
\frac{f^\eta_{\I(\s_q(i))}}{f^\eta_{\I(\s_q(j))}}=\eta^{p'-p}\frac{f_{\I(i)}}{f_{\I(j)}}\xrightarrow[\eta\rightarrow0^+]{}0,
\qquad \forall\, (i,j)\in\J_p\times\J_{p'}.
\]
This directly implies the expected result, which is that, for any $p\in\PP_q$
\[
\D^\eta_{\I(\s_q(i))\I(\s_q(j))}=\D_{\I(i)\I(j)}, \qquad \forall \, (i,j)\in\J_p^{(2,\nD)};
\]
and, for any $(p,p')\in\PP_q^{(2,<)}$
\begin{gather*}
\D^\eta_{\I(\s_q(i))\I(\s_q(j))}\xrightarrow[\eta\rightarrow0^+]{}S_q=\D_{\I(i)\I(j)}
\\
\Leftrightarrow \quad \D^\eta_{\I(\s_q(j))\I(\s_q(i))}\xrightarrow[\eta\rightarrow0^+]{}0=\D_{\I(j)\I(i)},
\qquad \forall\, (i,j)\in\J_p\times\J_{p'}.
\end{gather*}

The limit for $d$-coef\/f\/icients is trivially deduced from above, up to the multiplication by the $\nD$-closed $\nD$-multiplicative 2-form $(g_{ij})_{\!(i,j)\in\K_q^{(2,\!\!\nD\!\!,<)}}$, def\/ined as, for any $p\in\PP_q$
\[
g_{ij}=1, \qquad \forall\, (i,j)\in\J_p^{(2,\nD)};
\]
and, for any $(p,p')\in\PP_q^{(2,<)}$
\begin{gather*}
g_{ij}=\frac{\sqrt{\Si_q}}{B_q-S_q}, \qquad g_{ji}\frac{\sqrt{\Si_q}}{B_q}, \qquad \forall\, (i,j)\in\J_p\times\J_{p'}.\tag*{\qed}
\end{gather*}
\renewcommand{\qed}{}
    \end{proof}

\begin{remark}
 The construction of the scaled $R$-matrix in terms of the non-zero constants $(f_\I)_{\I\in\{\I(i),\; i\in\K_q\}}$ guarantees that the scaled $R$-matrix with multiple $\D$-classes still preserves the fundamental feature that a $\mc D$-class is inside a single $\D$-class, as shown in Proposition~\ref{subprop_3c}.

 We conclude that the solution built on any ordered partition of any subset $\K_q$ into $\D$-classes $\{\J_p,\; p\in\N^*_r\}$ is obtained as the result of scaling procedure applied to a solution built on a single $\D$-class $\K_q=\J_p$. The single $\D$-class solution is then generic.
 \end{remark}

\subsection[Re-parametrization of variables $(\l_k)_{k\in\N^*_n}$]{Re-parametrization of variables $\boldsymbol{(\l_k)_{k\in\N^*_n}}$} \label{subsec_5_3}

Here we propose an interpretation of the parameters $(f_\I)_{\I\in\{\I(i),\; i\in\N^*_n\}}$.
  \begin{proposition} \label{prop_13}
Let a $R$-matrix, solution of \ref{DQYBE2}, be parametrized $($among others$)$ by the set of constants $(f_\I)_{\I\in\{\I(i),\; i\in\N^*_n\}}$. Then, there exists a re-parametrization of the dynamical variable~$\l_k$, for any $k\in\N^*_n$, which eliminates this dependence.
  \end{proposition}

    \begin{proof}
From Theorem~\ref{subtheo_3a} (when there exists $q\in\N^*_s$ such that the $R$-matrix, restricted to the subset $\K_q$, has a trigonometric behavior, i.e.\ when $S_q\neq0$) or Theorem~\ref{subtheo_3b} (when there exists $q\in\N^*_s$ such that the $R$-matrix, restricted to the subset $\K_q$, has a rational behavior, i.e.\ when $S_q=0$), it is manifest that these parameters can be respectively re-absorbed in a re-def\/inition of the dynamical variables $(\l_k)_{k\in\N^*_n}$ as
  \[
\l_i\longrightarrow\l_i+\frac{\e_{\I(i)}}{|\I(i)|}\frac{\log_\mathbb S(f_{\I(i)})}{A_q} \qquad \textrm{or} \qquad \l_i\longrightarrow\l_i+\frac{\e_{\I(i)}}{|\I(i)|}f_{\I(i)}, \qquad \forall\, i\in\N^*_n.
  \]
Following the same argumentation as in the remark of Theorem~\ref{subtheo_3a} concerning the def\/inition of the quantity $A_q$, this re-parametrization of the dynamical variables $(\l_k)_{k\in\N^*_n}$ (when the $R$-matrix has a trigonometric behavior) is indeed independent from the choice of the determination of the logarithm function, and of the choice of $\log(f_{\I(i)})$, when $f_{\I(i)}\in\R^*_-$. This justif\/ies that the constants $(f_\I)_{\I\in\{\I(i),\; i\in\N^*_n\}}$ should be advantageously seen as belonging to the Riemann surface~$\mathbb S$, as well as the use of function $\log_\mathbb S$.
    \end{proof}

This re-parametrization of the dynamical variables $(\l_k)_{k\in\N^*_n}$ is the only one under which \ref{DQYBE2} is form-invariant, since it must preserve the translation $\l_i\longrightarrow\l_i+1$, for any $i\in\N^*_n$. This is the reason why it will be refereed as the canonical parametrization of the dynamical variables $(\l_k)_{k\in\N^*_n}$. Moreover, the family of signs $(\e_\I)_{\I\in\{\I(i),\; i\in\N^*_n\}}$ cannot be re-absorbed in such a way, and represents a set of genuine relevant parameters of a generic solution of \ref{DQYBE2}, to be interpreted in the next subsection.

{\sloppy  To summarize, any $R$-matrix, solution  of \ref{DQYBE2}, characterized by an ordered partition of the $\D$-class $\J_p$ into $\mc D$-classes $\{\I(i),\; i\in\J_p\}$, is built by juxtaposition following Theorem~\ref{theo_1} and Proposition~\ref{prop_10}, and by $\nD$-multiplicative covariance following Propositions~\ref{subprop_9a},~\ref{subprop_9b} and~\ref{subprop_9c}, of solutions obtained by
  \begin{itemize}\itemsep=0pt
\item limit following Propositions~\ref{subprop_11a} and~\ref{subprop_11b},
\item scaling following Proposition~\ref{prop_12},
\item re-parametrization of the dynamical variables following Proposition~\ref{prop_13};
  \end{itemize}
of a solution (parametrized by the non-zero constants $S_p$ and $\Si_p$, the family of signs $(\e_\I)_{\I \in \{\I(i),\;  i \in \J_p\}}$, and the $\nD$-multiplicative 2-form $(g_{ij})_{\!(i,j)\in\J_p^{(2,\!\!\nD\!\!,<)}}$) of \ref{DQYBE2} on a single $\D$-class $\J_p$.

}

\subsection{Commuting operators}

The form of a generic $R$-matrix, solution of \ref{DQYBE2} given by Theorem~\ref{theo_5}, allows to immediately bring to light a set of operators which commute with the $R$-matrix.
  \begin{proposition} \label{prop_14}
For any $R$-matrix, solution of \ref{DQYBE2}, the operator
  \[
R_0=\sum\limits_{q=1}^s\sum\limits_{p\in\PP_q}\sum\limits_{i\in\I^{(p)}_0}R^{(q,p)}_i=\sum_{i\in\I_0}\D_{ii}e^{(n)}_{ii}\otimes e^{(n)}_{ii}
  \]
together with the family of operators $(R^{(q,p)}_{\I(i)})_{i\in\J_p\setminus\I^{(p)}_0,\; p\in\PP_q|q\in\N^*_s}$ build a set of mutually commuting operators, commuting with the $R$-matrix.
\end{proposition}

    \begin{proof}
{\bf 1.~Mutual commutation.}
For any $q\in\N^*_s$, for any $p\in\PP_q$, we recall the formula
\[
R^{(q,p)}_{\I(i)}=\frac{\D_{\I(i)}}{|\I(i)|}\sum_{j,j'\in\I(i)}e^{(n)}_{jj'}\otimes e^{(n)}_{j'j}, \qquad \forall \, i\in\J_p.
\]
For any $(i,j)\in(\J_p\setminus\I^{(p)}_0)\times(\J_{p'}\setminus\I^{(p')}_0)\; |\; (i,j)\in(\N^*_n)^2$, with $q,q'\in\N^*_s$ and $(p,p')\in\PP_q\times\PP_{q'}$, it is straightforward to check that
\[
\big[R_0,R^{(q,p)}_{\I(i)}\big]=\big[R^{(q,p)}_{\I(i)},R^{(q',p')}_{\I(j)}\big]=0.
\]
This relies on the fact that the indices appearing in the sum def\/ining the operator $R_0$ or $R^{(q',p')}_{\I(j)}$ do not appear in the sum def\/ining the operator $R^{(q,p)}_{\I(i)}$, because of the partitioning of the set of indices $\N^*_n$.

{\bf 2.~Commutation with the $\boldsymbol{R}$-matrix.}
This implies in particular that any operators of the family $(R_0,(R^{(q,p)}_{\I(i)})_{i\in\J_p\setminus\I^{(p)}_0,\; p\in\PP_q|q\in\N^*_s})$ commute with their sum, being the operator
\[
R_0+\sum\limits_{q=1}^s\sum\limits_{p\in\PP_q}\sum\limits_{i\in\J_p\setminus\I^{(p)}_0}R^{(q,p)}_{\I(i)}
=\sum\limits_{q=1}^s\sum\limits_{p\in\PP_q}\sum\limits_{i\in\J_p}R^{(q,p)}_{\I(i)}=\sum\limits_{i\in\N^*_n}R^{(q,p)}_{\I(i)}.
\]
Remember now that a $R$-matrix, solution of~\ref{DQYBE2}, is essentially the sum of three kinds of terms, which are the previous sum (containing all terms $\D_{jj'}e^{(n)}_{jj'}\otimes e^{(n)}_{j'j}$, when $j\mc Dj'$), terms such as $\D_{jj'}e^{(n)}_{jj'}\otimes e^{(n)}_{j'j}$ and such as $d_{jj'}e^{(n)}_{jj}\otimes e^{(n)}_{jj'}$, when $j\nD j'$. It remains to check the commutativity of  any operators of the family $(R_0,(R^{(q,p)}_{\I(i)})_{i\in\J_p\setminus\I^{(p)}_0,\; p\in\PP_q|q\in\N^*_s})$ with the second and third kind of terms appearing in the expression of a generic solution of~\ref{DQYBE2}.

To this end, let $i\in\N^*_n$, $j,j'\in\I(i)$ and $(k,k')\in\N_n^{*(2,\nD)}$. Here, for convenience, we allow for once to have an equality between two indices distinctly labelled. However, there are compatibility conditions to fulf\/il. For example, we can have $j=k$, but not at the same time $j'\neq k'$, otherwise~$k\mc Dk'$. Such considerations give directly that
\[
\big(e^{(n)}_{jj'}\otimes e^{(n)}_{j'j}\big)
\big(e^{(n)}_{kk'}\otimes e^{(n)}_{k'k}\big)=\big(e^{(n)}_{kk'}\otimes e^{(n)}_{k'k}\big)\big(e^{(n)}_{jj'}\otimes e^{(n)}_{j'j}\big)=0
\]
and
\[
\big(e^{(n)}_{jj'}\otimes e^{(n)}_{j'j}\big)
\big(e^{(n)}_{kk}\otimes e^{(n)}_{k'k'}\big)=\big(e^{(n)}_{kk}\otimes e^{(n)}_{k'k'}\big)\big(e^{(n)}_{jj'}\otimes e^{(n)}_{j'j}\big)=0,
\]
which yields, for any $i\in\J_p\setminus\I^{(p)}_0$, with $q\in\N^*_s$ and $p\in\PP_q$
  \begin{gather*}
[R,R_0]=\big[R,R^{(q,p)}_{\I(i)}\big]=0.\tag*{\qed}
  \end{gather*}
  \renewcommand{\qed}{}
    \end{proof}

\subsection[(weak) Hecke and non-Hecke $R$-matrices]{(weak) Hecke and non-Hecke $\boldsymbol{R}$-matrices} \label{subsec_5_5}

We recall that we have dropped any Hecke or weak Hecke condition in our derivation Nevertheless such conditions will be shown to arise in connection with the choice of the family of signs $(\e_\I)_{\I\in\{\I(i),\; i\in\N^*_n\}}$, parametrizing any $R$-matrix.

Let us f\/irst give the def\/inition for the (weak) Hecke condition following \cite{EtVar_1}.
\begin{definition}[(weak) Hecke condition] \label{def_2}\qquad
    \begin{enumerate}[label={\bf\roman*}.]\itemsep=0pt
\item \looseness=-1 A $R$-matrix satisf\/ies the Hecke condition with parameters $\varrho,\k\in\C^*$, such that $\varrho\neq-\k$, if the eigenvalues of the permuted $R$-matrix $\check R=PR$, $P$ being the permutation operator of spaces $V_1$ and $V_2$, are $\varrho$ on the one-dimensional vector space $V_{ii}=\C e^{(n)}_i\otimes e^{(n)}_i$, for any index $i\in\N^*_n$, and $\varrho$, $-\k$ on the two-dimensional vector space $V_{ij}=\C e^{(n)}_i\otimes e^{(n)}_j\oplus \C e^{(n)}_j\otimes e^{(n)}_i$, for any pair of indices $(i,j)\in\N_n^{*(2,<)}$, where $(e^{(n)}_i)_{i\in\N^*_n}$ is the canonical basis of the vector space $V=\C^n$.
\item A $R$-matrix satisf\/ies the weak Hecke condition with parameters $\varrho,\k\in\C^*$, such that $\varrho\neq-\k$, if the minimal polynomial of the permuted matrix $\check R$ is $\m_{\check R}(X)=(X-\varrho)(X+\kappa)$.
    \end{enumerate}
  \end{definition}

 \begin{remark}
 A $R$-matrix, for which the set of indices $\N^*_n$ is reduced to a single $\mc D$-class is not strictly of (weak) Hecke-type with parameters $\varrho,\k\in\C^*$, such that $\varrho\neq-\k$, since in this case $\mu_{\check R}(X)=X-\D_{\I(1)}$. By language abuse, it can be considered as a degenerate weak Hecke-type $R$-matrix with parameters $\varrho,\k\in\C^*$, such that $\varrho=-\k=\D_{\I(1)}$.
\end{remark}

In the following, we will assume that the set of indices $\N^*_n$ is not reduced to a single $\mc D$-class, unless otherwise stated.

The classif\/ication of Hecke-type solutions of \ref{DQYBE2} is well known \cite{EtSchif}, whereas the classif\/ication of weak Hecke-type solutions of \ref{DQYBE2} remains unknown. Reference \cite{EtVar_1} presents two fundamental theorems treating separately trigonometric and rational cases. They stress that, according to the value of the parameter $\k$ and up to gauge transformations and to additional trivial transformations, such as scalar multiplication, or global linear re-parametrization of the dynamical variable $\l$, there essentially exist two distinct types of $R$-matrices satisfying the Hecke condition, the so-called basic trigonometric ($\varrho=1$ and $\k\neq1$) and rational ($\varrho=\k=1$) Hecke-type solutions.

Let $r\in\N^*_n$ and $\{\J_p,\; p\in\N^*_r\}$ be an ordered partition of the set of indices $\N^*_n$ into $r$ $\D$-classes. Using the canonical parametrization of the dynamical variables $(\l_k)_{k\in\N^*_n}$ of Proposition~\ref{prop_13}, by simple identif\/ication, these basic Hecke-type solutions are particular examples of solutions of \ref{DQYBE2} given by Theorems~~\ref{subtheo_3a},~\ref{subtheo_3b},~\ref{theo_4} and~\ref{theo_5}, as described in the following def\/initions.

\begin{definition}[basic trigonometric and rational Hecke-type behavior]\label{subdef_3a} \qquad
    \begin{enumerate}[label={\bf\roman*}.]\itemsep=0.5pt
\item The basic trigonometric Hecke-type $R$-matrix, solution of \ref{DQYBE2}, with parameter $\k\in\C^*\setminus\{1\}$, is, up to trivial transformations, a $R$-matrix, parametrized by
      \begin{itemize}\itemsep=0.5pt
\item $\I_0=\N^*_n$, i.e.\ $\N^*_n$ does not contain any $\mc D$-classes;
\item $s=1$, i.e.\ $\N^*_n=\K_1=\bigcup\limits_{p=1}^r\J_p$;
\item the non-zero constants $S_1=\k-1$, $\Si_1=\k$ and $T_1=\frac1\k$;
\item the signs $\e_{\I(i)}=\e_i=-1$, for any $i\in\N^*_n$;
\item the multiplicative 2-form $g_{ij}=-1$, for any $(i,j)\in\N_n^{*(2,\nD)}$.
      \end{itemize}
\item The basic rational Hecke-type $R$-matrix, solution of \ref{DQYBE2}, is, up to trivial transformations, a $R$-matrix, parametrized by
      \begin{itemize}\itemsep=0.5pt
\item $\I_0=\N^*_n$, i.e.\ $\N^*_n$ does not contain any $\mc D$-classes;
\item $s=r$, i.e.\ $\K_q=\J_q$, for any $q\in\N^*_s$;
\item the constants $S_q=0$ and the non-zero-constants $\Si_q=1$, for any $q\in\N^*_s$;
\item the non-zero constants $\Si_{qq'}=1$, for any $(q,q')\in\N_s^{*(2,<)}$;
\item the signs $\e_{\I(i)}=\e_i=-1$, for any $i\in\N^*_n$;
\item the multiplicative 2-form $g_{ij}=-1$, for any $(i,j)\in\N_n^{*(2,\nD)}$.
      \end{itemize}
    \end{enumerate}
  \end{definition}

\begin{remark} The trivial manipulations to get Def\/inition~\ref{subdef_3a} from~\cite{EtVar_1} are a multiplication of the $R$-matrix by~$-1$, and a global re-parametrization of the dynamical variables $(\l_k)_{k\in\N^*_n}$ as $\l\rightarrow-\l$, the rest relies on a simple identif\/ication.
\end{remark}

To be as exhaustive as possible, we propose an alternative formulation of the classif\/ication theorems of Hecke-type solutions of \ref{DQYBE2}~\cite{EtVar_1}, as well as a classif\/ication of the weak Hecke-type solutions of \ref{DQYBE2}.

\begin{proposition}[Hecke-type $R$-matrices]\label{subprop_15a}
Any Hecke-type $R$-matrix, solution of \ref{DQYBE2}, with parameters $\varrho,\k\in\C^*$, such that $\varrho\neq-\k$, is twist-gauge reducible to a basic trigonometric or rational Hecke-type $R$-matrix.
  \end{proposition}

    \begin{proof}
This is an obvious corollary of the classif\/ication theorems of Hecke-type solutions of~\ref{DQYBE2} \cite{EtSchif}, and of the two Def\/initions~\ref{subdef_3a}. However, we can explicit why no other $R$-matrix does satisfy the Hecke condition. Let $R$ be a matrix, solution of \ref{DQYBE2}, built on the ordered partition $\{\J_p,\; p\in\N^*_r\}$ of the set of indices $\N^*_n$, which additionally satisf\/ies the Hecke condition. The proof of the proposition relies on the fact that, by direct calculation, the permuted matrix $\check R$ is expressed as follows
\[
\check R=\sum_{i,j=1}^n\D_{ji}e^{(n)}_{ii}\otimes e^{(n)}_{jj}+\sum_{i\neq j=1}^nd_{ji}e^{(n)}_{ij}\otimes e^{(n)}_{ji},
\]
from which we deduce the characteristic polynomial of its restriction to the subspace $V_{ij}$, for any pair of indices $(i,j)\in\N_n^{*(2,<)}$, as
\[
P_{ij}(X)=
    \begin{vmatrix}
\D_{ji}-X	& d_{ji}
\\
d_{ij}		& \D_{ij}-X

    \end{vmatrix}=X^2-S_{ij}X-\Si_{ij}.
\]

By projecting the permuted matrix $\check R$ in the subspace $V_{ii}$, we deduce immediately that $\D_{\I(i)}=\D_{ii}=\varrho$, for any $i\in\N^*_n$.

If $\I_0\subsetneq\N^*_n$, then there exists a pair of indices $(i,j)\in\N_n^{*(2,<)}$ such that $i\mc Dj$. From Proposition~\ref{prop_8}, we have $\D_{ij}=\D_{ji}=\D_{\I(i)}$, i.e.\ $S_{ij}=2\D_{\I(i)}$ and $\Si_{ij}=-\D_{\I(i)}^2$, yielding that $P_{ij}(X)=(X-\D_{\I(i)})^2$. This leads to a contradiction. The matrix $\check R$ does not indeed sa\-tis\-fy the Hecke condition, since it has a single eigenvalue $\D_{\I(i)}$ on the subspace $V_{ij}$. One must then assume that $\I_0=\N^*_n$ and the set of indices only contains free indices.

For any pair of indices $(i,j)\in\N_n^{*(2,<)}$, Def\/inition~\ref{def_2} directly implies that $\varrho$ and $-\k$ are root of the polynomial $P_{ij}$, i.e.\ $S_{ij}=\varrho-\k$ and $\Si_{ij}=\varrho\k$. This means in particular that $S_q=\varrho-\k$, $\Si_q=\varrho\k$ and $T_q=\frac\k\varrho$, for any $q\in\N^*_s$, as well as that  $\Si_{qq'}=\varrho\k$, for any $(q,q')\in\N_s^{*(2,<)}$, if $s\geq2$. But, these equalities are more constraining than that
    \begin{itemize}\itemsep=0pt
\item If $\varrho\neq\k$, then $s=1$. Otherwise, if $s\geq2$, let $(q,q')\in\N_s^{*(2,<)}$ and a pair of indices $(i,j)\in\K_q\times\K_{q'}$. Then, by construction, $S_{ij}=\varrho-\k=0$, which leads to a contradiction. Hence any trigonometric solution has a single subset $\K_1$.
\item If $\varrho=\k$, then $r_q=1$, for any $q\in\N^*_s$. Otherwise, let $q\in\N^*_s$, such that $r_q\geq2$. By Theorem~\ref{theo_4}, this leads to a contradiction, since $S_q=0$. Hence any rational solution is a~decoupled $R$-matrix, for which any subset $\K_q$ is reduced to a single $\D$-class.
    \end{itemize}
Finally, in both cases, from expressions for the non-zero constants $(\D_{ii})_{i\in\N^*_n}$ in Theorem~\ref{theo_2}, this additionally imposes that $\e_{\I(i)}=\e_i=1$, for any $i\in\N^*_n$.

Since, for any $R$-matrix satisfying the Hecke condition with parameters $\varrho,\k\in\C^*$, the matrix $\frac1\varrho R$ satisf\/ies the Hecke condition with parameters $1$ and $\k'=\frac\k\varrho$, it is suf\/f\/icient to assume that $\varrho=1$ without loss of generality. This gives the expected results up to the trivial manipulations presented in the remark of Def\/inition~\ref{subdef_3a}, which particularly make the re-parametrizations $S_q\rightarrow S'_q=-S_q=\k'-1$ and  $T_q\rightarrow T'_q=\frac1{T_q}=\frac1{\k'}$, for any $q\in\N^*_s$ and $\e_i\rightarrow\e_i'=-1$, for any $i\in\N^*_n$.

Finally let us expose why each non-zero function of the family $(g_{ij})_{\!(i,j)\in\N_n^{*(2,\!\!\nD\!\!)}}$ can be brought to $-1$. For any Hecke-type solution, $\I_0=\N^*_n$, hence any pair of indices $(i,j)\in(\N^*_n)^2$ is a $\nD$-pair, so that, according to Propositions~\ref{subprop_9a},~\ref{subprop_9b} and~\ref{subprop_9c}, the family of functions $(-g_{ij})_{\!(i,j)\in\N_n^{*(2,\!\!\nD\!\!)}}$ is a~closed multiplicative 2-form, then exact under appropriate assumptions, multiplicative 2-form. This means that they are of gauge form, and can be factorized out thanks to Proposition~\ref{subprop_1a}, f\/inally leaving the wanted factor $-1$ in front of any $d$-coef\/f\/icient. It remains now to apply the trivial manipulations of the remark of Def\/inition~\ref{subdef_3a}.
    \end{proof}

\begin{remark} From Proposition~\ref{prop_10}, if $s\geq2$, a basic rational Hecke-type solution is by construction a decoupled $R$-matrix, whereas a basic trigonometric Hecke-type solution never is. These two dif\/ferent behaviors are unif\/ied as soon as the Hecke condition is dropped.

In the same spirit, Proposition~\ref{subprop_11a} generalizes the well-known property that basic rational Hecke-type solutions can be obtained as limits of basic trigonometric Hecke-type solutions of parameter $\k\in\C^*\setminus\{1\}$, when $\k\rightarrow1$.
\end{remark}

 By analogy with the terminology used for Hecke-type solutions, we will introduce the notion of basic trigonometric or rational weak Hecke-type solutions of \ref{DQYBE2} as follows. Let us particularly insist on the fact that, as we will see explicitly later, unlike the Hecke-type condition, the weak Hecke-type condition allow $\D$-classes and do not constrain the choice of signs $(\e_\I)_{\I\in\{\I(i),\; i\in\N^*_n\}}$.

  \begin{definition}[basic trigonometric and rational weak Hecke-type behavior]\label{subdef_3b}\qquad
      \begin{enumerate}[label={\bf\roman*}.]\itemsep=0pt
\item The basic trigonometric weak Hecke-type $R$-matrix, solution of \ref{DQYBE2}, with parameter $\k\in\C^*\setminus\{1\}$, is, up to trivial transformations, a $R$-matrix, parametrized by
      \begin{itemize}\itemsep=0pt
\item $s=1$, i.e.\ $\N^*_n=\K_1=\bigcup\limits_{p=1}^r\J_p$;
\item the non-zero constants $S_1=\k-1$, $\Si_1=\k$ and $T_1=\frac1\k$;
\item the multiplicative 2-form $g_{ij}=-1$, for any $(i,j)\in\N_n^{*(2,\nD)}$.
      \end{itemize}
\item The basic rational weak Hecke-type $R$-matrix, solution of \ref{DQYBE2}, is, up to trivial transformations, a $R$-matrix, parametrized by
      \begin{itemize}\itemsep=0pt
\item $s=r$, i.e.\ $\K_q=\J_q$, for any $q\in\N^*_s$; \vs{-.15cm}
\item the constants $S_q=0$ and the non-zero-constants $\Si_q=1$, for any $q\in\N^*_s$; \vs{-.15cm}
\item the non-zero constants $\Si_{qq'}=1$, for any $(q,q')\in\N_s^{*(2,<)}$; \vs{-.15cm}
\item the multiplicative 2-form $g_{ij}=-1$, for any $(i,j)\in\N_n^{*(2,\nD)}$.
      \end{itemize}
    \end{enumerate}
In both cases, the ordered partition of set of indices $\N^*_n$ remain free, and the family of signs $(\e_\I)_{\I\in\{\I(i),\; i\in\N^*_n\}}$ is arbitrary.
  \end{definition}

\begin{proposition}[weak Hecke-type $R$-matrices]\label{subprop_15b}
Any $R$-matrix, solution of \ref{DQYBE2}, which satisfies the weak Hecke condition with parameters $\varrho,\k\in\C^*$, such that $\varrho\neq-\k$, is $\nD$-multiplicatively reducible to a basic trigonometric or rational weak Hecke-type $R$-matrix.
  \end{proposition}
    \begin{proof}
Thanks to the zero-weight condition, the permuted matrix $\check R$ is block diagonal, up to a~permutation in $\mf S_{n^2}$, where the blocks are of the form $\D_{\I(i)}$ (in the subspace $V_{ii}$) or $
      \begin{pmatrix}
\D_{ji} & d_{ji}
\\
d_{ij} & \D_{ij}
      \end{pmatrix}$. By projecting the permuted matrix $\check R$ in the subspace $V_{ii}$, we deduce immediately either $\D_{\I(i)}=\varrho$ or $\D_{\I(i)}=-\k$, for any $i\in\N^*_n$.

Assuming that $\N^*_n$ is not reduced to a single $\mc D$-class, the restriction of the matrix $\check R$ to the subspace $V_{ij}$ has to satisfy the weak Hecke condition, for any pair of indices $(i,j)\in\N_n^{*(2,<)}$, i.e.\ $\mu_{ij}\; |\; \mu_{\check R}$, where $\mu_{ij}$ is the minimal polynomial of the restriction of the matrix $\check R$.

As above, this is trivially satisf\/ied for any pair of indices $(i,j)\in\N_n^{*(2,<)}$, such that $i\mc Dj$. Let then now a $\nD$-pair of indices $(i,j)\in\N_n^{*(2,\nD,<)}$. The rest of the proof is almost identical to the proof of Proposition~\ref{subprop_15a}, since we have that $\mu_{ij}=P_{ij}=\mu_{\check R}$. The major dif\/ferences are that the signs $(\e_\I)_{\I\in\{\I(i),\; i\in\N^*_n\}}$ are {\it no longer constrained to be all equal} (from Theorems~\ref{subtheo_3a} and~\ref{subtheo_3b}, we particularly deduce that $\e_i=1$, if $\D_{ii}=\varrho$, and $\e_i=-1$, if $\D_{ii}=-\k$), and that we have to use the $\nD$-multiplicative covariance instead of the twist covariance, when the set of indices~$\N^*_n$ contains $\mc D$-classes, which was excluded in the case of Hecke-type $R$-matrices. Let us note that the $\nD$-multiplicative covariance does not af\/fect the minimal polynomial $\mu_{\check R}$, since we have just proved that $\mu_{\check R}=P_{ij}$, for any pair of indices $(i,j)\in\N
 _n^{*(2,<)}$, where the polynomial $P_{ij}$ is obviously invariant under such transformation.

This occurrence only arises, when $n\geq3$, hence it could not arise in the classif\/ication of $R$-matrices, solution of $\mc Gl_2(\C)$, as seen for trigonometric $R$-matrices in \cite{Ju}.
    \end{proof}

 We now come to the main concluding statement.

  \begin{theorem}[decoupling theorem]\label{theo_6}
Any $R$-matrix, solution of \ref{DQYBE2}, parametrized $($among others$)$ by the family of constants $(S_q,\Si_q)_{q\in\N^*_s}$, is $\nD$-multiplicatively reducible to a decoupled $R$-matrix, whose constituting blocks are either $\mc D$-classes or satisfy the weak Hecke condition with the family of parameters $(\varrho_q,\k_q)_{q\in\N^*_s}$, such that $\varrho_q\neq-\k_q$ and
\[
S_q=\varrho_q-\k_q \qquad \textrm{and} \qquad \Si_q=\varrho_q\k_q, \qquad \forall\,  q\in\N^*_s.
\]
  \end{theorem}

      \begin{proof}
This is a corollary of Propositions~\ref{prop_10} and~\ref{subprop_15b} and Theorems~\ref{theo_1}, \ref{theo_2}, \ref{subtheo_3a}, \ref{subtheo_3b},
  \ref{theo_4} and~\ref{theo_5}.
    \end{proof}

 Beforing concluding this article, Fig.~\ref{figure_1} gives an example of a non-Hecke $R$-matrix, solution of $\mc Gl_4(\C)$-\ref{DQYBE2} with rational behavior. We have chosen the set of indices $\N^*_4=\K_1$ to be a~single $\D$-class $\J_1$, where the free subset is $\I_0=\I^{(1)}_0=\{1,2\}$ and where the remaining indices $\I^{(1)}_1=\{3,4\}$ form a $\mc D$-class. For any $i\in\I_0$ or for any $i\in\I^{(1)}_1$, we have respectively denoted the quantities depending on $\I(i)$, such as the sign $\e_{\I(i)}$ or the variable $\L_{\I(i)}$, by ``${}_i$'' or by ``$\_$'' instead of the previous notation ``${}_{\I(i)}$''. We have also dropped the index ``${}_1$'' for the constants $S_1$ and $\Si_1$. We have chosen the signs $\e_1=\e_2=1$ and $\e=-1$, and have f\/ixed the non-zero constants $f_1$, $f_2$ and $f$ to 0, and the $\nD$-multiplicative 2-form to 1. The $R$-matrix we present is non-Hecke thanks to the presence of dif\/ferent signs and the $\mc D$-class $\I^
 {(1)}_1$, but satisf\/ies the weak Hecke condition, because it is not decoupled.

  \begin{figure}[t!]
\rotatebox{90}{\mbox{%
\
  \begin{minipage}{20cm} \vs{3.5cm}
\footnotesize
  \begin{displaymath}
\textrm{\normalsize $R=$}\left(
      \begin{BMAT}{cccc.cccc.cccc.cccc}{cccc.cccc.cccc.cccc}
\blue1	& 0				& 0				& 0				& 0				& 0	& 0			& 0						& 0				& 0			& 0	& 0	& 0				 & 0					& 0	& 0
\\
0	& \blue{1-\tfrac1{\l_1-\l_2}}	& 0				& 0				& \blue{\tfrac1{\l_1-\l_2}}	& 0	& 0			& 0						& 0				 & 0			 & 0	& 0	& 0				& 0					& 0	& 0
\\
0	& 0				& \magenta{1-\tfrac1{\l_1+\l}}	& 0				& 0				& 0	& 0			& 0						& \magenta{\tfrac1{\l_1+\l}}	& 0			 & 0	& 0	& 0				& 0					& 0	& 0
\\
0	& 0				& 0				& \cyan{1-\tfrac1{\l_1+\l}}	& 0				& 0	& 0			& 0						& 0				& 0			& 0	& 0	& \cyan{\tfrac1{\l_1+\l}}	& 0					& 0	& 0
\\
0	& \blue{\tfrac{-1}{\l_1-\l_2}}	& 0				& 0				& \blue{1+\tfrac1{\l_1-\l_2}}	& 0	& 0			& 0						& 0				 & 0			 & 0	& 0	& 0				& 0					& 0	& 0
\\
0	& 0				& 0				& 0				& 0				& \blue1	& 0			& 0						& 0				& 0			& 0	& 0	& 0				 & 0					& 0	& 0
\\
0	& 0				& 0				& 0				& 0				& 0	& \cyan{1-\tfrac1{\l_2+\l}}	& 0			& 0				& \cyan{\tfrac1{\l_2+\l}}	& 0	& 0	& 0				& 0					& 0	& 0
\\
0	& 0				& 0				& 0				& 0				& 0	& 0							& \magenta{1-\tfrac1{\l_2+\l}}	& 0				& 0				 & 0	& 0	& 0				& \magenta{\tfrac1{\l_2+\l}}	& 0	& 0
\\
0	& 0				& \magenta{\tfrac{-1}{\l_1+\l}}	& 0				& 0				& 0	& 0			& 0						& \magenta{1+\tfrac1{\l_1+\l}}	& 0			 & 0	& 0	& 0				& 0					& 0	& 0
\\
0	& 0				& 0				& 0				& 0				& 0	& \cyan{\tfrac{-1}{\l_2+\l}}	& 0				& 0				& \cyan{1+\tfrac1{\l_2+\l}}	& 0	& 0	& 0				& 0					& 0	& 0
\\
0	& 0				& 0				& 0				& 0				& 0	& 0			& 0						& 0				& 0			& \blue{-1}	& 0	& 0				 & 0					& 0	& 0
\\
0	& 0				& 0				& 0				& 0				& 0	& 0			& 0						& 0				& 0			& 0	& \blue0	 & 0				 & 0					& \blue{-1}	& 0
\\
0	& 0				& 0				& \cyan{\tfrac{-1}{\l_1+\l}}	& 0				& 0	& 0			& 0						& 0				& 0			 & 0	& 0	& \cyan{1+\tfrac1{\l_1+\l}}	& 0					& 0	& 0
\\
0	& 0				& 0				& 0				& 0				& 0	& 0			& \magenta{\tfrac{-1}{\l_2+\l}}						& 0				 & 0			 & 0	& 0	& 0				& \magenta{1+\tfrac1{\l_2+\l}}	& 0	& 0
\\
0	& 0				& 0				& 0			& 0				& 0	& 0			& 0						& 0				& 0			& 0	& \blue{-1}	& 0				 & 0					& \blue0	& 0
\\
0	& 0				& 0				& 0			& 0				& 0	& 0			& 0						& 0				& 0			& 0	& 0	& 0			 & 0						 & 0	& \blue{-1}
      \end{BMAT}
    \right)\!\!
  \end{displaymath}
  \end{minipage}
}}

\vspace*{-9mm}

\caption{Example of a non-Hecke $R$-matrix, solution of $\mc Gl_4(\C)$-\ref{DQYBE2}.} \label{figure_1}
  \end{figure}

\section{Conclusion}
We have proceeded to the exhaustive classif\/ication of the non-af\/f\/ine ``non-Hecke''-type quantum $\mc Gl_n(\C)$ dynamical $R$-matrices obeying DQYBE. In particular, we have succeeded to fully characterize its space of moduli, and prove that weak Hecke-type $R$-matrices are the elementary constituting blocks of non-Hecke-type $R$-matrices. This classif\/ication then brings to light a wide range of new solutions to this equation, while the Hecke-type solutions appear as a~very particular type of solutions. As a matter of fact, the parametrization of a general solution of DQYBE involves a large number of objects of dif\/ferent mathematical natures, which is drastically restricted when the Hecke condition is considered.

These results may pave the way for the classif\/ication of af\/f\/ine non-Hecke-type quantum dynamical $R$-matrices obeying DQYBE, or less ambitiously may be a f\/irst step for the under\-stan\-ding of the Baxterization of non-af\/f\/ine non-Hecke-type quantum dynamical $R$-matrices, whose general case still remains an open problem nowadays. Occurrence of weak Hecke building blocks, for which Baxterization procedure is known \cite{AKR}, at least in the non-dynamical case, allows to be quite hopeful in this respect.

Moreover, the non-Hecke-type solutions of DQYBE are interesting by themselves. In recent developments of researches on the second Poisson structure of Calogero models emerge non-Hecke-type solutions of dynamical classical Yang--Baxter equation \cite{AR}, such as
\[
r=
          \begin{pmatrix}
0	& 0			& 0			& 0
\vspace{1mm}\\
0	& \frac{w_1}{\l_1-\l_2}	& \frac2{\l_1-\l_2}	& 0
\vspace{1mm}\\
0	& \frac2{\l_2-\l_1}	& \frac{w_2}{\l_2-\l_1}	& 0
\vspace{1mm}\\
0	& 0			& 0			& 0
      \end{pmatrix},
\]
where $w_1,w_2\in\C^*$. More precisely, this matrix is the solution for the matrix~$a$, occurring in a~general quadratic Poisson bracket algebra
\[
\{l_1,l_2\}=al_1l_2+l_1bl_2+l_2cl_1+l_1l_2d,
\]
and can be obviously obtained as a semi-classical limit of a non-Hecke solution of \ref{DQYBE2}.

\section*{Appendix}
Here we give an exhaustive lists of def\/initions, theorems, propositions and lemmas, along with their names or some keywords.
\subsection*{List of def\/initions}
  \begin{itemize} \itemsep=0pt
\item[]Def\/inition~\ref{def_1}~~Multiplicative 2-forms\dotfill\pageref{def_1}
\item[]Def\/inition~\ref{def_2}~~(weak) Hecke condition\dotfill\pageref{def_2}
\item[]Def\/initions~\ref{subdef_3a} and~\ref{subdef_3b}~~Basic trigonometric and rational\\ \hspace*{15mm}(weak) Hecke-type behavior\dotfill\pageref{subdef_3a}~\&~\pageref{subdef_3b}
  \end{itemize}
\subsection*{List of theorems}
  \begin{itemize} \itemsep=0pt
\item[]Theorem~\ref{theo_1}~~$\D$-incidence matrices\dotfill\pageref{theo_1}
\item[]Theorem~\ref{theo_2}~~inside a set $\K_q$ of $\D$-classes\dotfill\pageref{theo_2}
\item[]Theorems~\ref{subtheo_3a} and~\ref{subtheo_3b}~~Trigonometric and rational behavior\dotfill\pageref{subtheo_3a}~\&~\pageref{subtheo_3b}
\item[]Theorem~\ref{theo_4}~~Trigonometric behavior\dotfill\pageref{theo_4}
\item[]Theorem~\ref{theo_5}~~General $R$-matrices\dotfill\pageref{theo_5}
\item[]Theorem~\ref{theo_6}~~Decoupling theorem\dotfill\pageref{theo_6}
  \end{itemize}

\subsection*{List of propositions}
  \begin{itemize}\itemsep=0pt
\item[]Propositions~\ref{subprop_1a}, \ref{subprop_1b} and \ref{subprop_1c}~~Twist covariance, decoupled and contracted\\
 \hspace*{15mm}$R$-matrices\dotfill\pageref{subprop_1a}~\&~\pageref{subprop_1b}~\&~\pageref{subprop_1c}
\item[]Proposition~\ref{subprop_2a} and~\ref{subprop_2b} ~~$d$-indices\dotfill\pageref{subprop_2a}
\item[]Proposition~\ref{subprop_3a},~\ref{subprop_3b} and~\ref{subprop_3c}~~$\D$-indices\dotfill\pageref{subprop_3a}
\item[]Proposition~\ref{subprop_4a} and~\ref{subprop_4b}~~(Reduced) $\D$-incidence matrix\dotfill\pageref{subprop_4a}
\item[]Proposition~\ref{prop_5}~~Triangularity\dotfill\pageref{prop_5}
\item[]Proposition~\ref{subprop_6a}, \ref{subprop_6b} and \ref{subprop_6c}~~Comparability of $\D$-classes\dotfill\pageref{subprop_6a}
\item[]Proposition~\ref{prop_7}~~Block upper-triangularity\dotfill\pageref{prop_7}
\item[]Proposition~\ref{prop_8}~~Inside a $\mc D$-class\dotfill\pageref{prop_8}
\item[]Propositions~\ref{subprop_9a},~\ref{subprop_9b} and~\ref{subprop_9c}~~($\nD$-)multiplicative covariance\dotfill\pageref{subprop_9a}~\&~\pageref{subprop_9c}
\item[]Proposition~\ref{prop_10}~~Decoupling proposition\dotfill\pageref{prop_10}
\item[]Propositions~\ref{subprop_11a} and~\ref{subprop_11b}~~Continuity propositions\dotfill\pageref{subprop_11a}~\&~\pageref{subprop_11b}
\item[]Proposition~\ref{prop_12}~~Scaling proposition\dotfill\pageref{prop_12}
\item[]Proposition~\ref{prop_13}~~Re-parametrization of variables\dotfill\pageref{prop_13}
\item[]Proposition~\ref{prop_14}~~Commuting operators\dotfill\pageref{prop_14}
\item[]Proposition~\ref{subprop_15a} and \ref{subprop_15b}~~(Weak) Hecke-type $R$-matrices\dotfill\pageref{subprop_15a}~\&~\pageref{subprop_15b}
  \end{itemize}

\subsection*{List of lemmas}
  \begin{itemize} \itemsep=0pt
\item[]Lemma~\ref{lem_1}~~Order of $\D$-classes\dotfill\pageref{lem_1}
\item[]Lemmas~\ref{sublem_2a} and~\ref{sublem_2b}~~Dependences and forms of the $\D$-coef\/f\/icients\dotfill\pageref{sublem_2a}
\item[]Lemmas~\ref{sublem_3a} and~\ref{sublem_3b}~~Dependences and forms of the $d$-coef\/f\/icients\dotfill\pageref{sublem_3a}~\&~\pageref{sublem_3b}
\item[]Lemmas~\ref{sublem_4a} and~\ref{sublem_4b}~~Multiplicative and additive shifts\dotfill\pageref{sublem_4a}~\&~\pageref{sublem_4b}
  \end{itemize}

\subsection*{Acknowledgements}
This work was supported by CNRS, Universit\'e de Cergy Pontoise, and ANR Project DIADEMS (Programme Blanc ANR SIMI 1 2010-BLAN-0120-02). We wish to thank Dr.\ Thierry  Gobron for interesting discussions and comments on the structure of $\D$-incidence matrices, and Professor Laszlo Feh\'er for pointing out to us reference~\cite{Feh_1}. Finally we thank the anonymous referees for their careful study and their relevant contributions to improve the paper.

\pdfbookmark[1]{References}{ref}
\LastPageEnding

\end{document}